\documentclass[notitlepage,prd,twocolumn,showpacs,preprintnumbers,nofootinbib,superscriptaddress,floatfix,tightenlines, longbibliography]{revtex4-1}
\usepackage{mathrsfs}
\usepackage{amssymb}
\usepackage{amsmath}
\usepackage{graphicx}
\usepackage{amsfonts,amsbsy}
\usepackage{nicefrac}
\usepackage{xcolor}
\usepackage{array}
\usepackage{multirow}
\usepackage{siunitx}
\usepackage[breaklinks,colorlinks,urlcolor=blue,citecolor=citcolor,linkcolor=lcolor]{hyperref}
\usepackage{soul}

\definecolor{lcolor}{rgb}{0.5,0,0}
\definecolor{citcolor}{rgb}{0,0.3,0.0}
\definecolor{ao(english)}{rgb}{0.0, 0.5, 0.0}
\usepackage{comment}

\newcommand{\mrm}{\mathrm}

\newcommand{\dA}{d_{\mrm{A}}}
\newcommand{\CA}{C_{\mrm{A}}}
\newcommand{\CR}{C_{\mrm{R}}}
\newcommand{\NC}{N_{\mrm{C}}}
\newcommand{\nf}{n_{\mrm{f}}} 
\newcommand{\CF}{C_{\mrm{F}}}
\newcommand{\dF}{d_{\mrm{F}}}

\newcommand{\fig}{Fig.~}

\newcommand{\eq}{Eq.~}
\newcommand{\eqs}{Eqs.~}
\newcommand{\se}{Sec.~}
\newcommand{\ses}{Secs.~}
\newcommand{\re}{Ref.~}
\newcommand{\res}{Refs.~}
\newcommand{\tab}{Table~}
\newcommand{\app}{Appendix~}
\newcommand{\apps}{Appendices~}

\newcommand{\qperp}{q_\perp}

\newcommand{\lperp}{\Lambda_\perp} 
\newcommand{\qhat}{\hat q}
\newcommand{\qhatf}{\qhat_{\mathrm{f}}}
\newcommand{\qhatff}{\qhat_{\mathrm{ff}}}
\newcommand{\kmin}{k_{\mathrm{min}}}
\newcommand{\kmax}{k_{\mathrm{max}}}
\newcommand{\Gammael}{\Gamma_{\text{el}}}
\newcommand{\vecp}{\vec p}
\newcommand{\vb}{\vec}
\renewcommand{\vec}[1]{\mathrm{\mathbf{#1}}}
\newcommand{\dd}[2][]{\mathrm d^{#1}{#2}\,} 
\newcommand{\dv}[2][]{\frac{\dd{#1}}{\dd{#2}}}
\newcommand{\pdv}[2][]{\frac{\partial{#1}}{\partial{#2}}}

\newcommand{\qmin}{q_{\mathrm{min}}}
\newcommand{\qmax}{q_{\mathrm{max}}}
\newcommand{\Ejet}{E_{\mathrm{jet}}}

\newcommand{\Conetwo}{\mathcal {C}^{1\leftrightarrow 2}}
\newcommand{\Ctwotwo}{\mathcal{ C}^{2\leftrightarrow 2}}
\newcommand{\Cexp}{\mathcal C^{\mathrm{exp}}}

\newcommand{\NOg}{{N_+}} 
\newcommand{\NOq}{{N_-}} 
\newcommand{\Teps}{T_\varepsilon}

\newcommand{\Mhtl}{{M_{\mathrm{HTL}}}} 
\newcommand{\Mxi}{{M_\xi}} 
\newcommand{\tildeMhtl}{{\tilde{M}_{\mathrm{HTL}}}} 
\newcommand{\tildeMxi}{{\tilde{M}_\xi}} 

\newcommand{\ILO}{I^{\mathrm{LO}}}
\newcommand{\INLO}{I^{\mathrm{NLO}}}

\newcommand{\qhatLO}{\qhat^{\mathrm{LO}}}
\newcommand{\qhatNLO}{\qhat^{\mathrm{NLO}}}

\newcommand{\qhattherm}{\qhat^{\mathrm{therm}}}
\newcommand{\qhatftherm}{\qhatf^{\mathrm{therm}}}
\newcommand{\qhatfftherm}{\qhatff^{\mathrm{therm}}}
\newcommand{\qhatffthermimproved}{\qhat_{\mathrm{ff, im}}^{\mathrm{therm}}}
\newcommand{\qhatffimproved}{\qhat_{\mathrm{ff, im}}}
\newcommand{\qhatimproved}{\qhat_{\mathrm{im}}}

\newcommand{\phipq}{\phi_{pq}}
\newcommand{\phikq}{\phi_{kq}}

\newcommand{\PiLR}{{\Pi^{00}_R}}
\newcommand{\tildePiLR}{{\tilde{\Pi}^{00}_R}}
\newcommand{\PiTR}{{\Pi^{T}_R}}
\newcommand{\GLR}{{G^{00}_R}}
\newcommand{\GTR}{{G^{T}_R}}
\newcommand{\RE}{\mathrm{Re}}
\newcommand{\IM}{\mathrm{Im}}

\newcommand{\tform}{t^{\mathrm{form}}}

\newcommand{\lxi}{{\Lambda_\omega}}
\newcommand{\otild}{{\tilde\omega}}

\makeatletter
\newcolumntype{M}[1]{>{\centering\arraybackslash}m{#1}}
\makeatother

\begin{document}

\title{Jet quenching parameter in QCD kinetic theory}

\author{K.~Boguslavski} 
\affiliation{Institute for Theoretical Physics, TU Wien, Wiedner Hauptstrasse 8-10, 1040 Vienna,
Austria}

\author{A.~Kurkela} 
\affiliation{Faculty of Science and Technology, University of Stavanger, 4036 Stavanger, Norway}

\author{T.~Lappi} 
\affiliation{Department of Physics, University of Jyv\"{a}skyl\"{a}, P.O.~Box 35, 40014 University of Jyv\"{a}skyl\"{a}, Finland}
\affiliation{Helsinki Institute of Physics, P.O.~Box 64, 00014 University of Helsinki, Finland}

\author{F.~Lindenbauer} 
\email{florian.lindenbauer@tuwien.ac.at}
\affiliation{Institute for Theoretical Physics, TU Wien, Wiedner Hauptstrasse 8-10, 1040 Vienna,
Austria}

\author{J.~Peuron} 
\affiliation{Department of Physics, University of Jyv\"{a}skyl\"{a}, P.O.~Box 35, 40014 University of Jyv\"{a}skyl\"{a}, Finland}
\affiliation{Helsinki Institute of Physics, P.O.~Box 64, 00014 University of Helsinki, Finland}
\affiliation{Dept. of Astronomy and Theoretical Physics, Lund University, S\"{o}lvegatan 14A, S-223 62 Lund, Sweden}

\begin{abstract}
We study the jet quenching parameter $\qhat$ in a non-equilibrium plasma using the QCD effective kinetic theory.
We discuss subleading terms at large jet momentum $p$, show that our expression for $\qhat$ reproduces thermal results at small and large transverse momentum cutoffs for infinite $p$ and construct an interpolation between these limits to be used in phenomenological applications. 
Using simple non-equilibrium distributions that model pertinent features of the bottom-up thermalization scenario, we analytically assess how anisotropy, under- or overoccupation affect the jet quenching parameter.
Our work provides more details on the $\qhat$ formula used in our preceding work [{Phys.~Lett.~B 850 (2024) 138525}] and sets the stage for further numerical studies of jet momentum broadening in the initial stages of heavy-ion collisions from QCD kinetic theory.
\end{abstract}

\maketitle


\tableofcontents

\section{Introduction}
\label{sec_introduction}
The goal of relativistic heavy-ion collisions performed at RHIC and the LHC is to achieve a better understanding of the quark-gluon plasma (QGP) that is created in the collisions. In particular, much attention has been directed to the non-equilibrium initial stages of this state of matter. 
While in general the theoretical description of the early stages requires a solution to non-perturbative non-equilibrium quantum field theory, a description based on weak-coupling methods becomes appropriate in the limit of (asymptotically) high collision energies.
In this limit, the initial stages of central heavy-ion collisions follow the bottom-up scenario of \cite{Baier:2000sb}.
The time-evolution of the bottom-up scenario can be described through a set of different effective descriptions that capture the important aspects of the pre-equilibrium dynamics
after the collision. 
After a so-called glasma phase that can be described using a classical field description \cite{Gelis:2010nm}, 
the system can be described using quasi-particle degrees of freedom \cite{Kurkela:2016mhu, Boguslavski:2018beu, Boguslavski:2021buh, Boguslavski:2021kdd} within the QCD effective kinetic theory framework \cite{Arnold:2002zm, Baier:2000sb}, in which all leading order scattering processes are properly taken into account. 
It smoothly connects to relativistic hydrodynamics \cite{Kurkela:2015qoa, Romatschke:2017ejr}, which is naturally encompassed in kinetic theory~\cite{Denicol:2012cn, Rezzolla:2013, Ambrus:2022qya} and is the standard paradigm to describe the later spacetime evolution of the QGP.


The medium modification of jets during the non-equilibrium evolution has recently attracted much attention \cite{Citron:2018lsq, Ipp:2020mjc, Ipp:2020nfu, Carrington:2021dvw, Carrington:2022bnv, Avramescu:2023qvv, Andres:2022bql, Boguslavski:2023alu, Hauksson:2023tze}. The jet quenching parameter 
\begin{align}
    \qhat = \dv[\langle \qperp^2\rangle]{L}\label{eq:def_qhat1}
\end{align}
is an important quantity that determines 
the rate of change of transverse momentum of a hard parton traveling through a medium. This 
parameter $\qhat$ is used in many models to quantify medium effects on jet energy loss to compare with experimental data \cite{Baier:1998kq, Baier:1998yf, Zakharov:2000iz, Arnold:2008iy, Andres:2019eus, Adhya:2019qse,Huss:2020dwe, Huss:2020whe}.
In kinetic theory, the jet quenching parameter is determined by the elastic scattering collision kernel 
\begin{equation} 
C(\vb \qperp)=(2\pi)^2\frac{\dd\Gammael}{\dd[2]{\vb\qperp}},
\end{equation}
where $\Gammael$, the rate of elastic collisions, encodes the probability of the leading jet-parton to receive a transverse momentum kick with $\vb \qperp$ per unit time \cite{Arnold:2008vd, Caron-Huot:2008zna}. It is related to $\qhat$ via
\begin{align}
    \qhat = \int\frac{\dd[2]{\vb \qperp}}{(2\pi)^2} \, \qperp^2\,  C(\vb \qperp).\label{eq:def_qhat2}
\end{align}
The purpose of this paper---and a closely related paper \cite{Boguslavski:2023alu}---is to study the jet quenching parameter $\qhat$ far from equilibrium within QCD effective kinetic theory.

The jet quenching parameter has been calculated before analytically for a weakly coupled plasma in thermal equilibrium at leading \cite{Aurenche:2002pd, Arnold:2008vd} and next-to-leading order \cite{Caron-Huot:2008zna} in weak coupling perturbative QCD (pQCD). At strong couplings using AdS/CFT \cite{Maldacena:1997re, Witten:1998qj}, calculations have been performed at leading \cite{Liu:2006ug} and next-to-leading order \cite{Zhang:2012jd} in the inverse coupling. Computations also exist in lattice QCD \cite{Kumar:2020wvb}, dimensionally reduced EQCD \cite{Moore:2021jwe}, quasiparticle models in thermal equilibrium \cite{Song:2022wil, Grishmanovskii:2022tpb}, QCD effective kinetic theory with an equilibrium background \cite{Ghiglieri:2015ala, Schlichting:2020lef, Mehtar-Tani:2022zwf}. There are also extractions from experimental data by e.g. the JET \cite{JET:2013cls} and JETSCAPE \cite{JETSCAPE:2021ehl} collaborations. While these calculations have been done assuming (at least local) thermal equilibrium, which is only reached at later stages, recent studies include modifications of the jet evolution due to inhomogeneous, anisotropic, and flowing systems \cite{Romatschke:2004au, Romatschke:2006bb, Dumitru:2007rp, Hauksson:2021okc, Andres:2022ndd, Barata:2022krd, Barata:2022utc, Barata:2023qds, Kuzmin:2023hko}, and the extraction of the jet quenching parameter $\qhat$ during the glasma stage at the earliest times of heavy-ion collisions \cite{Carrington:2021dvw, Carrington:2022bnv, Ipp:2020mjc, Ipp:2020nfu, Avramescu:2023qvv}. 
The impact of pre-equilibrium dynamics on the related case of heavy-quark diffusion has also sparked interest in the field~\cite{Das:2015aga, Mrowczynski:2017kso, Sun:2019fud, Boguslavski:2020tqz, Carrington:2022bnv, Ruggieri:2022kxv, Avramescu:2023qvv, Boguslavski:2023fdm}.

In \re \cite{Boguslavski:2023alu} we have very recently extracted 
$\qhat$ during the anisotropic initial stages of the kinetic bottom-up scenario \cite{Baier:2000sb} using QCD effective kinetic theory (EKT) \cite{Arnold:2002zm, Kurkela:2015qoa}. We found that it smoothly connects the large values in the early glasma phase with the smaller values of the hydrodynamical evolution, is consistent with experimentally extracted values of $\qhat$ at late times, and leads to anisotropic jet quenching at early times. 
Our quantitative study of $\qhat$ goes beyond the parametric estimates of
\cite{Baier:2000sb, Kurkela:2011ti, Kurkela:2011ub, Berges:2013eia, Berges:2013fga}.

In this paper, we provide the explicit derivation of the leading-order formula for the jet quenching parameter $\qhat$ for an on-shell parton that 
we have used in our EKT simulations \cite{Boguslavski:2023alu}, and that encodes the anisotropy of the system. Our calculations, however, still neglect the effect of plasma instabilities by employing an isotropic approximation to the in-medium propagator.
It is valid for an arbitrary jet momentum and direction and for anisotropic particle distributions with azimuthal symmetry around the beam axis. Since $\qhat$ in the eikonal limit (infinite jet energy) is logarithmically ultraviolet divergent due to its Coulomb logarithm, our results are therefore often functions of an ultraviolet (UV) transverse momentum 
cutoff $\Lambda_\perp$. We discuss the behavior of our formula of $\qhat$ for 
large jet momentum and large UV cutoff, show explicitly that it reproduces the known analytic limits for small and large cutoffs in thermal equilibrium.  We also assess different (screening) approximations of the matrix elements that are also typically employed in EKT simulations of the time evolution and provide a new approximate form for $\qhat$ in thermal gluonic systems that interpolates between the analytic expressions.
We also discuss toy models of bottom-up  thermalization~\cite{Baier:2000sb}. Different stages of this scenario for passage from the initial state towards hydrodynamical evolution are characterized by either over- or underoccupation of gluonic modes, and by an anisotropy of the sytem related to the longitudinal expansion. Thus, to shed light on the effects of the pre-equilibrium stage on jets,  we calculate $\qhat$ for an effectively two-dimensional and a scaled thermal distribution, respectively.
Although we restrict ourselves to on-shell partons, our formula can also be used as an input for jet evolution models that include an initial large virtuality phase \cite{Cao:2021rpv,JETSCAPE:2023hqn}.

The paper is organized as follows. In \se\ref{sec:background}, we review the parts of the effective kinetic theory description of on-shell (massless) partons that we will need. In \se\ref{sec:kinetic_formula} we arrive at a formula of $\qhat$ that is useful for EKT simulations. 
We then apply it in \se \ref{sec:qhat_special_cases} to a thermal distribution and to toy models for bottom-up thermalization.
Finally, we conclude in \se \ref{sec:concl}. 
The Appendices contain details on the $\qhat$ formula, its derivation, properties and evaluation (\apps \ref{app:qhat_derivation_integral_boundaries} - \ref{app:monte_carlo}) and on the calculation of $\qhat$ in toy models (\app \ref{app:qhat_models}).


\section{Theoretical background and kinetic theory\label{sec:background}}

We use natural units $c=k_B=\hbar=1$, the mostly-plus metric convention, $\eta_{\mu\nu}=\text{diag}(-1,1,1,1)$, and denote 4-vectors with upper case letters, $Q^\mu=(\omega,\vb q)$, 3-vectors with bold upright symbols, e.g., $\vb q$, and similarly 2-vectors for transverse momenta $\vb \qperp$. A non-bold quantity denotes the length of the corresponding 3-vector, i.e., $q=|\vb q|$. For the analytic results, we leave the number of colors $\NC$ and the number of quark flavors $\nf$ arbitrary; for the numerical results, we specialize to $\NC=3$ for QCD and $\nf=0$, i.e., numerically, we consider a purely gluonic system. For the axis of our coordinate system, we use the letters $x$, $y$, and $z$.

We perform our calculation within the leading-order QCD effective kinetic theory formulated in Ref.~\cite{Arnold:2002zm} that describes the quark-gluon plasma in terms of phase-space densities or quasi-particle distribution functions $f_s(\vb p)$ for the particle species $s$. In general, $f_s(\vb p)$ depend on time and their time-evolution is governed by the Boltzmann equation
\begin{align}
-\pdv[f_{s}(\vb p)]{\tau}=\Conetwo_s[f_{s}(\vb p)]+ \Ctwotwo_s[f_{s}(\vb p)]+ \Cexp[f_s({\vb p})],\label{eq:boltzmann_equation}
\end{align}
where $\Ctwotwo$ includes elastic collisions, $\Conetwo$ summarizes inelastic interactions, 
and $\Cexp = - \frac{p_z}{\tau} \pdv[f_{\vb p}]{p_z}$ accounts for the longitudinal expansion of the plasma along the beam direction $z$ \cite{Mueller:1999pi}.
Here, we assume that the medium is homogeneous in the transverse plane and we are interested in the mid-rapidity region, where we assume boost-invariance in the longitudinal direction. Then, our quantities do not depend on the spatial coordinate $\vb x$.
\footnote{However, we note that the EKT approach \cite{Arnold:2002zm} is more general and allows for spatially inhomogeneous systems \cite{Keegan:2016cpi, Kurkela:2018vqr, Kurkela:2021ctp}.} 
Additionally, since the particle production is isotropic, and the longitudinal expansion singles out only one preferred direction, the distribution function is independent of the azimuthal angle,
and we often write $f(p,\cos\theta_p):=f(\vb p)$, with $\theta_p$ the angle between the $z$ axis and $\vb p$.

To calculate the jet quenching parameter $\qhat$, we only need the elastic collision term $\Ctwotwo$. It consists of a loss term and a gain term and reads
\begin{align}
\Ctwotwo_a[f_{s}(\vb p)]&=\frac{1}{4|\vb{p}|\nu_a}\sum_{bcd}\int_{\vb{kp'k'}}\left|\mathcal M^{ab}_{cd}(\vb{p},\vb{k};\vb{p'}\vb{k'})\right|^2 \nonumber\\
&\quad\times(2\pi)^4\delta^4(P+K-P'-K')\label{eq:c22_first}\\
&~~~\times\Big\{f_{a}(\vb p)f_{b}(\vb k)\left[1\pm f_{c}(\vb p')\right]\left[1\pm f_{d}(\vb k')\right]\nonumber\\
&~~\quad - f_{c}(\vb p')f_{d}(\vb k')\left[1\pm f_{a}(\vb p)\right]\left[1 \pm f_{b}(\vb k)\right]\Big\} \nonumber .
\end{align}
The number of spin times color states for a given particle species $a$ is denoted by $\nu_a = 2d_a$, where $d_a$ is the dimension of its representation. The external particles in the $2\leftrightarrow 2$ scattering are ultra-relativistic and on-shell, i.e., $P^2=0$ or $P^0=|\vb p|=p$, and the integral measure is defined as
\begin{align}
\int_{\vb k}:=\int\frac{\dd[3]{\vb k}}{(2\pi)^3 2k}.
\end{align}

The matrix elements $\left|\mathcal M^{ab}_{cd}\right|^2$ correspond to elastic 2-particle scattering processes summed over spins and colors of all incoming and outgoing particles. They are calculated in pQCD and can be found in \cite{Arnold:2002zm}. However, due to medium effects, for soft gluon or fermion exchange these have to be modified by including the Hard Thermal Loop (HTL) self-energy. In practice, one often uses an isotropic screening approximation \cite{AbraaoYork:2014hbk, Kurkela:2015qoa, Kurkela:2018oqw, Du:2020dvp, Du:2020zqg}. Screening is also important for $\qhat$, and we will discuss this approximation and the HTL-screened matrix elements in more detail in \se\ref{sec:screening}.

We will need several observables that can also be computed in a kinetic theory.
The energy density $\epsilon$ can be calculated by weighing the distribution function for a specific species with its energy,
\begin{align}
	\epsilon = \sum_s\nu_s\int\frac{\dd[3]{\vb p}}{(2\pi)^3} p f_s(\vb p).\label{eq:kinetic_energy_density}
\end{align}
We will frequently encounter the Debye mass $m_D$, which is an effective gluonic screening mass given by
\begin{align}
	m_D^2 = \sum_s 8d_sg^2\frac{C_s}{\dA}\int_{\vb p}f_s(\vb p).\label{eq:debye_mass_general}
\end{align}
In thermal equilibrium with $\nf$ quark flavors these definitions reduce to
\begin{subequations}
\begin{align}
    \epsilon(T)&=\frac{\pi^2T^4}{60}\left(4\dA + 7\nf\dF\right), \label{eq:kinetic_energy_T}\\
    m_D^2(T)&=\frac{g^2T^2}{3}\left(\NC + \frac{\nf}{2}\right).
    \label{eq:mD_T}
\end{align}
\end{subequations}
The indices $F$ and $A$ that we used in these expressions denote the fundamental and adjoint representation of fermions and gluons, respectively. In particular, their dimensions are $\dF = \NC$ and $\dA = \NC^2 - 1$ with the number of colors $\NC = 3$. Similarly, the quadratic Casimir reads $\CF=\dA/(2\NC)$ and $\CA=\NC$.

\begin{figure}
    \centering
    \includegraphics[width=\linewidth]{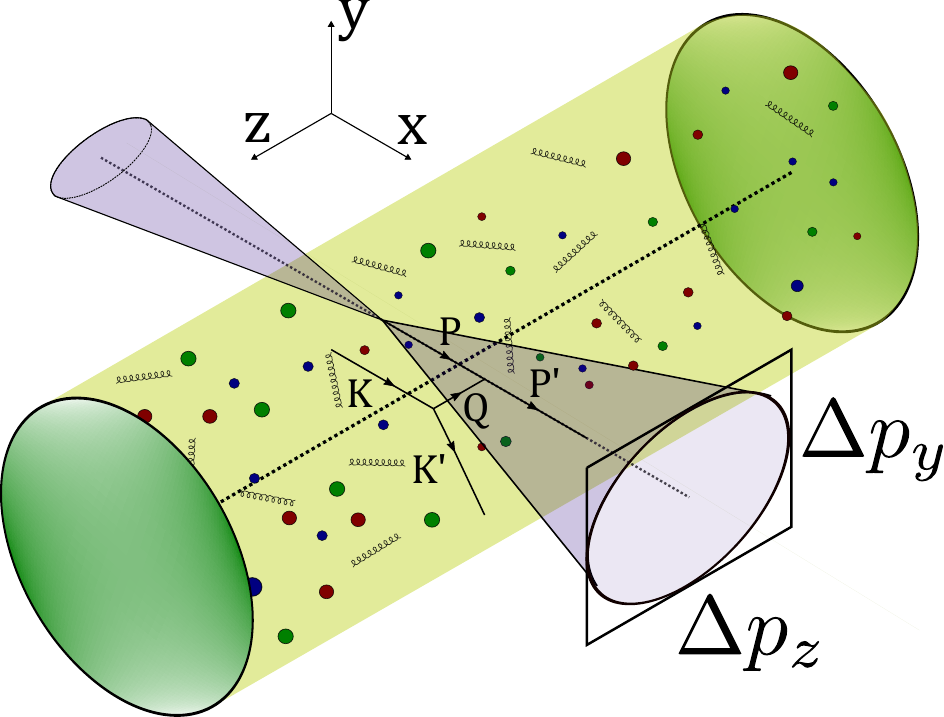}
    \caption{Geometry of jet quenching.
    The beam axis is the $z$ direction, the jet parton moves perpendicularly to the $y$ axis, in this graphic along the $x$ direction.}
    \label{fig:heavy-ion_overview}
\end{figure}

\section{Kinetic theory formula for $\qhat$\label{sec:kinetic_formula}}

In this section, we give a derivation of $\qhat$ in the quark-gluon plasma with fermionic and gluonic degrees of freedom. Our final formula is given by \eq \eqref{eq:qhat_formula} with the details described in \se \ref{sec:qhat-formula-finite-p} for a finite jet momentum, and those in \se \ref{sec:pinf_formula} for infinite momentum $p \to \infty$. The corresponding matrix elements and their screening prescriptions are discussed in detail in \ses \ref{sec:matrix-el} - \ref{sec:pinf_formula}. 

The geometry of the problem we consider here is motivated by its application to bottom-up thermalization and illustrated in \fig\ref{fig:heavy-ion_overview}, but is more generally applicable to any system with azimuthal symmetry in momentum space. We take $z$ to be the anisotropy direction (e.g., the beam axis) and the momentum of the jet to be within the $x-z$ plane.
We denote its polar angle $\theta_p$, which is usually set to $\theta_p=\pi/2$ for jet propagation perpendicular to the beam axis. 
We emphasize that we are not actually following the trajectory of a jet and letting it be deflected by the medium. Rather we measure the momentum transfer rate that such a jet would receive were it allowed to interact, performing the calculation for a jet momentum that is constant in magnitude and angle during the time evolution of the system.

\subsection{Derivation of $\qhat(p)$ from the scattering rate}

We first consider the leading hard parton of a jet with a large but finite momentum $\vb p$, which should be much larger than all momentum scales of the plasma such that $f(\vb p)\approx 0$.
We start with the definition \eqref{eq:def_qhat2} and extend it to account for anisotropies 
\begin{align}
	\qhat^{ij}(p)=\int\dd[2]{\vb\qperp}\,\qperp^i\qperp^j \frac{\dd\Gammael}{\dd[2]\qperp},\label{eq:def_qhat}
\end{align}
where $\Gammael$ is the rate of elastic collisions of a highly energetic jet parton with plasma particles and $\qperp$ is the transferred transverse momentum in such a single collision. For total transverse momentum broadening $\qhat$ we need to sum over the directions perpendicular to the jet direction, 
\begin{align}
    \label{eq:qhat_1122}
    \qhat = \qhat^{11}+\qhat^{22}\,.
\end{align}
It measures the average transverse momentum transfer squared to the jet particle per unit time (or length, as we are using units in which $c=1$). Here, $1$ and $2$ denote the directions perpendicular to the jet.

We can also think of the elastic scattering rate $\Gammael$ as the decay rate of a particle with a fixed momentum $\vb p$. It can be obtained~\cite{Laine:2016hma} by identifying it with the loss term in the elastic collision kernel \eqref{eq:c22_first}  that describes scatterings out of the state $\vb p$, leading to
\begin{align}
		\Gammael&=\frac{1}{4p\nu_a}\sum_{bcd}\int_{\vec k\vec p'\vec k'}(2\pi)^4\delta^4(P+K-P'-K')\label{eq:decay_rate}\\
			&\quad\times\left|\mathcal M^{ab}_{cd}(\vecp, \vec k;\vec p',\vec k')\right|^2f_b(\vec k)\left[1\pm f_d(\vec k')\right]\left[1\pm f_c(\vec p')\right]. \nonumber
\end{align}
{This formula is also valid out of equilibrium, provided that the spectral function, which describes the relevant excitations, is very narrow. In particular, the duration of the scattering process should be much smaller than the typical time between collisions.}
{
In fact, the same conditions as for QCD kinetic theory }{\cite{Arnold:1997gh, Arnold:2002zm}} {are required, which we shall briefly summarize here for the reader's benefit:
A large scale separation between the medium-dependent effective screening masses (such as the Debye mass {$m_D$} from Eq.~{\eqref{eq:debye_mass_general}}) and the momenta of the relevant excitations {$p_{\mathrm{hard}}$} is required,} $p_{\mathrm{hard}}\gg m_D$.
{Additionally, the effective masses should be large compared to the quark masses and {$\Lambda_{\mathrm{QCD}}$}, as well as to the small-angle scattering rate {$\tau^{-1}_{\mathrm{soft}}$}. Although Eq.~{\eqref{eq:decay_rate}} does not yet specify the concrete form of the matrix element, we assume that no other contributions need to be taken into account. For instance, the presence of plasma instabilities might lead to additional contributions to the scattering rate, which we neglect.
Additionally, the distribution functions {$f(\vb p)$} should not vary significantly with {$\mathcal O(m_D)$} changes in momentum, and they should not be non-perturbatively large for the typical momenta,}
{$f(p_{\mathrm{hard}}) \ll 1/\lambda$}.

{Equation {\eqref{eq:decay_rate}}} is symmetric under the exchange of the outgoing particles $\vecp' \leftrightarrow \vec k'$ and $c\leftrightarrow d$, but the inclusion of $\qperp^i\qperp^j$ as in Eq.~\eqref{eq:def_qhat} breaks this symmetry. We choose to define the harder outgoing particle to be the jet particle and label it $c$ with momentum $p'$, which is also done in previous studies, as e.g., in \cite{Arnold:2008vd, Caron-Huot:2008zna, He:2015pra} as well. In Ref.~\cite{Ghiglieri:2015ala} this arises naturally if one only considers soft momentum exchange processes.
This choice implies that $p'>k'$ is always valid.
With this we can rewrite Eq.~\eqref{eq:decay_rate} with a step function $\theta(p'-k')$ and an additional factor of $2$ using the identity
\begin{align}
\int_{\vb k'\vb p'} g(k',p')
= 2\int_{\vb k'\vb p'}g(k',p')\theta(p'-k'),
\end{align}
for symmetric functions $g(p',k')=g(k',p')$. We note that this leads to 
more matrix elements than in Ref.~\cite{Arnold:2002zm} because we now treat processes like $qg\leftrightarrow gq$ differently than $qg\leftrightarrow qg$. We will discuss this new complication in more detail and list the required matrix elements explicitly in \se\ref{sec:matrix-el}.

Due to the large jet momentum, we have
$f(\vecp')=0$, and with $p'>k'$ we obtain
\begin{align}
\hat q^{ij} &= \frac{1}{2p\nu_a}\sum_{bcd}\int_{\substack{\vb k\vb p'\vb k'\\p'>k'}}\qperp^i\qperp^j(2\pi)^4\delta^4(P+K-P'-K')\nonumber\\
&\qquad\times\left|\mathcal M_{cd}^{ab}(\vb p,\vb k;\vb p',\vb k')\right|^2 f^b(\vb k)\left[1\pm f^d(\vb k')\right]. \label{eq:qhat_general}
\end{align}
In \app\ref{app:qhat_derivation_integral_boundaries}, we show that kinematically one obtains the following restrictions of the integration variables
\begin{align}
    |\omega| < q, && p >\frac{q-\omega}{2}, && k >\frac{q+\omega}{2}, \label{eq:phase_space_relation_omega_q_k}
\end{align}
where $\vb q$ is the transferred momentum and $\omega$ the transferred energy, 
\begin{subequations}\label{eq:definition_Q}
  \begin{align}
\vec q=\vec p'-\vecp=\vec k-\vec k',\\
\omega = p'-p=k-k'.
\end{align}
\end{subequations}
Using these restrictions from energy-momentum conservation, 
we will reduce the phase space integration in \eq \eqref{eq:qhat_general} to five integrals in \se \ref{sec:qhat-formula-finite-p}, where we choose $k$, $q=|\vb q|$, $\omega$ and two angles as integration variables.
But before that, we describe the coordinate frames necessary for the angular part of the integral in the next section.

\subsection{Coordinate systems}
\label{sec:coordinate_systems}

\begin{figure*}
\centering
\includegraphics[width=\linewidth]{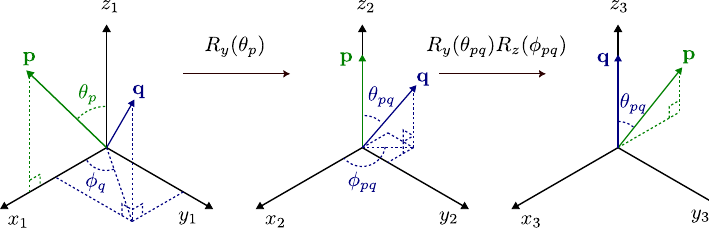}
\caption{The integration frames. {\em (Left:)} `Lab frame'. The jet momentum $\vb p$ lies in the $x-z$ plane.
{\em (Center:)} `p frame', obtained by rotating the `lab frame' around the $y$ axis, such that $\vb p$ points in the $z$ direction. 
{\em (Right:)} `q-frame'. Here, $\vb q$ points in the $z$ direction and $\vb p$ lies in the $x-z$ plane.
}
\label{fig:frames}
\end{figure*}

For the angular integrals, it is convenient to choose specific coordinate systems whose $z$-axis coincides with either $\vb p$ or $\vb q$. This is due to
our choice of the integration variables $k$, $q$ and $\omega$, which kinematically fixes the angles between $\vb p$ and $\vb q$, and between $\vb q$ and $\vb k$ by energy conservation, 
see \app\ref{app:qhat_derivation_integral_boundaries}. 
However, since the phase-space density $f$ in our kinetic program is stored in the `lab frame' where $z$ is the beam direction, 
we need to work out the precise rotations to connect it to the other integration frames. We denote the  vectors in this `lab frame' by a lower subscript $1$,
\footnote{Although we do not need the expressions for $\vb p$, $\vb q$ and $\vb k$ in this frame and in the following discussions, we list them here for completeness and convenience, since this illustrates the naming convention of the angles, out of which we will only later need $\phi_p$, $\phikq$ and $\phipq$.}
\begin{subequations}
\begin{align}
\vb p_1&=p(\sin\theta_p,0,\cos\theta_p),\\
\vb q_1&=q(\sin\theta_q\cos\phi_q,\sin\theta_q\sin\phi_q,\cos\theta_q),\\
\vb k_1&=k(\sin\theta_k\cos\phi_k,\sin\theta_k\sin\phi_k,\cos\theta_k).\label{eq:labframe_k}
\end{align}
\end{subequations}
Because of our azimuthal symmetry around the $z$ axis, we can always rotate this frame such that $\vb p $ lies in the $x-z$ plane.
For the angular integrals in Eq.~\eqref{eq:qhat_general} we need to choose appropriate coordinate systems, or frames, in which we perform them.
\footnote{In typical EKT implementations \cite{Kurkela:2015qoa, Kurkela:2018oqw, Kurkela:2018xxd, Du:2020dvp, Du:2020zqg}, the $\vb q$ integral 
is evaluated in the `lab frame' and all other integrals in a frame in which $\vb q$ points in the $z$ direction. In this case, because of the `wedge' function discretization \cite{AbraaoYork:2014hbk}, also an integral over $\vb p$ (in our case the jet momentum) is performed. Here, similarly to Ref.~\cite{Arnold:2003zc}, we choose to proceed differently.}
Here, the jet momentum defines a distinct direction. 
Therefore, we define a second frame, denoted by a subscript 2, in which $\vb p$ points in the $z$ direction and that is obtained via a rotation of the `lab frame' around the $y$ axis (see \fig\ref{fig:frames}). We refer to this frame as `p frame', 
\begin{subequations}
    \begin{align}
        \vb p_2&=p(0,0,1),\\
        \vb q_2&=q(\sin\theta_{pq}\cos\phi_{pq},\sin\theta_{pq}\sin\phi_{pq},\cos\theta_{pq}),\label{eq:pframe_q}\\
        \vb k_2&=k(\sin\theta_{pk}\cos\phi_{pk},\sin\theta_{pk}\sin\phi_{pk},\cos\theta_{pk}).
    \end{align}
\end{subequations}
In this `p frame' we perform the integral over $\vb q$.

The $\vb k$ integral is performed in a third frame, in which $\vb q$ points in the $z$ direction and $\vb p$ lies in the $x-z$ plane. We call this the `q frame' and denote it by a subscript 3:
\begin{subequations}
    \begin{align}
        \vb p_3&=p(\sin\theta_{pq},0,\cos\theta_{pq}),\label{eq:qframe_p}\\
        \vb q_3&=q(0,0,1), \label{eq:qframe_q}\\
        \vb k_3&=k(\sin\theta_{kq}\cos\phikq,\sin\theta_{kq}\sin\phikq,\cos\theta_{kq}). \label{eq:qframe_k}
    \end{align}
\end{subequations}

The components of the vectors transform between the frames according to the matrix relations
\begin{subequations}
\begin{align}
\vb v_2=A\vb v_1, && A&=R_y(\theta_p),\label{eq:def_A}\\
\vb v_3 = B\vb v_2, && B&=R_y(\theta_{pq})R_z(\phipq),\label{eq:def_B}
\end{align}
\end{subequations}
where $R_y(\alpha)$ and $R_z(\alpha)$ denote the matrices corresponding to a rotation with angle $\alpha$ around the $y$- and $z$-axis, respectively.
The transformation matrices read
\begin{subequations}
\begin{align}
A&=\begin{pmatrix}
\cos\theta_p&0&-\sin\theta_p\\
0&1&0\\
\sin\theta_p&0&\cos\theta_p
\end{pmatrix},\\
B&=\begin{pmatrix}
\cos\theta_{pq}\cos\phipq&\cos\theta_{pq}\sin\phipq & -\sin\theta_{pq}\\
-\sin\phipq&\cos\phipq&0\\
\cos\phipq\sin\theta_{pq}&\sin\theta_{pq}\sin\phipq&\cos\theta_{pq}
\end{pmatrix}.
\end{align}
\end{subequations}
For the calculation of $\qhat^{ij}$ we use the components $\qperp^{i}$ of $\vec q$ in the $p$-frame,
\begin{subequations}
\begin{eqnarray}
q^1 &= (\vb q_2)^1 \;&= q\sin\theta_{pq}\cos\phipq\\
q^2 &= (\vb q_2)^2 \;&= q\sin\theta_{pq}\sin\phipq\\
q^3 &= (\vb q_2)^3 \;&= q\cos\theta_{pq}\label{eq:qz_pframe}
\end{eqnarray}
\end{subequations}
In this way the components of $\hat q$ are defined relative to $\vec p$. Thus the components $1$, $2$ are perpendicular to $\vec p$ and  quantify the momentum broadening transverse to the jet.

Having taken the Dirac delta functions in \eq\eqref{eq:qhat_general} into account, 
we choose $\phipq$, $\phikq$, $k$, $\omega$, and $q$ as independent integration variables. 
Therefore, we need to express all other quantities in terms of them. The value of $\cos\theta_{pq}$, for example, is actually set by kinematic constraints (or the delta function in \app\ref{app:qhat_derivation_integral_boundaries}), as can be seen easily via $|\vec k'|^2=|\vec k-\vec q|^2=(k-\omega)^2$. Similarly,  together with $|\vec k' +\vec q|^2=(k'+\omega)^2$, we find
\begin{subequations}
\begin{align}
\cos\theta_{kq}&=\frac{\omega}{q}-\frac{\omega^2-q^2}{2kq},\\
\cos\theta_{pq}&=\frac{\omega}{q}+\frac{\omega^2-q^2}{2pq},\\
\cos\theta_{k'q}&=\frac{\omega}{q}+\frac{\omega^2-q^2}{2k'q}.
\end{align}
\end{subequations}
The distribution functions $f(\vb k)$ and $f(\vb k')$ in \eqref{eq:qhat_general} are numerically stored in the `lab-frame'. Thus we need a way to express their polar angles $\theta_k$ and $\theta_{k'}$ in terms of the integration variables. The azimuthal angles $\phi_k$ and $\phi_{k'}$ are not needed due to azimuthal symmetry.
Because we perform the $\vec k$-integral in the `q-frame' and need $\cos\theta_k$ in the `lab-frame', we need to work out their relation
via \eqref{eq:def_A} and \eqref{eq:def_B}, thus $\vec k_1=A^TB^T\vec k_3$. From $(\vec k_1)_z$ we can read off 
\begin{align}
&\cos\theta_k = \sin\phikq\sin\phipq\sin\theta_{kq}\sin\theta_p\\
& \quad -\cos\phikq\sin\theta_{kq}\left(\cos\phipq\cos\theta_{pq}\sin\theta_p+\cos\theta_p\sin\theta_{pq}\right)\nonumber\\
& \quad +\cos\theta_{kq}\left(\cos\theta_p\cos\theta_{pq}-\cos\phipq\sin\theta_p\sin\theta_{pq}\right),\nonumber
\end{align}
and a similar expression holds for $\cos\theta_{k'}$,
\begin{align}
&\cos\theta_{k'} = \sin\phikq\sin\phipq\sin\theta_{k'q}\sin\theta_p\\
& \quad -\cos\phikq\sin\theta_{k'q}\left(\cos\phipq\cos\theta_{pq}\sin\theta_p+\cos\theta_p\sin\theta_{pq}\right)\nonumber\\
& \quad +\cos\theta_{k'q}\left(\cos\theta_p\cos\theta_{pq}-\cos\phipq\sin\theta_p\sin\theta_{pq}\right).\nonumber
\end{align}
The azimuthal angle is $\phi_{k'q}=\phikq$ because $\vec k'=\vec k -\vec q$ and $\vec q$ points in the $z$ direction in the `q-frame'.

\subsection{Formula for $\qhat(p)$\label{sec:qhat-formula-finite-p}}

We are now ready to give the formula for the components of $\qhat$:
\begin{align}
    \qhat^{ij}=\frac{1}{2^9\pi^5\nu_a}\sum_{bcd}\int\dd{\Gamma}q^iq^j\frac{|\mathcal M^{ab}_{cd}|^2}{p^2}f_b(k,v_k)\left(1\pm f_d(k',v_{k'})\right)\,, \label{eq:qhat_formula}
\end{align}
where we use the abbreviation of $v_{\dots}=\cos\theta_{\dots}$.
The phase space integration measure can be written as a product of two angular integrals and three additional integrals that are different depending on the order of integration and integration variables,
\begin{align}
    \int\dd{\Gamma}&=\int_0^{2\pi}\dd{\phipq}\int_0^{2\pi}\dd{\phikq}\int\dd{\Gamma_3},\label{eq:qhat_phase_space_integral_splitting}
\end{align}
with the three equivalent versions (\app\ref{app:qhat_derivation_integral_boundaries})
\begin{subequations} \label{eq:qhat_finitep_measures}
\begin{align}
    \int\dd{\Gamma_3}&=\int_0^\infty\dd{k}\int_{-\frac{p-k}{2}}^k\dd{\omega}\int_{|\omega|}^{\text{min}(2p+\omega,2k-\omega)}\dd{q}, \label{eq:qhat_finitep_measure_qlast}\\
    \int\dd{\Gamma_3}&=\int_0^\infty\dd{k}\int_{0}^{\frac{k+p}{2}}\dd{k'}\int_{|k-k'|}^{\text{min}(p+p',k+k')}\dd{q}, \label{eq:qhat_finitep_measure_qlast2}\\
    \int\dd{\Gamma_3}&=\int_0^\infty \dd{q} \int^q_{\max\left(-q,q-2p,\frac{q-2p}{3}\right)}\!\!\!\dd{\omega}\int_{\frac{q+\omega}{2}}^{p+2\omega}\!\!\!\dd{k}. \label{eq:qhat_finitep_measure_klast}
\end{align}
\end{subequations}
The components $q^i$ of \eq \eqref{eq:qhat_formula} in the `p-frame' read
\begin{subequations}
\begin{align}
q^1 &= q\sqrt{1-v_{pq}^2}\cos\phipq\label{eq:qx_def},\\
q^2 &= q\sqrt{1-v_{pq}^2}\sin\phipq\label{eq:qy_def},\\
q^3 &= qv_{pq}\label{eq:qz_def}.
\end{align}
\end{subequations}
The angles between $\vb p$, $\vb q$, $\vb k$ and $\vb k'$ are then given by
\begin{subequations}
\begin{align}
v_{pq} &= \frac{\omega}{q} + \frac{t}{2pq}\label{eq:vpq},\\
v_{kq} &= \frac{\omega}{q} - \frac{t}{2kq}\label{eq:vkq}, \\
v_{k'q} &= \frac{\omega}{q} + \frac{t}{2k'q},
\end{align}
and the polar angles of $\vb k$ and $\vb k'$ in the `lab-frame' by
\begin{align}
v_k &= \sin\phikq\sin\phipq\sqrt{1-v_{kq}^2}\sqrt{1-v_p^2}\nonumber\\
&-\cos\phikq\sqrt{1-v_{kq}^2}\left(\cos\phipq v_{pq}\sqrt{1-v_p^2}+v_p\sqrt{1-v_{pq}^2}\right)\nonumber\\
&+v_{kq}\left(v_pv_{pq}-\cos\phipq\sqrt{1-v_p^2}\sqrt{1-v_{pq}^2}\right),\label{eq:vk_explicit}\\
v_{k'} &= \sin\phikq\sin\phipq\sqrt{1-v_{k'q}^2}\sqrt{1-v_p^2}\nonumber\\
&-\cos\phikq\sqrt{1-v_{k'q}^2}\left(\cos\phipq v_{pq}\sqrt{1-v_p^2}+v_p\sqrt{1-v_{pq}^2}\right)\nonumber\\
&+v_{k'q}\left(v_pv_{pq}-\cos\phipq\sqrt{1-v_p^2}\sqrt{1-v_{pq}^2}\right),
\end{align}
and
\begin{align}
k'&=k-\omega, \\
t &=\omega^2-q^2, \label{eq:mandelstam_t_qhat}\\
\begin{split}
s &= -\frac{t}{2q^2}\Big((p+p')(k+k')+q^2\\
	&\qquad\qquad-\sqrt{(4pp'+t)(4k'k+t)}\cos\phikq\Big),
\end{split} \label{eq:mandelstam_s_qhat}\\
\begin{split}
u &= \frac{t}{2q^2}\Big((p+p')(k+k')-q^2\\
	&\qquad\qquad-\sqrt{(4pp'+t)(4k'k+t)}\cos\phikq\Big),
\end{split}\label{eq:mandelstam_u_qhat} \\
p'&=p+\omega.
\end{align}
\end{subequations}
Recall that $\nu_a = 2d_R$, where $d_R$ is the dimension of the representation of the jet particle.
The upper sign in Eq.~\eqref{eq:qhat_formula} is to be used when the $d$ particle is a boson (gluon), and the lower sign if it is a fermion (quark). Here the Mandelstam variables $s$, $t$, and $u$ are defined as in Ref.~\cite{Arnold:2002zm} with respect to the momenta corresponding to the particles with labels $a,b,c,d$, 
\begin{align}
s=-(P+K)^2, && t=-(P'-P)^2, && u =-(K'-P)^2.\label{eq:def-mandelstam}
\end{align}
The expressions for $s,u$ as in Eqs.~\eqref{eq:mandelstam_s_qhat} and \eqref{eq:mandelstam_u_qhat} can also be found in \cite{Arnold:2003zc}.
Note that we defined the components $\qhat^{ij}$ with respect to the jet direction.
If the jet moves perpendicular to the beam-axis $z$ in the $x$-direction as in \fig \ref{fig:heavy-ion_overview}, then $v_p = \cos\theta_p = 0$ and $\qhat^{yy}=\qhat^{22}$ is the momentum broadening in the $y$ direction and $\qhat^{zz}=\qhat^{11}$ is the momentum broadening in the beam direction, which sum to the usual $\hat q = \hat q^{11} + \hat q^{22}$. We can also express momentum broadening along the jet direction, i.e., longitudinal momentum broadening, by $\qhat_L=\qhat^{33}$. If we replace $\qhat^i\qhat^j$ by $\omega$ in \eqref{eq:qhat_formula}, we obtain collisional energy loss.

\subsection{Symmetries of $\qhat^{ij}$}\label{sec:symmetries_of_qhat}

Obtaining the symmetries of the matrix $\qhat^{ij}$ is complicated by the fact that the angle $\phikq$ appears both in the matrix element (via $s$ and $u$), and in the distribution functions $f(\vb k)$ and $f(\vb k - \vb q)$ through $v_k$ and $v_{k'}$. They also depend on $\phipq$, which enters $q^i$. Nevertheless, in the case of a {\em spherically} symmetric phase space density $f(k)$ it is easy to see that 
\begin{align}
    \qhat^{12}=\qhat^{13}=\qhat^{23}=0\,, \qquad \qhat^{11}=\qhat^{22},
\end{align}
due to \eqref{eq:qx_def} and \eqref{eq:qy_def}.

For a phase space density that is {\em azimuthally} symmetric around the $z$-axis (beam direction), i.e., the most general case we are considering here with $f(k,v_k)$, we also find that
\begin{align}
    \qhat^{12}=\qhat^{23}=0\,.
\end{align}
If the jet is additionally moving in the $x$ direction, i.e., $v_p=0$, we also obtain $\qhat^{13} = 0$. For a jet moving in the beam direction, $v_p=1$, the quantity $v_k$ does not depend on $\phipq$ any longer and one has $\qhat^{13}=0$ as well. In summary, we have
\begin{align}
    \qhat^{13} = 0\,, \quad \text{if}~v_p=0~\text{or}~v_p=1\,.
\end{align}

The fact that $\qhat^{12}=\qhat^{23}=0$ can be seen by rewriting the angular integrals $\int_0^{2\pi}\dd{\phipq}\int_0^{2\pi}\dd{\phikq}g(\phipq,\phikq)=\int_{-\pi}^{\pi}\dd{\phipq}\int_{-\pi}^{\pi}\dd{\phikq}g(\phipq,\phikq)$ and then splitting the $\phipq$ integral into the integral from $(-\pi,0)$ and $(0,\pi)$ to arrive at
\begin{align*}
	&\int_0^{2\pi}\dd{\phipq}\int_0^{2\pi}\dd{\phikq}g(\phipq,\phikq)\\
&\quad=\int_0^{\pi}\dd{\phipq}\int_{-\pi}^{\pi}\dd{\phikq}\left[g(-\phipq,-\phikq)+g(\phipq,\phikq)\right]
\end{align*}
The angles $\phikq$ and $\phipq$ appear in $v_k$ and $v_{k'}$, which are not changed by simultaneously replacing $\phikq\to-\phikq$ and $\phipq\to-\phipq$. In the matrix element, $\phikq$ appears in $s$ and $u$ in the cosine argument, which is an even function. The only change happens in $q_2^2\to -q_2^{2}$, which results in $\qhat^{12}=\qhat^{23}=0$.

To see that $\qhat^{13}=0$ for $v_p=0$, we can look at $\phipq\to\phipq+\pi$, which, for $v_p=0$ changes $v_k\to-v_k$, but $f(k,-v_k)=f(k,v_k)$ and thus this only results in $q^1\to -q^1$. Thus we obtain $\qhat^{13}=0$.

\subsection{Matrix elements\label{sec:matrix-el}}

We started our derivation of $\qhat$ with the collision term $\Ctwotwo$ of Ref.~\cite{Arnold:2002zm} and thus started with the same matrix elements. They are symmetric under the exchanges $(abcd)\to(cdab)$, $(abcd)\to(bacd)$, and $(abcd)\to(abdc)$. Thus, the matrix element labelled there `$q_1g\leftrightarrow q_1 g$' also describes the processes `$q_1g \leftrightarrow g q_1$', `$gq_1 \leftrightarrow q_1 g$', and `$gq_1\leftrightarrow g q_1$'.

Due to our choice $p'>k'$, we break the symmetry of exchanging the outgoing particles, $(abcd)\to(abdc)$, which means that we now have to distinguish between `$q_1g\leftrightarrow q_1 g$' and `$q_1g \leftrightarrow g q_1$'.
This enlarges the number of matrix elements. We obtain them from \cite{Arnold:2002zm} by relabelling $p'\leftrightarrow k'$ and $c\leftrightarrow d$, 
which effectively means (see \eqref{eq:def-mandelstam})
\begin{align}
s\to s, && t\to u, && u\to t.
\end{align}
The resulting matrix elements are still symmetric under $(abcd)\to(badc)$ and are listed in \tab\ref{tab:qhat_matrix_el}. 

\subsection{Screening  using HTL}\label{sec:screening}
Screening becomes important when the Mandelstam variable $t$ becomes small $s \gg t \sim\mathcal O(m_D^2)$. This concerns the  underlined terms with inverse powers of $t$ in the matrix elements listed in \tab \ref{tab:qhat_matrix_el}, which need to be modified to account for medium screening \cite{Arnold:2002zm}.
Due to our our requirement $p'>k'$ we  could only have $|u|\ll s$  when $k\gg p$, which is highly suppressed by the fact that we are choosing $p$ to be hard and $k$ to be a medium particle. This we have also checked numerically. Thus,  unlike in Ref.~\cite{Arnold:2002zm}, we only need to consider for screening the  terms with $t$ in the denominator.

\begin{table*}
\begin{ruledtabular}
    \begin{tabular}{ M{0.2\linewidth} | m{0.65\linewidth} }
\vspace{5pt}
$ab\leftrightarrow cd$ & $\left|\mathcal M^{ab}_{cd}\right|^2/g^4$ \\[5pt] 
\hline 
$q_1q_2\leftrightarrow q_1q_2,$ & \multirow{4}{*}{$8\frac{\dF ^2\CF^2}{\dA }\left(\frac{s^2+u^2}{\underline{t^2}}\right)$}\\ $q_1\bar q_2\leftrightarrow q_1\bar q_2,$ &\\ $\bar q_1 q_2\leftrightarrow \bar q_1 q_2,$ &\\ $\bar q_1\bar q_2\leftrightarrow \bar q_1\bar q_2$ &  \\[10pt] 

$q_1q_2\leftrightarrow q_2q_1,$ & \multirow{4}{*}{$8\frac{\dF ^2\CF^2}{\dA }\left(\frac{s^2+t^2}{{u^2}}\right)$}\\ $q_1\bar q_2\leftrightarrow \bar q_2 q_1,$ &\\ $\bar q_1 q_2\leftrightarrow q_2\bar q_1,$ &\\ $\bar q_1\bar q_2\leftrightarrow \bar q_2\bar q_1$ &  \\[10pt] 
 
$q_1q_1\leftrightarrow q_1q_1,$ & \multirow{2}{*}{$8\frac{\dF ^2\CF^2}{\dA }\left(\frac{s^2+u^2}{\underline{t^2}}+\frac{s^2+t^2}{{u^2}}\right)+16 \dF \CF\left(\CF-\frac{\CA  }{2}\right)\frac{s^2}{tu}$}\\ $\bar q_1 \bar q_1\leftrightarrow \bar q_1 \bar q_1$ & \\ [10pt]
 
$q_1\bar q_1\leftrightarrow q_1\bar q_1$ &$8\frac{\dF ^2\CF^2}{\dA }\left(\frac{s^2+u^2}{\underline{t^2}}+\frac{t^2+u^2}{s^2}\right)+16 \dF \CF\left(\CF-\frac{\CA  }{2}\right)\frac{u^2}{st}$ \\[10pt]

$q_1\bar q_1\leftrightarrow \bar q_1 q_1$ &$8\frac{\dF ^2\CF^2}{\dA }\left(\frac{s^2+t^2}{{u^2}}+\frac{u^2+t^2}{s^2}\right)+16 \dF \CF\left(\CF-\frac{\CA  }{2}\right)\frac{t^2}{su}$ \\[10pt]

$q_1\bar q_1\leftrightarrow q_2\bar q_2,$ & \multirow{2}{*}{$8\frac{\dF ^2\CF^2}{\dA }\left(\frac{t^2+u^2}{s^2}\right)$}\\
$q_1\bar q_1\leftrightarrow \bar q_2 q_2$ &\\[10pt]

$q_1\bar q_1\leftrightarrow gg$ & $8 \dF  \CF^2\left(\frac{u}{\underline{\underline{t}}}+\frac{t}{{{u}}}\right) - 8\dF \CF \CA\left(\frac{t^2+u^2}{s^2}\right)$ \\[10pt] 

$q_1 g\leftrightarrow q_1 g,$ & \multirow{2}{*}{$-8\dF \CF ^2\left(\frac{u}{s}+\frac{s}{{{u}}}\right)+8\dF \CF \CA  \left(\frac{s^2+u^2}{\underline{t^2}}\right)$}\\
$\bar q_1 g\leftrightarrow \bar q_1 g$ & \\[10pt]

$q_1 g\leftrightarrow gq_1 ,$ & \multirow{2}{*}{$-8\dF \CF ^2\left(\frac{t}{s}+\frac{s}{\underline{\underline{t}}}\right)+8\dF \CF \CA  \left(\frac{s^2+t^2}{{u^2}}\right)$}\\
$\bar q_1 g\leftrightarrow g\bar q_1 $ & \\[10pt]

$gg\leftrightarrow gg$ & $16\dA \CA  ^2\left(3-\frac{su}{\underline{t^2}}-\frac{st}{u^2}-\frac{tu}{s^2}\right)$\\[5pt] 
\end{tabular}
\end{ruledtabular}
\caption{Matrix elements for $\hat q$, obtained from the matrix elements from Ref.~\cite{Arnold:2002zm} by breaking the symmetry of exchanging outgoing particles. They are obtained by replacing $c\leftrightarrow d$ and $t\leftrightarrow u$. Singly-underlined denominators indicate infrared-sensitive contributions from soft gluon exchange and double-underlined denominators from soft fermion exchange.
The group constants are given by $\dF =\CA=\NC$, $\dA = \NC^2-1$ and $\CF =\dA/(2\NC)$.}
\label{tab:qhat_matrix_el}
\end{table*}

We follow the prescription of Ref.~\cite{Arnold:2002zm} to include medium modifications by replacing
\footnote{It is easy to see that inserting the free propagator $G_0(Q)_{\mu\nu}=\frac{\eta_{\mu\nu}}{Q^2}$ yields $(s-u)^2/t^2$. As argued in Ref.~\cite{Arnold:2002zm}, due to the spin-independence of the matrix elements at leading order, one can use a theory with fictitious scalar quarks for the infrared screening of the matrix elements. Then this prescription arises naturally.} 
the singly underlined terms in the matrix elements in \tab \ref{tab:qhat_matrix_el} by
\begin{align}
    \label{eq:prescription_screening}
	M_0&=\frac{(s-u)^2}{t^2} \\
    &\to\left|G_R(P-P')_{\mu\nu}(P+P')^\mu (K+K')^\nu\right|^2 \equiv M_{\rm screen}, \nonumber
\end{align}
where $G_R$ denotes the retarded gluon propagator in the HTL approximation. 
It should be noted that, as also discussed in \cite{Arnold:2002zm}, for anisotropic systems, this prescription in general leads to instabilities \cite{Mrowczynski:1988dz, Romatschke:2003ms, Romatschke:2006bb, Hauksson:2021okc}. It is currently unknown how to properly treat these instabilities in kinetic theory. Note however that there is numerical evidence that suggests that the quantitative effect of the instabilities on the plasma evolution is less dramatic \cite{Berges:2013eia, Berges:2013fga} than expected from power counting arguments \cite{Kurkela:2011ub}.
Here, we will use two different approximations, such that these instabilities are not present: First, we will use the isotropic gluon propagator, which includes the isotropic HTL expressions for the self-energy. Second, we will use a simple screening prescription that is also commonly used in EKT simulations \cite{AbraaoYork:2014hbk}.

All the singly underlined terms in \tab\ref{tab:qhat_matrix_el} can be rewritten in terms of the same unscreened gluon propagator $M_0$,
\begin{align}
    \label{eq:s2u2Overt2}
    \frac{s^2+u^2}{t^2} = \frac{1}{2}+\frac{1}{2}M_0, && \frac{su}{t^2}=\frac{1}{4}-\frac{1}{4}M_0.
\end{align}
In the following, we will use different screening approximations for the retarded HTL propagator $M_{\rm screen}$.
First, we use the full isotropic HTL propagators, which can be expressed as
(see \app \ref{app:full_htl_matrix_el} for details)
\begin{align}
	\Mhtl=\frac{c_1^2}{A^2+B^2}+\frac{c_2^2}{C^2+D^2}-\frac{2c_1c_2(AC+BD)}{(A^2+B^2)(C^2+D^2)},\label{eq:full_HTL_finite_p_matrix_element_replacement}
\end{align}
where $A, B, C, D$ are obtained from the real and imaginary parts of the retarded HTL self-energies and are explicitly given by
\begin{subequations} \label{eq:parameters_for_full_HTL_matrix_element}
\begin{align}
    A&=q^2+m_D^2\left(1+\frac{\omega}{2q}\ln\frac{q-\omega}{q+\omega}\right),\\
    B&=-\frac{m_D^2\omega}{2q}\pi,\\
    C&=q^2-\omega^2+\frac{m_D^2}{2}\left(\frac{\omega^2}{q^2}+\left(\frac{\omega^2}{q^2}-1\right)\frac{\omega}{2q}\ln\frac{q-\omega}{q+\omega}\right),\\
    D&=\frac{\pi m_D^2\omega}{4q}\left(1-\frac{\omega^2}{q^2}\right),
\end{align}
\end{subequations}
and
\begin{subequations}
\begin{align}
    c_1&=(2p+\omega)(2k-\omega),\\
    c_2&=4pk\sin\theta_{pq}\sin\theta_{kq}\cos\phikq.
\end{align}
\end{subequations}
Note that, for isotropic distributions, 
the last term  in \eq~\eqref{eq:full_HTL_finite_p_matrix_element_replacement} can be dropped, since it is proportional to $\cos\phikq$ and will thus vanish in the angular integration of $\qhat$.

The doubly underlined terms in \tab\ref{tab:qhat_matrix_el} correspond to soft fermionic exchange. We do not consider them here explicitly because, as we will discuss in \se\ref{sec:pinf_formula}, they  are subleading in $1/p$.

\begin{figure}
	\centering
	\includegraphics[width=\linewidth]{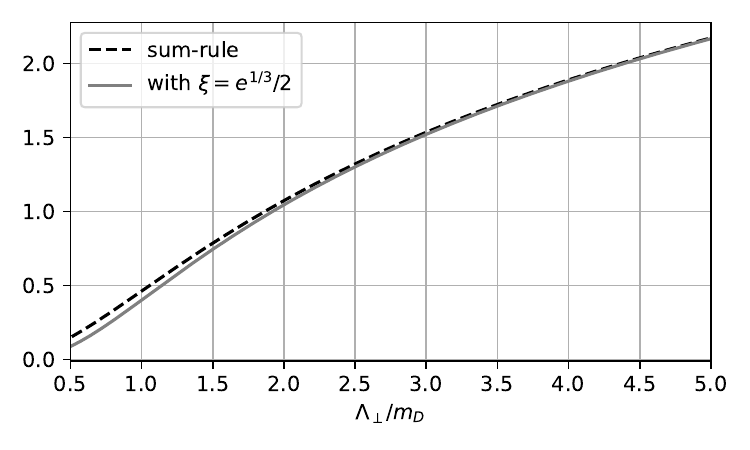}
	\caption{Shown are the HTL sum rule result on the left-hand side of \eq \eqref{eq:regularization_eq_to_solve} as a dashed curve and the values from the approximated expression on its right-hand side with the parameter $\xi$ given by \eqref{eq:xi_analytic} as a continuous line.
    The plot shows that for $\Lambda_\perp \gtrsim 4m_D$ the screening approximation of the full HTL matrix element provides accurate results.}
	\label{fig:regularization}
\end{figure}

There is an approximation to the isotropic HTL matrix element $\Mhtl$ that is commonly used in numerical simulations of the time evolution in EKT \cite{AbraaoYork:2014hbk, Kurkela:2015qoa, Kurkela:2018oqw, Kurkela:2018xxd, Du:2020dvp, Du:2020zqg} and that is also computationally more efficient. This approximation, which we will refer to as the $\xi$-screening prescription,  amounts to the replacement \cite{AbraaoYork:2014hbk}
\begin{align}
\Mhtl \to\Mxi=\frac{(s-u)^2}{t^2}\frac{q^4}{(q^2+\xi^2m_D^2)^2}
.
\label{eq:isotropic_screening_approximation_xi}
\end{align}
This replacement can be justified when
we are not directly interested in the matrix element but in the (weighted) integral over it, as in computations of  $\qhat$ or $\Ctwotwo$. 
The approximate matrix element $\Mxi$ agrees with $\Mhtl$ at large $q$, but behaves differently in the small $q$ region. It includes a constant $\xi$ that is fixed such that the integral over the approximated matrix element matches the result of the full isotropic HTL matrix element.  For transverse momentum broadening, this integral needs to taken in the high energy limit $p\to\infty$, be weighted with $\qperp^2$ and integrated over $\dd[2]{\vb \qperp}$ to obtain $\qhat$. 
Thus we fix $\xi$ by requiring such that
\begin{align} \label{eq:ximatching}
    \int_{0}^\infty\!\! \dd{\qperp}\qperp^3 \int_{-\infty}^\infty\!\! \frac{\dd{\omega}}{\sqrt{\qperp^2+\omega^2}}\int_0^{2\pi}\!\!\!\dd\phikq\left(\Mhtl - \Mxi\right)=0.
\end{align}
In \cite{AbraaoYork:2014hbk} one is matching the longitudinal momentum transfer rather than the transverse one, which leads to a different value for $\xi$.

To evaluate these integrals we take both matrix elements in the limit $p\to\infty$, and additionally consider the soft limit $\omega\ll k$, $\qperp\ll k$. 
We then first integrate over $\omega$. For $\Mhtl$ this can, in the soft limit, be done analytically using a sum rule \cite{Aurenche:2002pd}, which we discuss in more detail in \app\ref{app:sum-rule}.
Then we perform the $\qperp$ integral up to a cutoff $\Lambda_\perp$ and obtain the following condition 
\begin{align}
	&\frac{2}{3}\ln\left(1+\frac{\Lambda_\perp^2}{m_D^2}\right)\label{eq:regularization_eq_to_solve}\\
	&=4\ln\frac{\lperp}{2\,\xi m_D}-\frac{\lperp^2}{(\xi m_D)^2}\nonumber\\
    &-\frac{(\lperp^4+2\lperp^2(\xi m_D)^2+4(\xi m_D)^4)\ln\frac{\lperp}{\xi m_D+\sqrt{\lperp^2+(\xi m_D)^2}}}{(\xi m_D)^3\sqrt{\lperp^2+m_D^2}}\nonumber
\end{align}
where the left-hand side comes from $\Mhtl$.
Expanding both sides of the equation for large cutoff $\Lambda_\perp\gg \xi m_D$ leads to
\begin{align}
\xi=\frac{e^{1/3}}{2}\approx 0.6978 \label{eq:xi_analytic}.
\end{align}
We show in \fig \ref{fig:regularization} that both sides of \eq \eqref{eq:regularization_eq_to_solve} are indeed in very good agreement for $\Lambda_\perp \gtrsim 4m_D$, justifying the validity of the approximation for sufficiently large momentum cutoffs.
We note that the value for $\xi$ in \eq \eqref{eq:xi_analytic} entering the matrix element in $\qhat$ is slightly different from the one typically used in the elastic collision kernel $\Ctwotwo$ in kinetic theory simulations of the thermalization dynamics of the quark-gluon plasma \cite{AbraaoYork:2014hbk, Kurkela:2015qoa, Kurkela:2018oqw,Kurkela:2018xxd, Du:2020dvp, Du:2020zqg}. While the matrix element is approximated in a similar manner as here, for the thermalization dynamics $\xi$ is fixed by demanding that longitudinal momentum broadening agrees with the one from HTL matrix elements entering $\Ctwotwo$. In contrast, $\qhat$ requires that the transverse momentum broadening agrees instead. 

With the $\xi$-screening prescription,
the gluonic matrix element in \tab\ref{tab:qhat_matrix_el} becomes\footnote{The unusual value of the constant $11/4$  stems from rewriting $su/t^2$ according to \eq \eqref{eq:s2u2Overt2}.}
\begin{align}
\frac{\left|\mathcal M^{gg}_{gg}\right|^2}{16\dA \CA  ^2g^4}\to \frac{11}{4}-\frac{(s-u)^2}{4t^2}\frac{q^4}{(q^2+ \xi^2m_D^2)^2}- \frac{st}{u^2} - \frac{tu}{s^2}.\label{eq:gluonic_matrix_el_simple_screening}
\end{align}
We will investigate this approximation in \se\ref{sec:qhat_special_cases} numerically by comparing it to the HTL screened results. 
For instance, we find that the largest differences occur for a  small cutoff $\Lambda_\perp$ or  a large coupling $\lambda$. For the physically motivated values $\lambda=10$ and $\Lambda_\perp=T$ we obtain a 30\% deviation, showing that the choice of the screening prescription can be important for the evaluation of $\qhat$. To be safe from this effect, we have used  the full  HTL screening prescription for $\qhat$ in our study of the jet quenching parameter during bottom-up thermalization in \re \cite{Boguslavski:2023alu}.

\subsection{Towards the limit $p\to\infty$: NLO terms in $1/p$}
\label{sec:pfinite}

In the derivation of $\qhat$, we have considered the jet momentum $p$ to be much larger than all other momentum scales of the plasma. However, in the strict limit $p\to \infty$ the momentum diffusion coefficient $\qhat$ has a logarithmic divergence, unless the exchanged momentum is limited by a cutoff. We will first, in this subsection, discuss the limit of $p$ being large, but not infinite. Then, in \se \ref{sec:pinf_formula}, we will introduce a cutoff on $\qperp$ and take $p\to \infty$.
In the limit of large $p$, only the terms $su/t^2$ and $(s^2+u^2)/t^2$ in the matrix elements (\tab\ref{tab:qhat_matrix_el}) are $\sim p^2$ and thus contribute.

For example, let us consider the screened gluonic matrix element \eqref{eq:gluonic_matrix_el_simple_screening}
\begin{align}
\frac{\left|\mathcal M^{gg}_{gg}\right|^2}{p^2}&=16 \dA  \CA  ^2\frac{\left(2k-\omega-\sqrt{(2k-\omega)^2-q^2}\cos\phikq\right)^2}{(q^2+ \xi^2m_D^2)^2} \nonumber\\
&\qquad\qquad\qquad\times\left(1+\frac{\omega}{p}+\mathcal{O}\left(\frac{1}{p^2}\right)\right).\label{eq:matrixelement_gluonic_limit}
\end{align}
Here $k$ is a medium momentum scale (the collision integral is proportional to $f(\vb k)$) and we can thus assume that $k\ll p$ even if formally $k$ is integrated over up to infinity.
Na\"ively, we could assume that $\frac{\omega}{p}\sim\mathcal O\left(\frac{1}{p}\right)$. However, $p$ appears also in the lower integration limit of $\omega$ 
(see Eq.~\eqref{eq:qhat_finitep_measure_qlast}), and we therefore
consider the term proportional to $\omega/p$ more carefully. 
For positive $\omega$ one has $\omega < k \ll p$, such that it becomes indeed negligible. For negative $\omega$,
however, one has
\begin{align}
\left|\frac{\omega}{p}\right|=\frac{-\omega}{p}<\frac{p-k}{2p} = \frac{1}{2} - \frac{k}{2p},
\end{align}
which does not vanish for $p \to \infty$. 
However, a more careful calculation (carried out explicitly in  \app \ref{app:behav_large_p}) shows that   the leading large $p$ contribution in \eq\eqref{eq:matrixelement_gluonic_limit}
diverges logarithmically $\sim\ln p$, whereas the $\mathcal{O}(\omega/p)$ contribution becomes constant in $p$. Thus indeed the leading behavior is obtained by assuming $p\gg \omega$ term in the matrix element.

In summary, we now know, that for large jet energies $\Ejet =p$, the jet quenching parameter is given by
\begin{align}
\qhat(p\gg \Teps) \simeq a_p\ln p + b_p. \label{eq:qhat_largep_behavior}
\end{align}
Here, $\Teps$ is the characteristic momentum scale of plasma particles, e.g., the temperature in thermal equilibrium. For non-equilibrium systems, such a scale can be obtained, for example, as the temperature of an equilibrium system with the same energy density.
The coefficient $a_p$ for isotropic distributions is derived explicitly in  \app \ref{app:behav_large_p} as
\begin{align}
\begin{split}
    a_p/\CR&=\frac{\CA g^4}{4\pi^3}\int_0^\infty \dd{k}k^2f_g(k)\\
    &\quad+\sum_{f}\frac{\dF\CF g^4}{4\pi^3\dA}\int_0^\infty \dd{k} k^2 f_f(k),
    \end{split}\label{eq:qhat_ap_coefficient}
\end{align}
where $f_g$ is the gluon distribution function and the subscript $f$ in $f_f$ labels different quark species.
$q_\perp < \Lambda_\perp$, then for $p\to\infty$ it becomes indeed sufficient to only consider the leading order contribution in \eq \eqref{eq:matrixelement_gluonic_limit} since the other terms are then suppressed.

\subsection{Limit $p\to\infty$ with a momentum cutoff\label{sec:pinf_formula}}

\begin{figure}
    \includegraphics[width=0.9\linewidth]{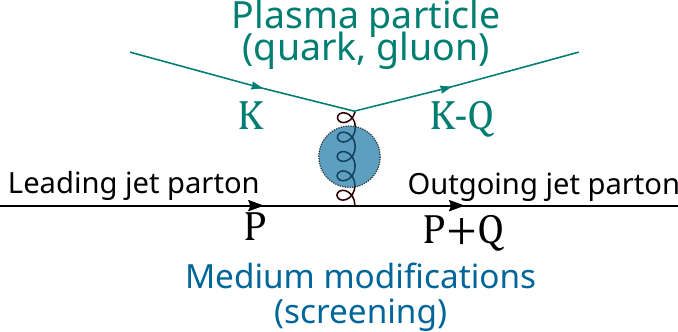}
    \caption{Feynman diagram for the $t$-channel gluon exchange processes that dominate the matrix element for $\qhat$ in the high-energy limit $p\to\infty$. In the internal gluon propagator, medium effects should be included as explained in \se\ref{sec:screening}.}
    \label{fig:feynmandiag}
\end{figure}

Let us now introduce a cutoff for the transverse component of the momentum exchange in the scattering, 
\begin{align}
	\qperp^2=q^2-\omega^2<\Lambda_\perp^2. \label{eq:qperp_cutoff}
\end{align}  Compared to the case without the cutoff, $\qperp$ is now not restricted by the large $p$ but by $\Lambda_\perp$ and  we do not need to worry about the $\omega \sim p$ region as in Sec.~\ref{sec:pfinite}.
The behavior of the phase space integral with a cutoff $q_\perp < \Lambda_\perp$ is also analyzed in more detail in \app \ref{app:behav_large_p}.

We now directly take the limit $p\to\infty$ in the matrix element, 
 which considerably simplifies the calculation. 
The required matrix elements are written down in \tab \ref{tab:p-inf_matrix_el}. Now only $t$-channel gluon exchanges contribute,  as depicted in \fig\ref{fig:feynmandiag}. 
Apart from that, few changes need to be made to the formula of $\qhat$ presented in \se\ref{sec:qhat-formula-finite-p}. Equation \eqref{eq:qhat_formula} remains valid, i.e.,
\begin{align}
    \qhat^{ij}=\frac{1}{2^9\pi^5\nu_a}\!\sum_{bcd}\!\int\!\!\dd{\Gamma}q^iq^j\frac{|\mathcal M^{ab}_{cd}|^2}{p^2}f_b(k,v_k)\left(1\pm f_d(k',v_{k'})\right),
    \label{eq:qhat_formula_pinf}
\end{align}
with the integration measure $\int\dd{\Gamma}$ in \eqref{eq:qhat_phase_space_integral_splitting}. However, the term $\int\dd{\Gamma_3}$ of the measure given by \eq \eqref{eq:qhat_finitep_measures} and the kinematic variables need adjustments.
As before, the upper and lower signs in the term $(1 \pm f_d)$ denote bosonic particles (gluons) and fermionic particles (quarks), respectively.

In particular, in \eqref{eq:qhat_finitep_measure_qlast} and \eqref{eq:qhat_finitep_measure_qlast2} we just need to adjust the upper integration limit of the $q$ integral. For \eqref{eq:qhat_finitep_measure_klast} we implement the condition \eqref{eq:qperp_cutoff} in the $\omega$ integral, $\omega^2 > q^2-\Lambda_\perp^2$, which only changes the integral boundaries for $q > \Lambda_\perp$. We can thus write the integration measures as
\begin{subequations}\label{eq:qhat_pinf_measures}
\begin{align}
\int\dd{\Gamma_3}&=\int_0^\infty\dd{k}\int_{-\infty}^k\dd{\omega}\int_{|\omega|}^{\min\left(2k-\omega,\sqrt{\omega^2+\Lambda_\perp^2}\right)}\!\!\!\!\!\!\!\!\!\dd{q} \label{eq:qhat_pinf_measure_klast}\\
&= \int_0^\infty\dd{k}\int_{0}^\infty\dd{k'}\int_{|k-k'|}^{\min\left(k+k',\sqrt{(k-k')^2+\Lambda_\perp^2}\right)}\!\!\!\!\!\!\!\!\!\dd{q}
\label{eq:qhat_pinf_measure_klast2} \\
     &= \left(\int_{\Lambda_\perp}^\infty\dd{q}\left[\int_{-q}^{-\sqrt{q^2-\Lambda_\perp^2}}\dd{\omega}+\int_{\sqrt{q^2-\Lambda_\perp^2}}^q\dd{\omega}\right] \right. \nonumber\\
     & \qquad + \left. \int_0^{\Lambda_\perp}\dd{q}\int_{-q}^q\dd{\omega} \right)\times\int_{\frac{q+\omega}{2}}^\infty \dd{k}\label{eq:qhat_pinf_measure_qlast}
\end{align}
\end{subequations}
In this limit, one needs to replace \eqref{eq:vpq} by $v_{pq} = \omega / q$ (see \app\ref{app:behav_large_p}) and the few nonvanishing matrix elements
for $\lim_{p\to\infty}\frac{|\mathcal M|^2}{p^2}$ that are given in \tab \ref{tab:p-inf_matrix_el} are expressed in terms of the same screening matrix element. Thus we do not need the explicit expressions \eqref{eq:mandelstam_s_qhat} and \eqref{eq:mandelstam_u_qhat} for $s,u$ in terms of our phase space integration variables. 
From the matrix elements in \tab\ref{tab:p-inf_matrix_el} and \eqref{eq:qhat_formula}, one finds Casimir scaling
\begin{align}
    \frac{\qhat^{\mathrm{gluon}}}{\CA}=\frac{\qhat^{\mathrm{quark}}}{\CF}.\label{eq:Casimir_scaling}
\end{align}

\begin{table}
\begin{ruledtabular}
\begin{tabular}{M{0.4\linewidth}| M{0.6\linewidth}}
$ab\leftrightarrow cd$ & $\lim_{p\to\infty}|\mathcal M^{ab}_{cd}|^2/(p^2g^4)$ \\ [10pt]
\hline 
$q_1q_i\leftrightarrow q_1q_i$ & \multirow{4}{*}{$4\frac{\dF^2\CF ^2}{\dA }\tilde M_{\rm screen}$}\\ $\bar q_1q_i\leftrightarrow \bar q_1 q_i$ &\\ $q_1\bar q_i\leftrightarrow q_1\bar q_i$ &\\ $\bar q_1\bar q_i\leftrightarrow \bar q_1\bar q_i$ &  \\ [10pt]
$q_1g\leftrightarrow q_1g$ & \multirow{2}{*}{$4 \dF\CF \CA   \tilde M_{\rm screen}$}\\ $\bar q_1 g\leftrightarrow \bar q_1 g$ & \\ [10pt]
$gg\leftrightarrow gg$ & $4\dA \CA^2 \tilde M_{\rm screen}$\\ 
\end{tabular}
\end{ruledtabular}
\caption{The matrix elements for $\hat q$ as in \tab \ref{tab:qhat_matrix_el} in the limit $p\to\infty$. Here $\tilde M_{\rm screen}=\lim_{p\to\infty}M_{\rm screen}/p^2$ denotes the appropriate screening terms $\tildeMhtl$ or $ \tildeMxi$ as explained in \se\ref{sec:screening} and \app\ref{app:full_htl_matrix_el}, and quark flavors are labeled by the index $i$. 
}
\label{tab:p-inf_matrix_el}
\end{table}

The screening in  $\tilde M_{\rm screen}$ in the matrix elements in \tab\ref{tab:p-inf_matrix_el} is implemented as detailed in \se\ref{sec:screening}. In the $p\to \infty$ limit, the isotropic HTL matrix element from Eq.~\eqref{eq:full_HTL_finite_p_matrix_element_replacement}  becomes
\begin{align}
	\tildeMhtl=\frac{\tilde c_1^2}{A^2+B^2}-\frac{\tilde c_2^2}{C^2+D^2}+\frac{2\tilde c_1\tilde c_2(AC+BD)}{(A^2+B^2)(C^2+D^2)},\label{eq:full_htl_matrix_element}
\end{align}
with the parameters $A,\, B,\, C,$ and $D$ given by \eqref{eq:parameters_for_full_HTL_matrix_element} 
and
\begin{subequations} \label{eq:c_parameters_for_pinf_matrix_element_HTL}
\begin{align}
    \tilde c_1 &= \lim_{p\to\infty} \frac{c_1}{p}=2(2k-\omega),\label{eq:ctilde1}\\
    \tilde c_2 &= \lim_{p\to\infty}\frac{c_2}{p}=4k\sin\theta_{pq}\sin\theta_{kq}\cos\phikq. \label{eq:ctilde2}
\end{align}
\end{subequations}
Again, for isotropic systems, we do not need to include the last term in \eqref{eq:full_htl_matrix_element}, since it vanishes in the angular integral when calculating $\qhat^{ij}$. We refer to \app \ref{app:full_htl_matrix_el} for details of the derivation.

For the $\xi$-screening approximation, we obtain (c.f.~\eqref{eq:isotropic_screening_approximation_xi})
\begin{align}
   \tildeMxi = 4\frac{\left(2k-\omega -\sqrt{(2k-\omega)^2-q^2}\cos\phikq\right)^2}{(q^2+\xi^2m_D^2)^2},\label{eq:approximated_matrix_element}
\end{align}
with $\xi=e^{1/3}/2$ as before.

\subsection{Limiting behavior for large cutoff}
\label{sec:limiting_behavior_large_cutoff}

The jet quenching parameter $\qhat$ exhibits a logarithmic behavior when the cutoff $\Lambda_\perp$  exceeds the typical hard momenta $\Teps$ 
of the plasma constituents.
\begin{equation}
    \qhat(\Lambda_\perp \gg \Teps) \simeq a_{\lperp}\ln\Lambda_\perp + b_{\lperp},\label{eq:qhat_behavior_large_cutoff},
\end{equation}
where for isotropic distributions
\begin{align}
    \begin{split}
    a_{\lperp}/\CR&\simeq\frac{\CA g^4}{2\pi^3}\int_0^\infty \dd{k}k^2f_g(k)\\
    &\quad+\sum_{f}\frac{\dF\CF g^4}{2\pi^3\dA}\int_0^\infty \dd{k} k^2 f_f(k).
    \end{split}\label{eq:alperp}
\end{align}
 This is the same logarithmic behavior as in Eq.~\eqref{eq:qhat_ap_coefficient}, keeping in mind that now the phase space is limited by $\Lambda_\perp^2$ rather than $p$, and thus $\ln p$ gets replaced by $2\ln \Lambda_\perp$.

For thermal equilibrium, this yields 
\begin{align}
    &\qhat^{\mathrm{therm}}(\Lambda_\perp \gg T) \nonumber \\
    &~~\simeq \frac{\CR}{\pi^3} g^4\zeta(3) T^3\left(\NC + \frac{3}{4}\nf\right)\ln\Lambda_\perp + \mathrm{const},
\end{align}
which is \eq \eqref{eq:qhat_a_coefficient_equilibrium} in \app \ref{sec:qhat_behavior_large_cutoff} and agrees 
with \cite{Arnold:2008vd}, as we will later discuss around \eq \eqref{eq:qhat_hard_arnold}.

As discussed further in \app \ref{sec:qhat_behavior_large_cutoff}, for anisotropic systems Eq.~\eqref{eq:alperp} only gives  a rough estimate for the coefficient $a_{\lperp}$. For anisotropic distributions one still expects a logarithmic behavior $\qhat^{ii}(\Lambda_\perp\gg Q)\simeq \left(a_{\lperp}\right)_i\ln\Lambda_\perp + \left(b_{\lperp}\right)_i$, but with different coefficients depending on the direction.

\subsection{Interpreting the momentum cutoff}
\label{sec:momentum_cutoffs}

A peculiar feature of the jet quenching parameter $\qhat$ is its dependence on a transverse momentum cutoff $\lperp$. Let us discuss here briefly  how to interpret this cutoff in physical terms and how its value could be chosen.
In our kinetic picture, the cutoff stems from employing the eikonal limit, which means taking the jet momentum $p\to\infty$. In this case the jet particle can inject unrestricted amount of transverse momentum in the collision, leading to a logarithmic divergence that has to be regulated by introducing a cutoff $\lperp$, which restricts transverse momentum transfer $\qperp<\lperp$. Practically all analytic calculations that rely on quasi-particles or hard-thermal loop frameworks, but even with different interaction potentials, need to employ this cutoff (as, e.g., in \cite{Arnold:2008vd, Caron-Huot:2008zna, He:2015pra, JET:2013cls, Iancu:2018trm, JETSCAPE:2021ehl, Barata:2023qds}).

A simple way of setting the cutoff is to use the relation between the  
coefficient $a_{\lperp}$ for large cutoff $\Lambda_\perp$ and the coefficient $a_p$ for large (finite) jet energy $p$ (see Eqs.~\eqref{eq:qhat_ap_coefficient} and \eqref{eq:alperp}). Requiring that the dynamics of jet quenching calculated with a cutoff in the $p\to\infty$ approximation would have the same leading logarithmic behavior as a kinematically more accurate one with a finite $p$, we should choose the cutoff such that
\begin{align}
    \Lambda_\perp^{\mathrm{kin}}\sim \sqrt{pT}, \label{eq:kinematic_cutoff}
\end{align}
where $p$ is the energy of the jet parton and $T$ is an additional dimensionful scale (e.g., the temperature in equilibrium).
This kinematic cutoff $\Lambda_\perp^{\mathrm{kin}}$ is widely used in the literature \cite{Qin:2009gw, JET:2013cls, Xu:2014ica, He:2015pra, Cao:2021rpv, JETSCAPE:2021ehl,JETSCAPE:2022jer, Mehtar-Tani:2022zwf}.

While this is a straightforward result of our definition for $\qhat$ in \eq \eqref{eq:def_qhat1}, it encodes only the momentum diffusion due to elastic $2\leftrightarrow 2$ scattering processes, and competing inelastic effects like splittings or gluon emissions are neglected. Which effects one needs to include, and thus, which cutoff to use, depends in fact on the type of process where the value of $\qhat$ is used.
For radiative energy loss calculations, one can restrict the cutoff by considering the rate of momentum exchange processes and comparing it with the `life-time' of the leading parton under consideration. During an LPM splitting process this corresponds to the `formation time' $\tform$.
We are therefore interested in the accumulated transverse momentum until a splitting occurs. To calculate radiative energy loss of the leading parton, typical calculations (e.g., within the BDMPS-Z formalism \cite{Baier:1998kq, Baier:1998yf, Zakharov:2000iz} or related approaches \cite{Arnold:2008iy}) use the so-called harmonic oscillator approximation, in which the jet quenching parameter $\qhat$ naturally appears in the expansion of the interaction potential in position space, $v(\vb b)\simeq \frac{1}{4} \qhat \vb b^2$. In the leading-log approximation, it is sufficient to use a momentum cutoff $\lperp$ of the order of the typical total momentum transfer $Q_\perp$ during the formation time \cite{Arnold:2008iy}. By definition, it is given by $Q_\perp^2\sim \qhat \tform$, where for a small medium with length $L<\tform$ one should replace $\tform$ by $L$. The formation time of the splitting $p\to p_1+p_2$ can be estimated as $(\tform)^2\sim E_i/\qhat$, with $E_i$ being the energy of the emitted gluon. It has been argued \cite{Arnold:2008vd, Arnold:2008zu} that energy loss is dominated by processes in which both daughters share a similar energy fraction $p_1\sim p_2\sim p$, which enables us to use the leading-parton energy in the formation time estimate. With the parametric relation $\qhat\sim g^4 T^3$, we obtain for a large medium $L>\tform$ the expression
\begin{align}
    \lperp^{\mathrm{LPM}}\sim g\,(pT^3)^{1/4}. \label{eq:LPM_cutoff}
\end{align}

In order to present our results in a form that can be applied in different pictures of energy loss, we 
give our results as functions of $\lperp$. In our companion paper \cite{Boguslavski:2023alu} we study the value of $\qhat$ with a time-dependent cutoff during bottom-up thermalization, and present results using both scaling choices \eqref{eq:kinematic_cutoff} and \eqref{eq:LPM_cutoff}.

\subsection{Numerical implementation
\label{sec:numerical_implementation}}

Numerically, we obtain $\qhat$ in the limit $p\to\infty$ with a momentum cutoff $\lperp$ according to \eqref{eq:qhat_formula_pinf} with the integration measure
\footnote{We have checked that also \eqref{eq:qhat_pinf_measure_klast2} gives the same result, but in our implementation \eqref{eq:qhat_pinf_measure_qlast} showed a faster convergence in the Monte Carlo evaluation.} 
\eqref{eq:qhat_pinf_measure_qlast} using Monte Carlo integration with importance sampling. For more details we refer to \app\ref{app:monte_carlo}. The error bars in the figures correspond to the statistical uncertainty of the Monte Carlo integration.

All our numerical results are obtained for a purely gluonic plasma, i.e., $\nf=0$.
Although we extract the jet quenching parameter for a gluonic jet, its value for a quark jet is related via Casimir scaling, i.e., $\qhat^{\mathrm{quark}}=\frac{\CF}{\CA}\qhat^{\mathrm{gluon}}$, as can be seen easily from \tab\ref{tab:p-inf_matrix_el}. Note that the number of colors $\NC$ enters $\qhat^{\mathrm{gluon}}$ only via $\lambda$, but for $\qhat^{\mathrm{quark}}$ we need to specify it to $\NC=3$ for QCD.
In the following, for numerical results we only show $\qhat^{\mathrm{quark}}=:\qhat$; for analytical results we keep the Casimir factor $\CR$ explicitly.

Our datasets and analysis scripts can be found in Ref.~\cite{lindenbauer_2023_10419945}.

\section{Evaluation of $\qhat$ in special cases\label{sec:qhat_special_cases}}

In our paper~\cite{Boguslavski:2023alu} we have evaluated $\qhat$ during the bottom-up thermalization process in heavy-ion collisions. Here, in order to shed more light on the qualitative features of the jet quenching parameter in different equilibrium and off-equilibrium situations, we study $\qhat$ in some special cases. 
In \se \ref{sec:qhat-thermal}, we first review the derivation of $\qhat$ for thermal systems \cite{Aurenche:2002pd, Arnold:2008vd}, and compare the results with numerical evaluations of Eq.~\eqref{eq:qhat_formula_pinf}. 
We also provide an interpolation formula that reproduces the numerically obtained values of the quenching parameter in thermal equilibrium $\qhattherm$ for different couplings and momentum cutoffs.

In \se \ref{sec:toy_models}, we then consider toy models for the bottom-up thermalization process in heavy-ion collisions \cite{Baier:2000sb}. We first study an effectively two-dimensional distribution to model the large momentum-space anisotropy encountered in the initial stages in heavy-ion collisions, and then generalize the thermal results of \se \ref{sec:qhat-thermal} to a scaled thermal distribution to model over- and underoccupied systems that are typically encountered in the pre-equilibrium evolution of the quark-gluon plasma.

We also study the different contributions to $\qhat$ that are linear or quadratic in the distribution function, by splitting it into its individual components,
\begin{align}
    \qhat = \qhatf+\qhatff.\label{eq:qhat_f_ff_splitting}
\end{align}
Similarly as in Ref.~\cite{Kurkela:2021ctp}, we can refer to $\qhatf$ as the \emph{classical} and $\qhatff$ as the \emph{Bose-enhanced} part of $\qhat$.
\footnote{
This Bose-enhanced term can also be considered to be a \emph{classical field} contribution because it is dominant in 
highly-occupied systems $f\gg 1$ that can be studied numerically using classical-statistical simulations. This can be seen in the 
limit of $\lambda\to 0$ with $\lambda f$ held constant, in which only $\qhatff$ survives.
}

\subsection{Thermal distribution\label{sec:qhat-thermal}}

The equilibrium form of the particle distributions is given by
\begin{align}
f_{\pm}(k; T)=\frac{1}{\exp(k/T)\mp1},\label{eq:thermal_bose_fermi_combined}
\end{align}
where $T$ is the temperature. The upper signs $f_+$ denote the Bose-Einstein distribution and $f_-$ is the Fermi-Dirac distribution.

In thermal equilibrium, $\qhat$ has already been calculated for the limiting cases of small and large cutoffs $\lperp$ in \cite{Arnold:2008vd, Caron-Huot:2008zna}, which we briefly summarize here.
In \se\ref{sec:scaled-thermal-distribution}, we will generalize this derivation to the case of a scaled thermal distribution, which is obtained by rescaling a thermal distribution.

For the evaluation of $\qhat$, we work in the $p\to \infty$ limit with a transverse momentum cutoff $\lperp$, as discussed in \se\ref{sec:pinf_formula}. Since the phase-space density is spherically symmetric, one has $\qhat^{11}=\qhat^{22}$ and we can restrict to $\qhat=\qhat^{11}+\qhat^{22}$. Our starting point is Eq.~\eqref{eq:qhat_pinf_measure_qlast}, where  we integrate over the modulus of $\vec q=(\vec q_\perp, q^3)$. For $p\to\infty$ we obtain $q^3=\omega$ and thus
\begin{align}
q^2=q_\perp^2+\omega^2.
\end{align}

It will be useful to change the integration variables from $(q,\phi_{pq},\omega)$ to $(q^1,q^2,\omega)$. This yields a factor $q$ from the Jacobian 
\begin{align}
\begin{split}
&\int_0^{2\pi}\dd{\phipq}\int_0^\infty\dd{q}\int_{-q}^q\dd{\omega} \theta(\Lambda_\perp^2 + \omega^2- q^2)\\
&\quad=\int_{\qperp <\Lambda_\perp}\dd[2]{\vec q_\perp}\int_{-\infty}^\infty\frac{\dd{\omega}}{\sqrt{\qperp^2+\omega^2}}. \label{eq:coordinate_trafo_differentials_to_qperp}
\end{split}
\end{align}

The matrix elements in \tab\ref{tab:p-inf_matrix_el} do not allow for identity-changing processes, which means that the leading parton $a$ and the outgoing parton $c$ are of the same type, $a=c$, and similarly $b=d$.
Therefore, we can scale out the Casimir of the jet $\CR$, and the prefactors in front of $\tilde M_{\rm screen}$ in \tab\ref{tab:p-inf_matrix_el} neatly combine with $1/\nu_a$ for the degrees of freedom of the jet particle to 
\begin{subequations}
\label{eq:XiPlMin}
\begin{eqnarray}
    \Xi_+ \,& &= 2\NC \\
    \Xi_- \,&= 4\nf\frac{\dF\CF}{\dA} &= 2\nf
\end{eqnarray}
\end{subequations}
for scattering off a gluon and off a quark/anti-quark, respectively, which leads to Casimir scaling (c.f.~\eq \eqref{eq:Casimir_scaling}).

There are two limiting cases in which the result for $\qhat$ can be found analytically, for small and large momentum cutoffs, which we will study in the following.

\subsubsection{Small momentum cutoff}

For small $\qperp < \lperp \ll T$, the expression for $\qhat$ in \eq \eqref{eq:qhat_formula} with the integration measure \eqref{eq:qhat_pinf_measure_qlast}, the integrals \eqref{eq:coordinate_trafo_differentials_to_qperp} and the prefactors \eqref{eq:XiPlMin} becomes
\begin{align}
\label{eq:qhat_factorized}
&\hat q(\Lambda_\perp)=C_R\sum_{\pm}\Xi_\pm\frac{g^4}{2^9\pi^5}
\int_0^\infty\dd{k}\,f_\pm(k)\left(1\pm f_\pm(k)\right) \nonumber\\
&\qquad \times \int_0^{\Lambda_\perp}\dd[2]{\vb q_\perp}\qperp^2\int_0^{2\pi}\dd{\phikq}\int_{-\infty}^\infty\frac{\dd{\omega}\tildeMhtl}{\sqrt{q_\perp^2+\omega^2}}.
\end{align}
We have extended the lower boundary
\footnote{
The largest error of this approximation comes from the $f_+^2$ term. It can by estimated by $\int_0^{\frac{q+\omega}{2}}k^2f_+^2 < \frac{q+\omega}{2}\lim_{k\to 0}\left(k^2f_+^2\right)$, where the $k^2$ factor stems from $\tildeMhtl$ and we approximated the integral by the maximum value of the integrand at $k=0$. This yields the error estimate $\frac{T^2(q+\omega)}{2}$, which for $\qperp < \lperp \ll T$ is much smaller than the leading-order contribution $\int_0^{\infty}k^2f_+^2 = 2T^3(\zeta(2)-\zeta(3))$ (see \eq\eqref{eq:thermal_distributionfunctions_integrals})
for $q,\,\omega \ll T$.}
of the $k$-integral to $0$ and approximated $f(k-\omega)\approx f(k)$. This is appropriate because large values of $\omega$ are suppressed by the matrix element $\tildeMhtl$, as can be seen from Eq.~\eqref{eq:matrixelement_gluonic_limit}. 

The last two integrals
can be evaluated analytically using a sum rule \cite{Aurenche:2002pd} as discussed in \app\ref{app:sum-rule},
\begin{align}
\begin{split}
\hat q(\Lambda_\perp)&=\frac{C_Rg^4}{(2\pi)^3}2\int_{\qperp < \Lambda_\perp}\frac{\dd[2]{\vb\qperp}}{2\pi}\qperp^2\frac{1}{\qperp^2(\qperp^2+m_D^2)}
\\
&\quad \times \sum_{\pm}\Xi_{\pm}\int_0^\infty\dd{k}k^2\,f_\pm(k)\left(1\pm f_\pm(k)\right).
\end{split}\label{eq:qhat_soft_distribution_function_split_off}
\end{align}
Note that until now we have not used a specific form for the distribution function $f(k)$ and assumed only spherical symmetry. The thermal form of $\qhat$ for a small cutoff is then obtained by performing the integrals over the distribution function,
\begin{subequations}\label{eq:thermal_distributionfunctions_integrals}
    \begin{align}
        \int_0^\infty\dd{k}k^2f_\pm(k)&=2T^3\zeta_\pm(3),\\
        \int_0^\infty\dd{k}k^2 \left(f_\pm(k)\right)^2&=\pm 2T^3(\zeta_\pm(2)-\zeta_\pm(3)),
    \end{align}
\end{subequations}
where $\zeta_+(s)=\zeta(s)$ is the Riemann Zeta function and $\zeta_-(s)=(1-2^{1-s})\zeta(s)$ denotes its fermionic counterpart as in \re \cite{Arnold:2008vd}.
Using $\zeta(2)=\pi^2/6$, we obtain
\begin{align}
    \qhat=\int_{\qperp<\Lambda_\perp}\dd[2]{\vb \qperp} \qperp^2 \times\frac{g^2 C_R T }{(2\pi)^2}\frac{m_D^2}{ \qperp^2(\qperp^2+m_D^2)},\label{eq:qhat_from_collision_kernel_explicit}
\end{align}
from which we can read off the elastic scattering rate $\frac{\dd\Gammael}{\dd[2]{\qperp}}$ as in \eqref{eq:def_qhat2}.
This leads us to the thermal form of $\qhat$ for a small cutoff,
\footnote{This form is actually valid in general for any isotropic distribution $f(k)$ with the replacement of 
$T\to T_\ast=\frac{\sum_s \nu_s C_s\int\dd[3]{\vb p}f_s(\vb p)\left[1\pm f_s(\vb p)\right]}{2\sum_s\nu_s C_s\int\dd[3]{\vb p}f_s(\vb p)/p}$
and the more generally evaluated Debye mass $m_D$ as in \eq \eqref{eq:debye_mass_general}.}
\begin{align}
    \qhattherm(\Lambda_\perp\ll T)=\frac{g^2}{4\pi} \CR T m_D^2\ln\left(1+\frac{\Lambda_\perp^2}{m_D^2}\right).\label{eq:qhat_thermal_equlibrium_soft}
\end{align}

For a thermal system, the terms containing $\zeta_\pm(3)$ cancel if we consider the total $\qhat$, but are important if one considers the \emph{Bose-enhanced} part separately, as in \eqref{eq:qhat_f_ff_splitting}.
Splitting off the \emph{Bose-enhanced} term as in \eqref{eq:qhat_f_ff_splitting}, we obtain
\begin{subequations}
\begin{align}
	\qhatftherm(\Lambda_\perp\ll T)&= \zeta(3)\left(12\NC +9\nf\right)C_L\label{eq:qhat_thermal_f}\\
	\qhatfftherm(\Lambda_\perp\ll T)&=\left[2\NC(\pi^2-6\zeta(3))\right. \nonumber \\
    &\qquad + \left.\nf(\pi^2-9\zeta(3))\right]C_L,\label{eq:qhat_thermal_ff}
\end{align}
\end{subequations}
with $C_L=\frac{g^4T^3 \CR}{24\pi^3}\ln\left(1+\frac{\Lambda_\perp^2}{m_D^2}\right)$ and the thermal Debye mass given by \eq \eqref{eq:mD_T}.

\subsubsection{Large momentum cutoff\label{sec:thermal_qhat_large_momentum_cutoff}}

The jet quenching parameter in thermal equilibrium has been calculated for large cutoffs in Ref.~\cite{Arnold:2008vd}. In order to generalize this later to a scaled thermal distribution, we briefly review the derivation here. It relies on constructing an interpolating formula for the elastic scattering rate,
\begin{align}
    \frac{\dd{\Gammael}}{\dd[2]{\qperp}}\simeq \frac{\CR}{(2\pi)^2}\times\frac{g^4T^3 F(q_\perp/T)}{\qperp^2(\qperp^2+m_D^2)},\label{eq:def_F_function}
\end{align}
where the function $F(\qperp/T)$ interpolates between the known limits of this quantity (for small $\qperp/T$ see Eq.~\eqref{eq:qhat_from_collision_kernel_explicit}) and can be calculated in the approximation $q\gg m_D$. It is then split into gluonic ($I_+$) and fermionic ($I_-$) contributions,
\footnote{In principle,
we could take Eq.~\eqref{eq:qhat_factorized} instead and relax the assumption of small momentum transfer, i.e., keep $f_{\pm}(k)(1\pm f_\pm(k-\omega))$. However, the strategy employed in \re\cite{Arnold:2008vd} (scaling out this factor $F(\qperp/T)$ in \eqref{eq:def_F_function}) allows us to evaluate the expression analytically in the large $\qperp$ limit, where the matrix element does not need to be screened, and we can use the simpler form $su/t^2$ instead.}
\begin{align}
    F(\qperp/T)=\frac{1}{\pi^2}\left(\Xi_+ I_+(\qperp/T)+\Xi_- I_-(\qperp/T)\right).    
\end{align}
Following the notation in \re \cite{Arnold:2008vd}, we write these contributions to the elastic scattering rate in the limit $p\to\infty$ and $\qperp\gg m_D$ as
\begin{align}
    \label{eq:Ipm}
    I_\pm\left(\frac{\qperp}{T}\right) =& \frac{\pi^2}{T^3}\int\frac{\dd{q_z}}{2\pi}\int\frac{\dd[3]{\vb k}}{(2\pi)^3} 2\pi\delta( q_z+|\vb k-{\vb q}|-k) \nonumber\\
    &\quad \times \frac{(k-k_z)^2}{k|\vb k-{\vb{q}}|}f_\pm(\vb k)\left[1\pm f_\pm(\vb k - {\vb q})\right].
\end{align}
This formula follows directly from the t-channel matrix element in \tab\ref{tab:qhat_matrix_el}, i.e., $su/t^2$, with $t^2=\qperp^4$ being scaled out in \eqref{eq:def_F_function} and $s=-u=2p(k-k_z)$.

\begin{figure*}
    \centerline{
        \includegraphics[width=0.45\linewidth]{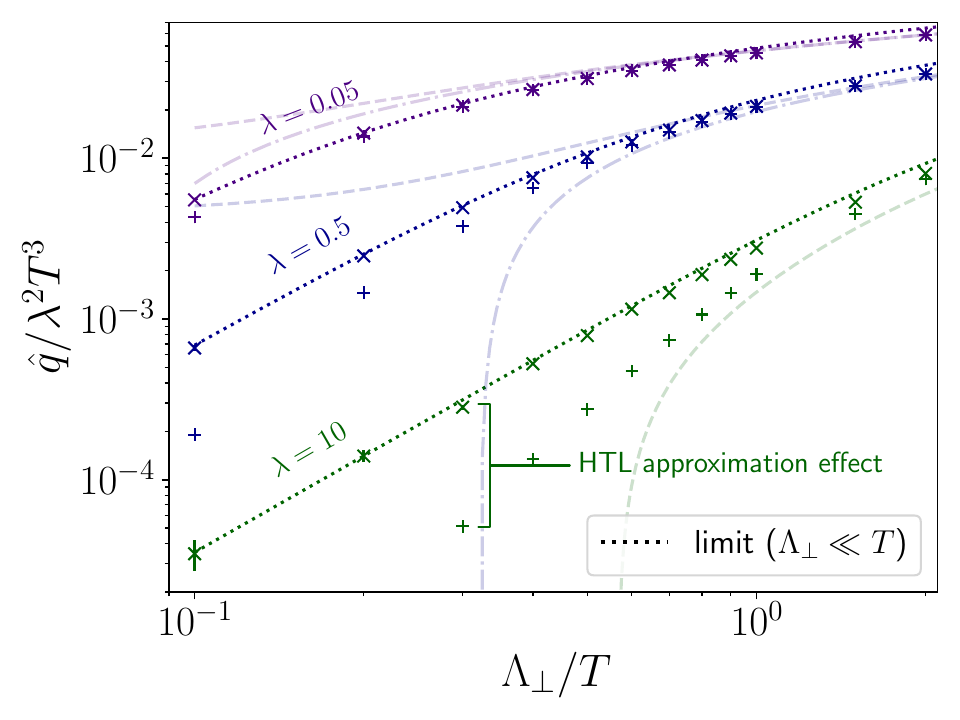}
    	\includegraphics[width=0.45\linewidth]{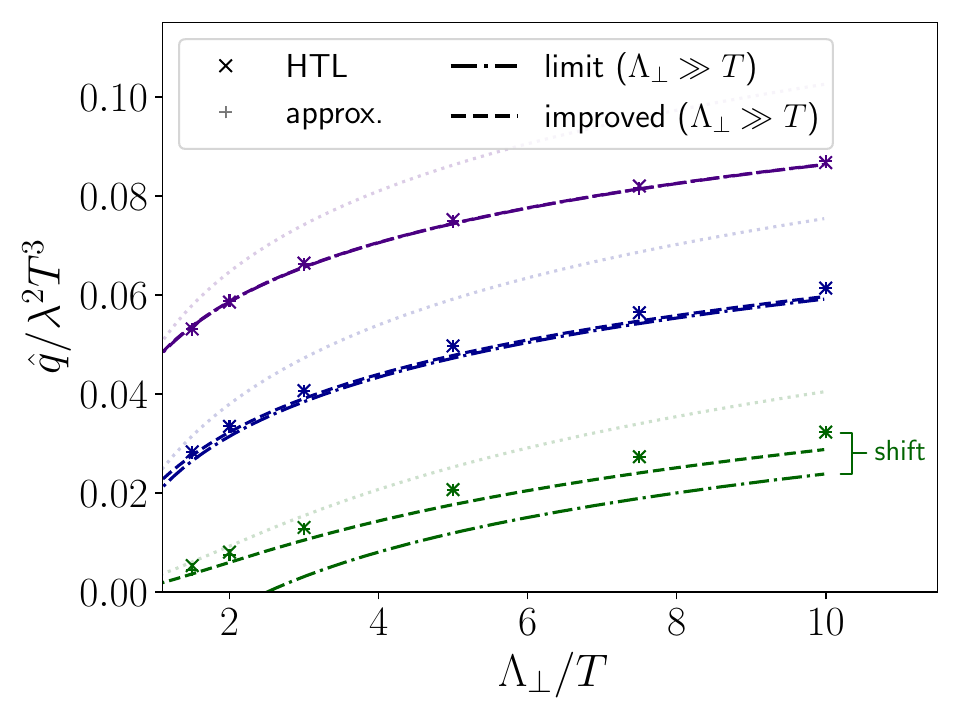}
    }
    \caption{The coefficient $\qhat$ for a quark jet in thermal equilibrium for different 't Hooft couplings $\lambda=g^2\NC$ and transverse momentum cutoffs $\Lambda_\perp$, with {\em (left:)}  small $\Lambda_\perp \lesssim T$ (logarithmically scaled axis), and {\em (right:)} large $\Lambda_\perp \gtrsim T$. The dotted curve, labeled `limit ($\Lambda_\perp \ll T$)' shows the analytical limiting expression of \eqref{eq:qhat_thermal_equlibrium_soft}, the dash-dotted curve `limit ($\Lambda_\perp\gg T$)' illustrates \eqref{eq:qhat_hard_arnold}, and `improved ($\Lambda_\perp\gg T)$' denotes \eqref{eq:qhat_hard_arnold-improved}. The small $+$-symbols show our numerical results with the approximated matrix element \eqref{eq:approximated_matrix_element}, whereas the $\times$-symbols show our results with the full HTL matrix element \eqref{eq:full_htl_matrix_element}. The curves that are not valid in the respective limit are 
    displayed with lighter color
    but still shown, because \eqref{eq:qhat_thermal_equlibrium_soft} is often also used at large $\Lambda_\perp$ as an approximation.
    }
	\label{fig:thermal_equilibrium}
\end{figure*}

As in \eqref{eq:qhat_f_ff_splitting} we can identify the contributions coming from the $f$ and $f^2$ parts via
\begin{align}
    I_\pm(\qperp/T)=I_\pm^{\mathrm{f}}+I_\pm^{\mathrm{ff}}(\qperp/T),
\end{align}
where $I_\pm^{\mathrm{f}}$ will turn out to be a constant.
To evaluate them, the thermal functions are written as
\begin{align}
    f_\pm(p)=\sum_{m=1}^\infty\left(\pm1\right)^{m-1}e^{-m p/T}.
\end{align}
This can then be inserted into \eq \eqref{eq:Ipm} to rewrite the equation as a double sum,
\begin{align}
    I_\pm^{\mathrm{f}}\left(\frac{\qperp}{T}\right)&=\sum_{m=1}^\infty (\pm 1)^{m-1}
I_{m0}(\qperp/T)\\
    I_\pm^{\mathrm{ff}}\left(\frac{\qperp}{T}\right)&=\sum_{m=1}^\infty\sum_{n=1}^\infty(\pm1)^{m+n-1}I_{mn}(\qperp/T)
\end{align}
with
\begin{align}
    \label{eq:I_{mn}}
    I_{mn}\left(\frac{\qperp}{T}\right) =& \frac{\pi^2}{T^3}\int\frac{\dd{q_z}}{2\pi}\int\frac{\dd[3]{\vb k}}{(2\pi)^3}\, 2\pi\delta( q_z+|\vb k-{\vb{q}}|-k) \nonumber\\
    &\qquad \times\frac{(k-k_z)^2}{k|\vb k-{\vb{q}}|}e^{-m k/T}e^{-n|\vb k-\vb q|/T}\,.
\end{align}
In \cite{Arnold:2008vd} $I_\pm$ was split in a similar way isolating the $n=0$ term $I_\pm(\infty)$, which is exactly the constant $I_\pm^{\mathrm{f}} = I_\pm(\infty)=\zeta_\pm(3)$. This is a consequence of 
the fact that for large momentum transfer only the $f$ part contributes, as discussed in  \se \ref{sec:limiting_behavior_large_cutoff}.

Performing the remaining integrals over $\vb\qperp$ as in \cite{Arnold:2008vd} leads to a $\qhat$ formula for large cutoffs $\Lambda_\perp\gg T$,
\begin{subequations}
\label{eq:qhat_thermal_hard_analytic}
\begin{align}
        \qhattherm(\Lambda_\perp\gg T)&= \CR \frac{g^4T^3}{\pi^2}\sum_\pm\Xi_\pm \mathcal I_\pm(\Lambda_\perp)\label{eq:qhat_hard_arnold}\\
	\mathcal I_\pm(\Lambda_\perp)&=\frac{\zeta_\pm(3)}{2\pi}\ln\left(\frac{\Lambda_\perp}{m_D}\right)+\Delta\mathcal I_\pm,\label{eq:qhat_thermal_hard_analytic_Ipm}\\
    \Delta\mathcal I_\pm &= \frac{\zeta_\pm(2)-\zeta_\pm(3)}{2\pi} \label{eq:qhat_thermal_hard_analytic_DeltaI}\\
    &~~\times\left[\ln\left(\frac{T}{m_D}\right)+\frac{1}{2}-\gamma_E+\ln 2\right] -\frac{\sigma_\pm}{2\pi}\nonumber\\
	\sigma_+&=0.386043817389949\\
	\sigma_-&=0.011216764589789
\end{align}
\end{subequations}
where $\gamma_E$ is the Euler-Mascheroni constant.

This formula \eqref{eq:qhat_thermal_hard_analytic}, as opposed to the one for small cutoffs \eqref{eq:qhat_thermal_equlibrium_soft}, has the (unphysical) feature that the logarithm $\ln T/m_D$ 
becomes negative for $T\leq m_D$.\footnote{In the literature, the small cutoff form \eqref{eq:qhat_thermal_equlibrium_soft} is also often written just as logarithm $\ln\lperp/m_D$ instead of the form we obtain.}
Normally in perturbation theory one has $T\gg m_D$ so that in the large cutoff regime $\lperp \gg T$ the form $\ln \Lambda_\perp/m_D$ is not a problem. However, to get an analytical expression that is well-behaved also for larger couplings, we propose to add a constant to the argument of the logarithm, which still preserves the leading order accuracy at weak coupling.
To be explicit, we replace $2\ln x \to \ln(1+x^2)$ in both logarithms, and we will denote the resulting `improved' analytic expressions for $\qhat$ by $\qhatimproved$. Although the replacement does not change the result at leading order, we find that this choice of regularization significantly improves the agreement with numerical evaluations of \eqref{eq:qhat_formula}, as we will discuss in \se \ref{sec:qhat_thermal_numerical}.
Moreover, the Bose-enhanced part $\qhatff$ of \eqref{eq:qhat_f_ff_splitting} comes solely from $\Delta\mathcal I_\pm$ in \eqref{eq:qhat_thermal_hard_analytic_Ipm}. With these replacements in the logarithm, the contribution $\qhatf$ has the same form as for small momentum cutoffs \eqref{eq:qhat_thermal_f}, $\qhatf(\Lambda_\perp\ll T)=\qhatf(\Lambda_\perp\gg T)$.

With this procedure, the improved version of Eq.~\eqref{eq:qhat_hard_arnold} becomes 
\begin{subequations}
\begin{align}
		\qhattherm_{\mathrm{im}}(\Lambda_\perp\gg T)&=
		\qhatftherm(\Lambda_\perp\ll T) + \qhatffthermimproved\label{eq:qhat_hard_arnold-improved}
\end{align}
with
\begin{align}
{\qhatffthermimproved}&={\CR g^4T^3}\sum_\pm\Xi_\pm \left\lbrace
 \frac{\zeta_\pm(2)-\zeta_\pm(3)}{4\pi^3}\right. \\
 & \times \left.\left[\ln\left(1+\frac{T^2}{m_D^2}\right) +1-2\gamma_E+2\ln 2\right] -\frac{\sigma_\pm}{2\pi^3}\right\rbrace . \nonumber
\end{align}
\end{subequations}

\subsubsection{Comparison with numerical results\label{sec:qhat_thermal_numerical}}

Let us now compare the analytical small and large cutoff limits of $\qhat$ given by \eqref{eq:qhat_thermal_equlibrium_soft} and \eqref{eq:qhat_hard_arnold} or the improved version \eqref{eq:qhat_hard_arnold-improved} to a numerical evaluation of $\qhat$ using \eqref{eq:qhat_formula_pinf}. For simplicity we consider a purely gluonic plasma, i.e., $\nf = 0$.
In particular, we want to study how well these analytic formulae 
describe the full numerical evaluation of the $\qhat$ integral, although being only valid for asymptotic regions of the cutoff $\Lambda_\perp$. We also want to compare the expressions using the isotropic HTL matrix element \eqref{eq:full_htl_matrix_element} with the simpler screened matrix element \eqref{eq:approximated_matrix_element} and study the impact of the approximation, which is also widely used in studies of the thermalization dynamics \cite{Kurkela:2015qoa, Kurkela:2018oqw, Kurkela:2018xxd,Du:2020dvp, Du:2020zqg}.

In \fig \ref{fig:thermal_equilibrium} we show $\qhat$ for various momentum cutoffs $\Lambda_\perp$ and different 't Hooft couplings $\lambda = g^2 \NC$. 
The prefactor $\lambda^2T^3$ is scaled out in the plots, leaving a nontrivial coupling dependence that enters via the Debye mass $m_D$ in the logarithms originating from the matrix element.
The curves show the analytical expressions for small (dotted, \eq \eqref{eq:qhat_thermal_equlibrium_soft}), large cutoffs (dash-dotted, \eq \eqref{eq:qhat_hard_arnold}) and the improved large-cutoff version (dashed, \eqref{eq:qhat_hard_arnold-improved}), while the numerical evaluation of $\qhat$ is depicted by crosses for the HTL matrix elements \eqref{eq:full_htl_matrix_element} and plus signs for the approximated screened ones \eqref{eq:approximated_matrix_element}. 
In the left panel of \fig \ref{fig:thermal_equilibrium}, we observe that the small-cutoff form of $\qhat$ 
accurately agrees with our numerical evaluation using the full HTL matrix element 
in the corresponding region $\Lambda_\perp\ll T$, even for $\Lambda_\perp\to 0$. 
We note in passing that the frequently employed form of $\qhat$ in this limit with the approximation $\ln\left(1+\frac{\Lambda_\perp^2}{m_D^2}\right)\to2\ln\frac{\Lambda_\perp}{m_D}$ (not shown in the figure)
would become negative at too small cutoffs $\lperp \sim m_D$.

In the region $\Lambda_\perp\gg T$ (right panel of \fig\ref{fig:thermal_equilibrium}), we observe that 
for small couplings $\lambda \sim 0.05$ both analytical large-cutoff expressions 
agree very well with the numerical values. However, they start to differ when increasing the coupling $\lambda \gtrsim 0.5$.
This is denoted as `shift' in \fig\ref{fig:thermal_equilibrium}.
We find that the values from our improved formula \eqref{eq:qhat_hard_arnold-improved} are closer to the numerical values 
than from the original formula \eqref{eq:qhat_hard_arnold}. However, for large couplings $\lambda \sim 10$, our improved analytical expression still seems to underestimate $\qhat$, with the difference being a constant.

Turning now to a comparison of the matrix elements, 
we observe in \fig \ref{fig:thermal_equilibrium} that for small values of the coupling $\lambda \sim 0.05$ (left panel) as well as for large cutoffs $\Lambda_\perp \gg T$ (right panel), the results with the screening approximation \eqref{eq:approximated_matrix_element} agree well with the full HTL matrix element \eqref{eq:full_htl_matrix_element}.
However, they start deviating with growing coupling
at small cutoffs $\Lambda_\perp \lesssim T$ (left panel). 
To guide the eye, for $\Lambda_\perp = 0.3 T$ we explicitly denote this difference as `HTL approximation effect'.
For $\Lambda_\perp = T$ 
and $\lambda=10$
the deviation between the approximated 
and the HTL matrix elements is of the order of 
30\%.

\begin{table*}
\begin{ruledtabular}
\begin{tabular}{c c c c} 
 $\lambda$ & $\tilde b$ & $\tilde d$ & $\tilde e$ \\ [0.5ex] 
 \hline
 $0.5$ & $0.0011944 \pm 0.0000020$ & $4.114 \pm 0.013$ & $-0.76919 \pm 0.00058$\\
$1.0$ & $0.0037772 \pm 0.0000062$ & $2.4910 \pm 0.0029$ & $-0.24707 \pm 0.00041$\\
$1.5$ & $0.007379 \pm 0.000013$ & $2.0956 \pm 0.0018$ & $0.03349 \pm 0.00032$\\
$2.0$ & $0.011905 \pm 0.000021$ & $1.9636 \pm 0.0014$ & $0.20498 \pm 0.00029$\\
$2.5$ & $0.017295 \pm 0.000031$ & $1.8987 \pm 0.0012$ & $0.32796 \pm 0.00028$\\
$3.0$ & $0.023563 \pm 0.000042$ & $1.8653 \pm 0.0010$ & $0.42226 \pm 0.00026$\\
$3.5$ & $0.030716 \pm 0.000054$ & $1.84570 \pm 0.00096$ & $0.49864 \pm 0.00025$\\
$4.0$ & $0.038770 \pm 0.000067$ & $1.83331 \pm 0.00088$ & $0.56271 \pm 0.00024$\\
$4.5$ & $0.047761 \pm 0.000082$ & $1.82484 \pm 0.00080$ & $0.61789 \pm 0.00023$\\
$5.0$ & $0.057714 \pm 0.000099$ & $1.81902 \pm 0.00075$ & $0.66626 \pm 0.00022$\\
$5.5$ & $0.06864 \pm 0.00012$ & $1.81444 \pm 0.00071$ & $0.70960 \pm 0.00021$\\
$6.0$ & $0.08061 \pm 0.00014$ & $1.81130 \pm 0.00069$ & $0.74868 \pm 0.00020$\\
$6.5$ & $0.09362 \pm 0.00015$ & $1.80845 \pm 0.00067$ & $0.78441 \pm 0.00020$\\
$7.0$ & $0.10772 \pm 0.00017$ & $1.80584 \pm 0.00066$ & $0.81733 \pm 0.00020$\\
$7.5$ & $0.12296 \pm 0.00020$ & $1.80380 \pm 0.00065$ & $0.84781 \pm 0.00019$\\
$8.0$ & $0.13933 \pm 0.00022$ & $1.80168 \pm 0.00064$ & $0.87635 \pm 0.00019$\\
$8.5$ & $0.15687 \pm 0.00024$ & $1.80026 \pm 0.00064$ & $0.90313 \pm 0.00019$\\
$9.0$ & $0.17562 \pm 0.00026$ & $1.79871 \pm 0.00064$ & $0.92836 \pm 0.00019$\\
$9.5$ & $0.19569 \pm 0.00029$ & $1.79776 \pm 0.00063$ & $0.95195 \pm 0.00019$\\
$10.0$ & $0.21701 \pm 0.00031$ & $1.79691 \pm 0.00063$ & $0.97442 \pm 0.00019$\\
$10.5$ & $0.23960 \pm 0.00034$ & $1.79628 \pm 0.00063$ & $0.99579 \pm 0.00019$\\
$11.0$ & $0.26361 \pm 0.00036$ & $1.79589 \pm 0.00063$ & $1.01605 \pm 0.00019$\\
$11.5$ & $0.28894 \pm 0.00039$ & $1.79532 \pm 0.00063$ & $1.03544 \pm 0.00019$\\
$12.0$ & $0.31570 \pm 0.00042$ & $1.79489 \pm 0.00063$ & $1.05399 \pm 0.00019$\\
$12.5$ & $0.34386 \pm 0.00045$ & $1.79432 \pm 0.00062$ & $1.07188 \pm 0.00019$\\
$13.0$ & $0.37349 \pm 0.00048$ & $1.79405 \pm 0.00062$ & $1.08902 \pm 0.00019$\\
$13.5$ & $0.40461 \pm 0.00052$ & $1.79343 \pm 0.00062$ & $1.10557 \pm 0.00019$\\
$14.0$ & $0.43722 \pm 0.00054$ & $1.79316 \pm 0.00062$ & $1.12149 \pm 0.00019$\\
$14.5$ & $0.47134 \pm 0.00058$ & $1.79241 \pm 0.00062$ & $1.13694 \pm 0.00019$\\
$15.0$ & $0.50704 \pm 0.00061$ & $1.79162 \pm 0.00061$ & $1.15192 \pm 0.00019$\\
$15.5$ & $0.54431 \pm 0.00066$ & $1.79053 \pm 0.00061$ & $1.16651 \pm 0.00019$\\
$16.0$ & $0.58324 \pm 0.00070$ & $1.78988 \pm 0.00060$ & $1.18054 \pm 0.00020$\\
$16.5$ & $0.62381 \pm 0.00074$ & $1.78933 \pm 0.00060$ & $1.19420 \pm 0.00019$\\
$17.0$ & $0.66613 \pm 0.00078$ & $1.78922 \pm 0.00060$ & $1.20734 \pm 0.00019$\\
$17.5$ & $0.71013 \pm 0.00081$ & $1.78864 \pm 0.00059$ & $1.22017 \pm 0.00019$\\
$18.0$ & $0.75590 \pm 0.00085$ & $1.78805 \pm 0.00059$ & $1.23280 \pm 0.00019$\\
$18.5$ & $0.80337 \pm 0.00090$ & $1.78734 \pm 0.00058$ & $1.24516 \pm 0.00019$\\
$19.0$ & $0.85257 \pm 0.00094$ & $1.78699 \pm 0.00058$ & $1.25721 \pm 0.00019$\\
$19.5$ & $0.90367 \pm 0.00099$ & $1.78674 \pm 0.00058$ & $1.26879 \pm 0.00019$\\
$20.0$ & $0.9565 \pm 0.0010$ & $1.78563 \pm 0.00058$ & $1.28029 \pm 0.00019$\\[1ex]
\end{tabular}
\end{ruledtabular}
\caption{Fitted coefficients for the interpolation formula for $\qhat$ in \eq \eqref{eq:empirical_qhat3}. The values were obtained by numerically integrating \eqref{eq:qhat_formula_pinf} with measure \eqref{eq:qhat_pinf_measure_qlast} and the HTL-screened matrix element \eqref{eq:full_htl_matrix_element}, then numerically fitting the coefficient $\tilde b$ in the region $\Lambda_\perp \gg T$, and finally fitting $\tilde d$ and $\tilde e$ in the range $0.1 T < \Lambda_\perp < 15 T$ using SciPy \cite{Virtanen:2019joe}.}
\label{tab:fitted_empirical3}

\end{table*}

\begin{figure*}
    \centerline{
    \includegraphics[width=0.45\linewidth]{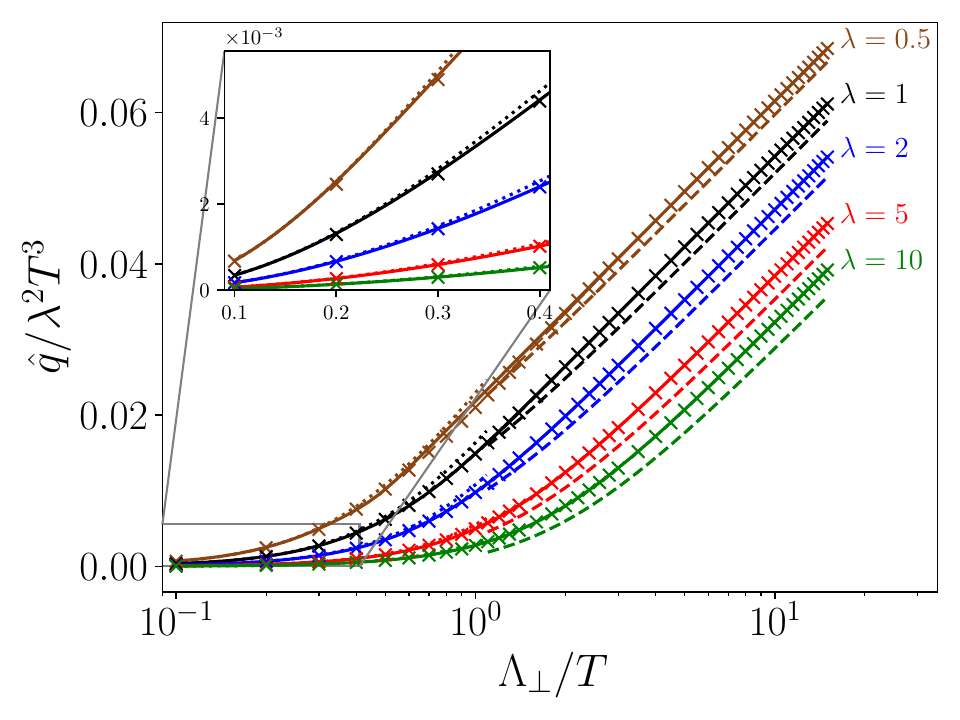}
    \includegraphics[width=0.45\linewidth]{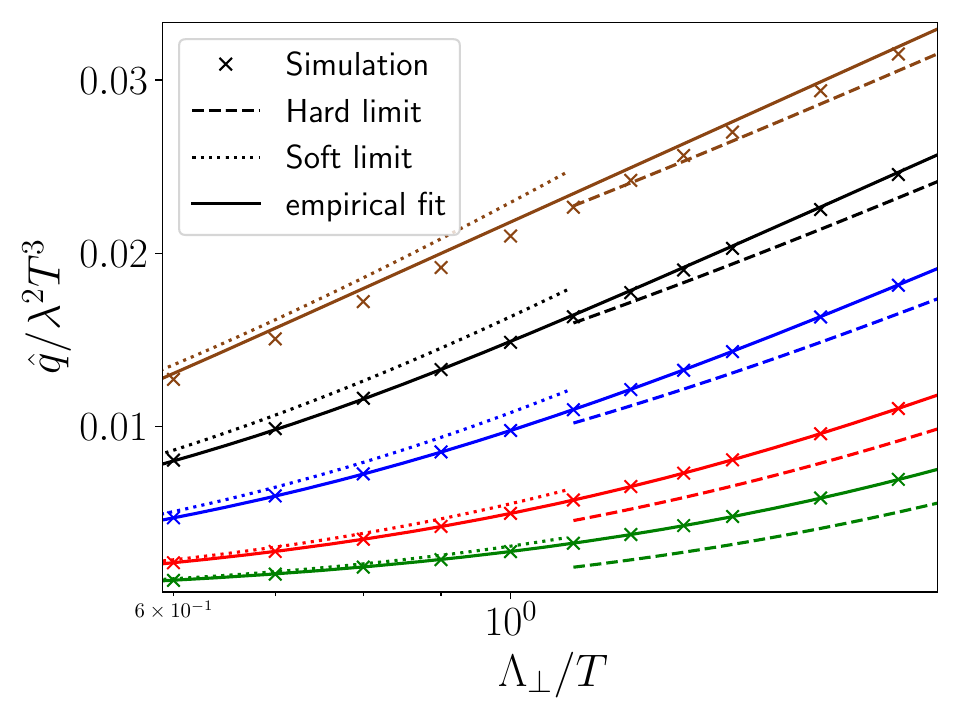}
}
    \caption{{\em (Left:)} The interpolation formula \eqref{eq:empirical_qhat3} for $\qhat$ with the fitted coefficients listed in \tab \ref{tab:fitted_empirical3} is shown as continuous lines for different couplings $\lambda$. The numerical results from \eqref{eq:qhat_formula} are shown as crosses, the limiting expressions for soft \eqref{eq:qhat_thermal_equlibrium_soft} and hard cutoffs \eqref{eq:qhat_hard_arnold} as dotted and dashed curves. The inset shows the behavior at small momentum cutoffs. 
    {\em (Right:)} Focus on the interpolation region $\Lambda_\perp \sim T$. For $\lambda \geq 1$ we see very good agreement with the numerical results.
    }\label{fig:qhat_empirical_fit3_combined}
\end{figure*}

\subsubsection{Interpolation formula for $\qhat$ in thermal equilibrium \label{sec:Empirical_fit_formula}}

We have now observed that the analytical expressions \eqref{eq:qhat_thermal_equlibrium_soft} and \eqref{eq:qhat_hard_arnold} describe $\qhat$ only in certain limits and \eq \eqref{eq:qhat_hard_arnold} only holds for small couplings $\lambda\lesssim 0.5$. For phenomenological calculations, a general formula for $\qhat$ in thermal equilibrium
may be useful without the need of performing the high-dimensional integral \eqref{eq:qhat_formula} numerically for the required value of the coupling $\lambda$ and transverse momentum cutoff $\Lambda_\perp$.
We thus look for an interpolation formula that reproduces the analytical results in the limits $\Lambda_\perp \ll T$ and for $\Lambda_\perp\gg T$ and agrees with our numerical evaluation.

From \eqref{eq:qhat_thermal_equlibrium_soft} and \eqref{eq:qhat_hard_arnold} we know the behavior of $\qhat$ for small $\Lambda_\perp\ll T$ and large cutoffs $\Lambda_\perp\gg T$. As discussed before, \eqref{eq:qhat_hard_arnold} differs from the numerical evaluation of $\qhat$ by a constant shift for larger values of the coupling $\lambda \gtrsim 0.5$. Our strategy is to find an empirical fit function 
that smoothly interpolates in between,
\begin{align}
    \frac{\qhat^{\mathrm{emp}}}{\CR T^3}=
    \begin{cases}
        \tilde c\ln(1+\Lambda_\perp^2/m_D^2)\,, & \text{for } \Lambda_\perp \ll T \\
        \tilde a\ln(\Lambda_\perp/m_D) + \tilde b\,, & \text{for } \Lambda_\perp \gg T\,.
    \end{cases}\label{eq:qhat_empirical_fit_philosophy}
\end{align}
The switching between those two cases will be done using a hyperbole tangent that smears out a step function with width parameter $\tilde d$,
\begin{align}
    \theta_{\tilde d}(x)=\frac{1+\tanh\left(\tilde d x\right)}{2}\,,
\end{align}
which approaches the usual step function for $\tilde d\to\infty$.

This leads to the following form for the fit formula
\begin{align}
\begin{split}
    \frac{\qhat^{\mathrm{emp}}}{\CR T^3}&=\tilde c\ln\left(1+\frac{\Lambda_\perp^2}{m_D^2}\right)\theta_{\tilde d}\left(\tilde e-\ln\frac{\Lambda_\perp}{T}\right)\\
    &+\left(\tilde a\ln\frac{\Lambda_\perp}{m_D} + \tilde b\right)\theta_{\tilde d}\left(\ln\frac{\Lambda_\perp}{T}-\tilde e\right).\label{eq:empirical_qhat3}
\end{split}
\end{align}
For the coefficients $\tilde c$ and $\tilde a$, we use the prefactors of \eqref{eq:qhat_thermal_equlibrium_soft} and \eqref{eq:qhat_hard_arnold}, which 
read for a gluonic plasma 
\begin{align}
    \label{eq:tilde_ac_empfit}
    \tilde c = \frac{\lambda^2}{12\pi \NC}, && \tilde a =\frac{\lambda^2\zeta(3)}{\pi^3 \NC}\,.
\end{align}
This leaves only three fit parameters: The constant $\tilde b$ encodes the linear shift in the large $\Lambda_\perp/T$ region,
while $\tilde d$ and $\tilde e$ describe the width and position of the switching between the two limiting cases in \eqref{eq:qhat_empirical_fit_philosophy}.
We first fit the coefficient $\tilde b$, such that it correctly reproduces $\qhat \simeq \tilde a\ln\Lambda_\perp/T + \tilde b$ in the large $\Lambda_\perp/T$ region. 
We then determine the coefficients $\tilde d$ and $\tilde e$ by fitting them to our numerical data. 

Our results for the remaining fit parameters in \eq \eqref{eq:empirical_qhat3} are listed in \tab\ref{tab:fitted_empirical3} for the couplings $\lambda = 0.5 - 20$. The resulting $\qhat$ are shown
in \fig \ref{fig:qhat_empirical_fit3_combined} as continuous lines. For comparison, we have included the numerically evaluated values as crosses, and the limiting expressions for hard and soft cutoffs, \eqs \eqref{eq:qhat_hard_arnold} and \eqref{eq:qhat_thermal_equlibrium_soft}, respectively, as dashed and dotted lines.
Consistently with the construction of the fit formula, its values are seen to agree well with our numerically evaluated $\qhat$ in the left panel of \fig \ref{fig:qhat_empirical_fit3_combined} and in the inset showing the small cutoff behavior at $\Lambda_\perp \ll T$. 
The right panel of \fig \ref{fig:qhat_empirical_fit3_combined} shows the interpolation region $\Lambda_\perp \sim T$. We find a very good agreement with our numerics for $\lambda \geq 1$, while for smaller couplings $\lambda \lesssim 0.5$ deviations grow in this region. Note that our fit formula provides a smooth interpolating expression for $\qhat$ with improved accuracy in this region as compared to the previous limiting forms.

Our expression \eqref{eq:empirical_qhat3} together with the coefficients in \tab \ref{tab:fitted_empirical3} can thus be used to obtain $\qhat$ in thermal equilibrium for any transverse momentum cutoff $\Lambda_\perp$ and the listed couplings $\lambda$ in the weak coupling leading-order pQCD limit.


\subsection{Toy models for bottom-up thermalization}
\label{sec:toy_models}

Our current weak-coupling understanding of 
how the non-equilibrium quark-gluon plasma created in heavy-ion collisions reaches local thermal equilibrium is based on the bottom-up thermalization scenario of \re \cite{Baier:2000sb}. Strictly speaking, it is only valid in the extremely weak coupling limit, where soft gluon radiation and the LPM effect play an important role and need to be included in the analysis. However, we believe that this picture can also shed light on what happens at intermediate couplings at least at a qualitative level.
It consists of several stages:
The first stage is characterized by a large anisotropy in momentum space as well as an overoccupation  of hard gluon modes. Due to the longitudinal expansion along the beam axis, the anisotropy further increases. As the occupancy of these hard gluons drops below unity, we enter the second stage, in which the momentum anisotropy remains roughly constant, while producing soft gluons through branching, which form a thermal bath. A significant amount of the total energy is still carried by the remaining small number of hard gluons, which, in the third stage, lose energy through multiple hard branchings, until equilibrium is reached.

As toy models for this thermalization process, we consider first an effectively two-dimensional 
distribution in \se \ref{sec:aniso_dist}. We then compute $\qhat$ analytically in \se \ref{sec:scaled-thermal-distribution} using an isotropic scaled thermal distribution, which can be understood as modeling key features of the over- and under-occupied bottom-up stages.

\subsubsection{Effectively two-dimensional distribution}
\label{sec:aniso_dist}

As a model for the large anisotropies encountered at early times in the bottom-up thermalization scenario, let us calculate $\qhat$ in a system brought to its extreme anisotropic limit with vanishing $k_z$ momentum,
\begin{align}
    f(\vb k)=B(k_x,k_y)\delta(k_z/Q), \label{eq:distribution_extremely_anisotropic}
\end{align}
where $B$ is an arbitrary function of $k_x$ and $k_y$, and $Q$ is an energy scale.
Due to its vanishing momentum in beam direction $k_z = 0$, such a state is similar in spirit to the glasma, which is often studied within classical-statistical simulations due to its large field values \cite{Ipp:2020mjc, Ipp:2020nfu, Carrington:2021dvw, Carrington:2022bnv}.

Let us focus on the \emph{Bose-enhanced} part $\qhatff$ in kinetic theory, which agrees with $\qhat$ in a classical-statistical framework since there is no $\qhatf$ contribution in the classical field limit.
By inserting the extremely anisotropic distribution \eqref{eq:distribution_extremely_anisotropic} into the $\qhat$ integral \eqref{eq:qhat_formula} with the measure \eqref{eq:qhat_pinf_measure_klast}, one immediately finds 
\begin{align}
\qhatff^{zz}=0,
\end{align}
due to its proportionality to $\int(q^z)^2\delta(k_z)\delta(k_z')$. 
Note that this is true regardless of the precise form of the matrix element or screening prescription. 
Thus, a purely two-dimensional momentum distribution remains two-dimensional 
in the classical field limit of kinetic theory, if only elastic processes are considered.

We can also consider a special case of \eqref{eq:distribution_extremely_anisotropic} that we can solve analytically: if additionally all particles have a specific momentum $\tilde k$,
\begin{align}
	f(\vec k)= A\,\delta\left(\frac{k_x^2+k_y^2-\tilde k^2}{Q^2}\right) \delta(k_z/Q)\,.\label{eq:special_distribution}
\end{align}
With a jet 
perpendicular to the beam direction and using the approximated gluonic matrix element \eqref{eq:approximated_matrix_element}, one obtains (see \app \ref{app:qhatff_special_dist} for details)
\begin{align}
	\qhatff^{zz} &= 0\\
	\qhatff^{yy} &=\frac{\dA \CA^2 A^2 g^4}{2^7\pi^5d_R\tilde k^3}Q^6\Bigg\{4\tilde k^2\left(\frac{2}{\xi^2m_D^2}-\frac{1}{4\tilde k^2+\xi^2m_D^2}\right)\nonumber\\
	&\qquad\qquad\qquad\qquad\qquad+\ln\frac{\xi^2m_D^2}{4\tilde k^2+\xi^2m_D^2}\Bigg\},\label{eq:qhatffyy_special}
\end{align}
where $m_D^2=A\frac{g^2Q^3}{\pi^2\tilde k}$ according to \eqref{eq:debye_mass_general}.

Indeed we have observed in Ref.~\cite{Boguslavski:2023alu} that in the overoccupied and anisotropic earliest stage of bottom-up thermalization {one has $\qhat^{zz} < \qhat^{yy}$ and that this is due to $\qhatff^{zz}<\qhatff^{yy}$, which is consistent with the simple toy model presented here. In Ref.~{\cite{Boguslavski:2023alu}}, we then showed that $\qhatff^{yy}$ quickly becomes similar to $\qhatff^{zz}$ and that both become much smaller than $\qhatf$.}

\subsubsection{Scaled thermal distribution\label{sec:scaled-thermal-distribution}}

Let us now study another aspect encountered during bottom-up thermalization: over- and underoccupied systems. For simplicity, we use an isotropic toy model and consider a scaled thermal distribution, i.e., we scale the amplitude of the thermal distribution \eqref{eq:thermal_bose_fermi_combined} with $N_\pm$. Here $N_+$ denotes the scaling parameter of the Bose-Einstein distribution and $N_-$ the scaling parameter of the Fermi-Dirac distribution,
\begin{align}
f_{\pm}(k; T)=\frac{N_\pm}{\exp(k/T)\mp1}\,.\label{eq:scaled_bose_fermi_combined}
\end{align}
This allows us to easily
generalize the results obtained in \se\ref{sec:qhat-thermal} for $\qhat$ in a thermal medium, and we start with $\qhat$ given by \eq \eqref{eq:qhat_soft_distribution_function_split_off}. Splitting the $f$ and $ff$ contributions 
and using the integrals \eqref{eq:thermal_distributionfunctions_integrals} over thermal distributions, 
we obtain for small cutoff 
\begin{align}
	&\qhat(\Lambda_\perp\ll T, N_\pm)=\frac{g^4T^3\CR}{24\pi^3}\ln\left(1+\frac{\Lambda_\perp^2}{m_D^2}\right) \nonumber\\
	&\qquad \times\Big(\pi^2(2\NC(\NOg)^2  +\nf (\NOq)^2 )\label{eq:qhat_scaled_soft}\\
	&\qquad +\zeta(3)\left[9\nf \NOq(1-\NOq)-12\NC\NOg(\NOg-1)\right]\Big), \nonumber
\end{align}

which generalizes
the equilibrium ($N_\pm=1$) result in \eq \eqref{eq:qhat_thermal_equlibrium_soft}. Similarly, we can generalize the large-cutoff 
formula \eqref{eq:qhat_hard_arnold} to
\begin{subequations}
\label{eq:qhat_thermal_hard_analytic_combined_notimproved}
\begin{align}
	\begin{split}
		\qhat(\Lambda_\perp\gg T, N_\pm)&=
		\CR \frac{g^4T^3}{\pi^2}\sum_\pm\Xi_\pm
		\mathcal I_\pm(\Lambda_\perp, N_\pm)\label{eq:qhat_hard_arnold_scaled_not_improved}
	\end{split}\\
	\mathcal I_\pm(\Lambda_\perp, N_\pm)&=\frac{N_\pm\zeta_\pm(3)}{2\pi}\ln\left(\frac{\Lambda_\perp}{m_D}\right)+(N_\pm)^2\Delta\mathcal I_\pm,\label{eq:qhat_thermal_hard_analytic_Ipm_scaled_notimproved}
\end{align}
with $\Delta\mathcal I_\pm$ given by \eq \eqref{eq:qhat_thermal_hard_analytic_DeltaI}, which is entirely determining $\qhatff$,
\begin{align}
    \qhatff(\lperp\gg T, N_\pm)=\CR\frac{g^4T^3}{\pi^2}\sum_{\pm}\left(N_\pm\right)^2\Xi_\pm \Delta\mathcal I_\pm.\label{eq:scaled_qhatff_notimproved}
\end{align}
\end{subequations}

Furthermore, similarly to our discussion for the thermal result in \se \ref{sec:thermal_qhat_large_momentum_cutoff}, by replacing $2\ln (\Lambda_\perp/m_D) \to \ln(1+(\Lambda_\perp/m_D)^2)$ in \eqref{eq:qhat_thermal_hard_analytic_Ipm_scaled_notimproved},
we obtain an `improved' formula valid for large cutoffs that is finite even at small $\Lambda_\perp$ and generalizes
\eq \eqref{eq:qhat_hard_arnold-improved}.
Then we can again split off the Bose-enhanced contribution as in \eqref{eq:qhat_f_ff_splitting},
    $\qhat = \qhatf+\qhatff$, 
and realize that
$\qhatf(\lperp)$ has the same form for small and large cutoffs, 
\begin{subequations}
\label{eq:qhat_scaled_individual_components_combined}
\begin{align}
	\qhatf(\Lambda_\perp, N_\pm)&= \zeta(3)\left(12\NOg\NC +9\nf\NOq\right)C_L(\Lambda_\perp)\,, \label{eq:qhat_scaled_f}
\end{align}
with, as before, $C_L(\Lambda_\perp)=\frac{g^4T^3 \CR}{24\pi^3}\ln\left(1+\frac{\Lambda_\perp^2}{m_D^2}\right)$.
In contrast, the Bose-enhanced terms differ 
\begin{align}
	\qhatff(\lperp\ll T,N_\pm)&=\Big[2\NC(\NOg)^2(\pi^2-6\zeta(3))\label{eq:qhatff_scaled_soft}\\
    &\quad~+\nf(\NOq)^2(\pi^2-9\zeta(3))\Big]C_L(\Lambda_\perp)\nonumber\\
    {\qhatffimproved(\lperp\gg T,N_\pm)}&={\CR g^4T^3}\sum_\pm\Xi_\pm(N_\pm)^2\label{eq:qhatff_hard_scaled_improved}\\
&
\times \left\lbrace
 \frac{\zeta_\pm(2)-\zeta_\pm(3)}{4\pi^3}\left[\ln\left(1+\frac{T^2}{m_D^2}\right) \right.\right. \nonumber \\
 & \quad\left.\left.+1-2\gamma_E+2\ln 2\right]
    -\frac{\sigma_\pm}{2\pi^3}\right\rbrace. \nonumber
\end{align}
\end{subequations}
The Debye mass entering these expressions for the scaled thermal distributions is given by (see \eq \eqref{eq:debye_mass_general})
\begin{align}
	m_D^2=\frac{g^2T^2}{3}\left(\NOg\NC  +\frac{\NOq\nf }{2}\right). \label{eq:debye_mass_scaled_thermal}
\end{align}
Thus, $m_D$ scales with $\sqrt{\lambda N_\pm}$. For large occupancies $N_+$, this may pose a problem for the validity of perturbation theory that assumes $m_D\ll T$ and that our arguments and the derivations in \cite{Arnold:2008vd} were based on.
The occupation of fermions $N_-$ cannot become large due to Pauli blocking.
 We can estimate the breakdown scale by requiring $m_D\ll T$, which leads to 
\begin{align}
    \NOg \ll \frac{1}{\NC}\left(\frac{3}{g^2}- \frac{\NOq\nf}{2}\right).\label{eq:qhat_scaled_hard_limitation}
\end{align}
This is consistent with the usual limitations of perturbation theory, which breaks down at nonperturbatively large occupation numbers $f \gtrsim 1/g^2$.

\begin{figure*}
    \centerline{
        \includegraphics[width=0.45\linewidth]{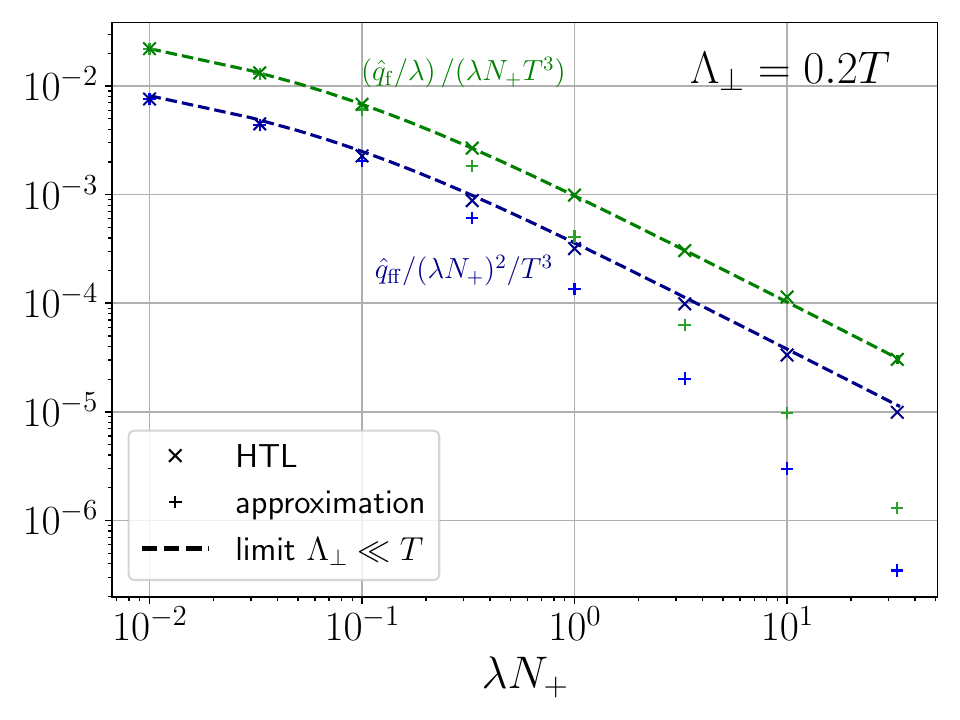}
        \includegraphics[width=0.45\linewidth]{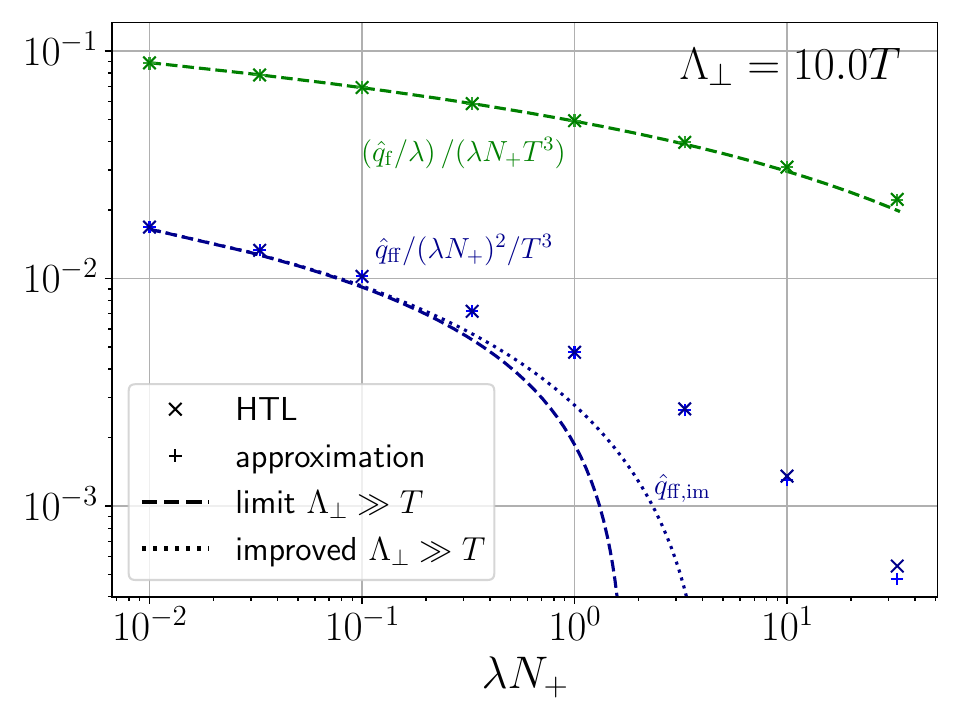}
    }
    \caption{\label{fig:qhat_components_combined}The individual components $\qhatf$ (green) and $\qhatff$ (blue) as defined in \eqref{eq:qhat_f_ff_splitting} and rescaled according to their parametric estimates in \eqref{eq:qhat_components_naive_behavior} as functions of their only argument $\lambda N_+$ for $\Lambda_\perp/T=0.2$ (left) and $\lperp/T=10$ (right). 
   The numerical HTL values from our calculation are shown with $\times$-symbols, and the numerical values using the $\xi$-approximated matrix element \eqref{eq:gluonic_matrix_el_simple_screening} are labeled ``approximation'' and shown by $+$-symbols. Our analytic estimate for $\qhatf$, Eq.~\eqref{eq:qhat_scaled_f},  and for $\qhatff$, Eqs.~\eqref{eq:qhatff_scaled_soft} for $\lperp \ll T$  and~\eqref{eq:scaled_qhatff_notimproved} for $\lperp \gg T $.
In the left panel, we show   the small-cutoff form \eqref{eq:qhatff_scaled_soft} for $\qhatff$ and Eq.~\eqref{eq:qhat_scaled_f} for $\qhatf$ labeled as ``$\lperp\ll  T$''.
 In the right panel, we show both high-cutoff expressions \eqref{eq:qhat_scaled_f} for $\qhatf$ and \eqref{eq:scaled_qhatff_notimproved} for $\qhatff$
 labeled ``limit $\lperp \gg T$''  and its improved version \eqref{eq:qhatff_hard_scaled_improved}
 labeled as ``improved $\lperp \gg T$''.
    }
\end{figure*}
\begin{figure*}
    \centerline{
        \includegraphics[width=0.45\linewidth]{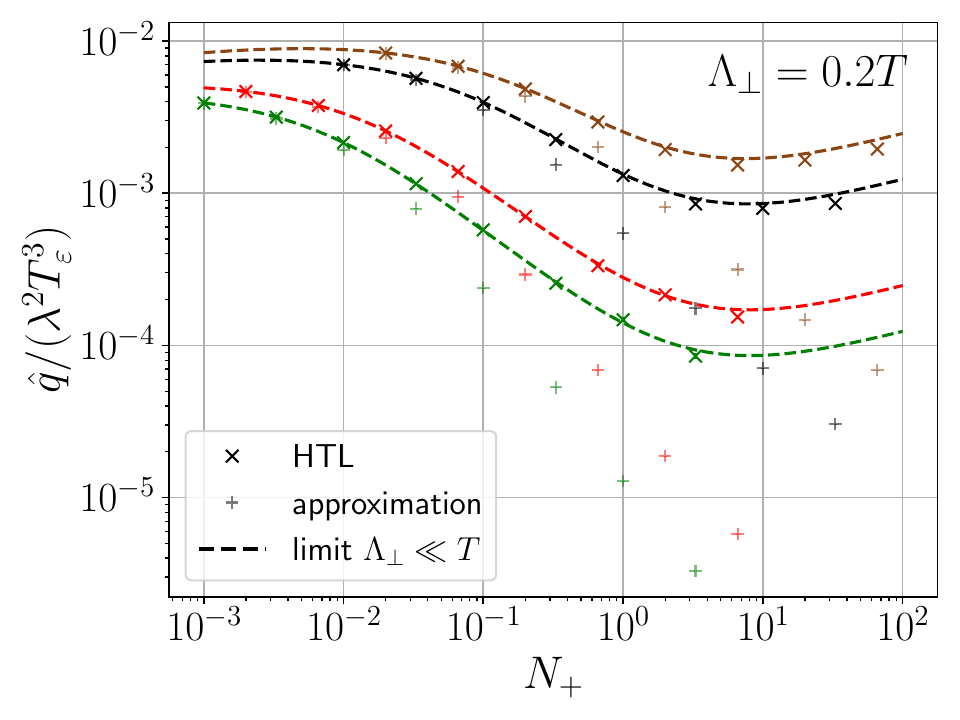}
        \includegraphics[width=0.45\linewidth]{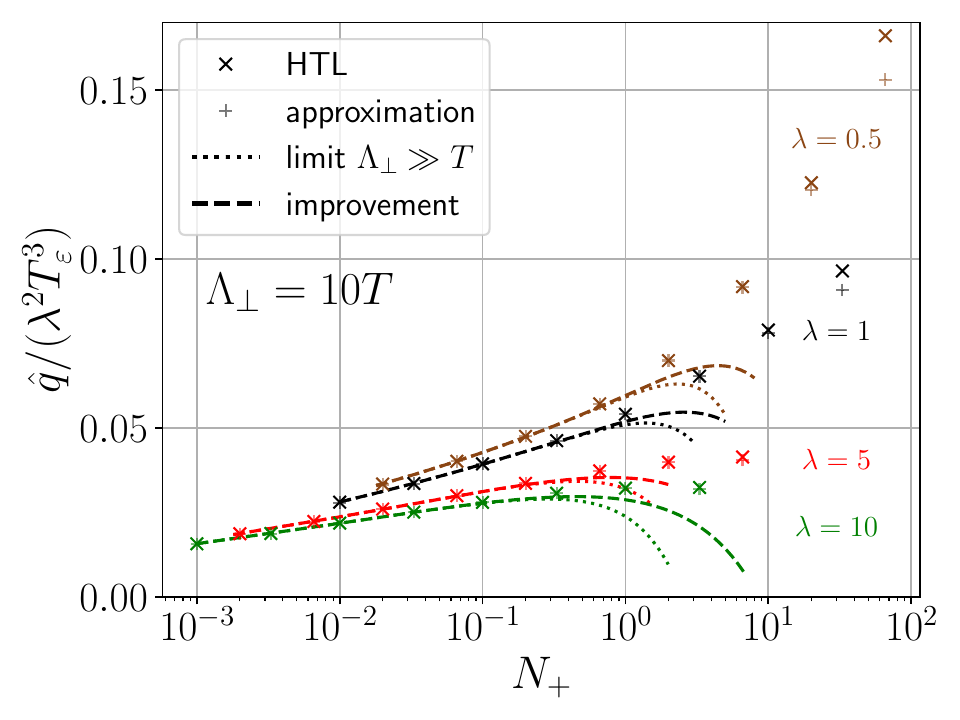}
    }
    \caption{\label{fig:qhat_scaled_combined}
The momentum broadening coefficient $\qhat$ for a scaled thermal distribution
\eqref{eq:scaled_bose_fermi_combined} as a function of $N_+$ for different values of $\lambda$ for $\Lambda_\perp/T=0.2$ (left) and $\Lambda_\perp/T=10$ (right). As in Fig.~\ref{fig:qhat_components_combined}, the numerical HTL values from our calculation are shown with $\times$-symbols, and the values using the $\xi$-approximated matrix element \eqref{eq:gluonic_matrix_el_simple_screening} are shown as $+$-symbols.
In the left panel, we show   the small-cutoff form \eqref{eq:qhat_scaled_soft}   labeled as ``$\lperp\ll  T$''.
In the right panel, we show both the high-cutoff expression
Eq.~\eqref{eq:qhat_hard_arnold_scaled_not_improved} labeled as ``limit $\lperp \gg T$'' and the improved version obtained by summing Eqs.~\eqref{eq:qhat_scaled_f} and \eqref{eq:qhatff_hard_scaled_improved} labeled as ``improvement'.
    }
\end{figure*}

Let us now assess the expressions derived above by comparing them to the numerical evaluation of $\qhat$ using \eq \eqref{eq:qhat_formula}, before applying the formulae to initial stages in heavy-ion collisions. 
We start with $\qhatf/\lambda$ and $\qhatff$ in \eqs \eqref{eq:qhat_scaled_individual_components_combined}, which are functions of the combination $\lambda N_+$ (we refer to \app\ref{app:scaled_thermal_technical_details} for details), i.e., 
\begin{align}
    \qhat(\Lambda_\perp, N_+, \lambda) = \lambda\left(\frac{\qhatf}{\lambda}\right)(\Lambda_\perp,\lambda N_+)+ \qhatff (\Lambda_\perp,\lambda N_+). \label{eq:scaled_thermal_splitting}
\end{align}
These contributions are plotted in \fig \ref{fig:qhat_components_combined} for small (left) and large (right) cutoffs $\Lambda_\perp = 0.2 T$ and $10 T$, respectively, divided by the prefactor
\begin{align}
    \qhatf \sim \lambda (\lambda N_+) T^3, && \qhatff \sim (\lambda N_+)^2T^3.\label{eq:qhat_components_naive_behavior}
\end{align}
We observe that 
their values deviate significantly from the simple estimates in \eq \eqref{eq:qhat_components_naive_behavior}.
This is a consequence of 
screening effects and the scaling of the Debye mass. In particular, one finds for sufficiently small cutoffs $\Lambda_\perp \lesssim m_D, T$ and large occupancies that
\begin{align}
    \label{eq:invLNplus}
    \frac{\qhatf}{\lambda^2 N_+ T^3} \sim \frac{\qhatff}{\lambda^2 N_+^2 T^3} \sim \frac{\Lambda_\perp^2}{m_D^2} \sim (\lambda 
 N_+)^{-1}\,,
\end{align}
which is visible in the left panel of \fig \ref{fig:qhat_components_combined} for sufficiently large $\lambda N_+$. Note that for a sufficiently large $\lambda N_+ \gg 1$ the effective kinetic theory description used here ceases to be valid. 
Similarly to the equilibrium case discussed in \se \ref{sec:qhat-thermal} and particularly in \fig \ref{fig:thermal_equilibrium}, the expression for small cutoffs \eqref{eq:qhat_scaled_soft} nicely agrees with the numerical values in the small cutoff asymptotic region, plotted in the left panel of \fig\ref{fig:qhat_components_combined}. In the right panel, for large cutoffs, 
we observe that our analytical form for $\qhatf$ in Eq.~\eqref{eq:qhat_scaled_f} remains a very good description coinciding with our numerical values, whereas the analytical estimate for $\qhatff$ in Eq.~\eqref{eq:scaled_qhatff_notimproved} (and its improvement Eq.~\eqref{eq:qhatff_hard_scaled_improved}) ceases to describe the data for nonperturbatively large occupancies $\lambda N_+ \gtrsim 1$. This is expected from the condition \eqref{eq:qhat_scaled_hard_limitation}, and we see sizable deviations already at $\lambda N_+ \gtrsim 0.1$.

The full HTL screening and the approximation with a constant $\xi m_D$ regulator \eqref{eq:approximated_matrix_element} nicely agree with each other at large cutoffs for the whole $\lambda N_+$ range despite the aforementioned limitations concerning $\qhatff$.
On the other hand, for small cutoffs (left panel), the $\xi m_D$ screening approximation shows large deviations from the full HTL screening, albeit in the large $\lambda N_+$  region that should be taken with caution, as discussed above. 
The resemblance to the thermal case here is of course no coincidence since by setting $N_+=1$ we recover the thermal results.

Recombining the contributions from $\qhatf$ and $\qhatff$, 
we show $\qhat$ in \fig\ref{fig:qhat_scaled_combined} for the couplings $\lambda = 0.5$, $1$, $2$, $5$ and $10$ as functions of the occupancy $N_+$, for the small cutoff $\Lambda_\perp/T=0.2$ in the left panel and the large cutoff $\Lambda_\perp/T=10$ in the right panel.
The values are shown scaled by the effective temperature $\Teps$ that represents the temperature of a thermal system with the same energy density, $\varepsilon = \nu_g \pi^2\Teps^4/30$ (c.f., \eqref{eq:kinetic_energy_T}),
and thus
\begin{align}
    \Teps = (\NOg)^{1/4}T. \label{eq:effective_temperature}
\end{align}
For comparison, we plot 
the analytic predictions for small \eqref{eq:qhat_scaled_soft} and large cutoffs \eqref{eq:qhat_thermal_hard_analytic_combined_notimproved} as well as its improved expression \eqref{eq:qhat_scaled_individual_components_combined}.
Similarly as for $\qhatf$ and $\qhatff$, we observe for $\qhat$ in \fig \ref{fig:qhat_scaled_combined} that the small-cutoff expression agrees well with our (HTL-)screened data points while the large-cutoff expressions describe the data points until $N_+ \lesssim 1/\lambda$. 
Moreover, the improved formula for large cutoffs increases the validity of the analytic result only to slightly larger occupancy $N_+$.
This plot emphasizes the importance of screening effects that prevent the na\"ive scaling with $N_+$ or $N_+^2$. 
We therefore have to be cautious when we want to use such analytic expressions to describe over-occupied systems with typical occupancies $N_+ \sim 1/\lambda$. 
Instead, transport coefficients in such systems can be studied using classical statistical lattice simulations \cite{Boguslavski:2020tqz, Ipp:2020mjc, Ipp:2020nfu, Avramescu:2023qvv}. In particular, it has been shown \cite{Boguslavski:2020tqz} that nonperturbative corrections can be substantial.
Interestingly, as visible in \fig\ref{fig:qhat_scaled_combined}, increasing the occupancy $N_+$ does not appear as drastically increasing the value of the jet quenching parameter $\qhat$. In particular, for small cutoff $\lperp/T=0.2$ visible in the left panel, we observe that the scaled $\qhat$ in fact decreases with increasing occupancy.
Even for large cutoffs (right panel), increasing the occupancy by several orders of magnitude only leads to a slight increase in the jet quenching parameter.
This behavior is due to a combination of two effects. 
The first effect is that the increasing occupation number also increases the Debye mass $m_D$. Thus one conclusion of our analysis is that a detailed understanding of screening effects is particularly important for a quantitative analysis of $\qhat$.
The second effect is that we are dividing the value of $\qhat$ with the third power of the effective temperature, which increases with the occupancy $\Teps \sim N_+^{1/4}$ when the hard momentum scale $T$ is kept fixed.
\begin{figure}
	\centering
	\includegraphics[width=\linewidth]{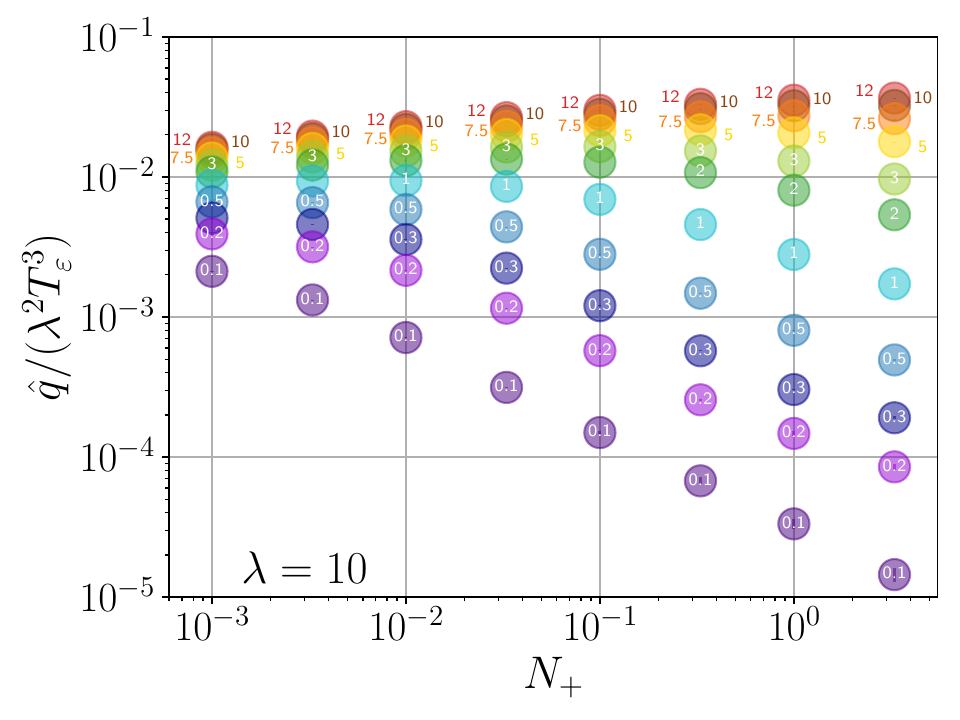}
	\caption{Numerical values of $\qhat$ for a scaled Bose-Einstein distribution \eqref{eq:scaled_bose_fermi_combined} as a function of its amplitude $\NOg$ for different momentum cutoffs $\Lambda_\perp$ (numbers in circle markers) and coupling $\lambda=10$.}
	\label{fig:qhat_scaled_all}
\end{figure}

In \fig \ref{fig:qhat_scaled_all} we provide an overview of the numerical values of $\qhat$ for the phenomenologically relevant coupling $\lambda = 10$ in heavy-ion collisions. 
Different values of the cutoff $\Lambda_\perp$
are color-coded and written in the circle markers in the figure. 
We observe the same behavior at small and large cutoffs that we have found in \fig \ref{fig:qhat_scaled_combined}. This involves a fast (power-law) decrease with growing occupancy $N_+$ at small cutoffs as $\qhat / (\lambda^2 \Teps^3) \sim N_+^{-3/4}$, and a slow growth at high cutoffs.
We additionally see how $\qhat$ interpolates smoothly between these two behaviors at small and large cutoffs. 
From a physical point of view, this confirms the observation that for small cutoffs, jet quenching in an over-occupied (isotropic) system similar to a scaled thermal distribution may be strongly suppressed. However, we repeat our note of caution below \eqref{eq:invLNplus} that these parameters may lie beyond the range of applicability of our original integral formula for $\qhat$ \eqref{eq:qhat_formula}.

\begin{figure}
    \includegraphics[width=\linewidth]{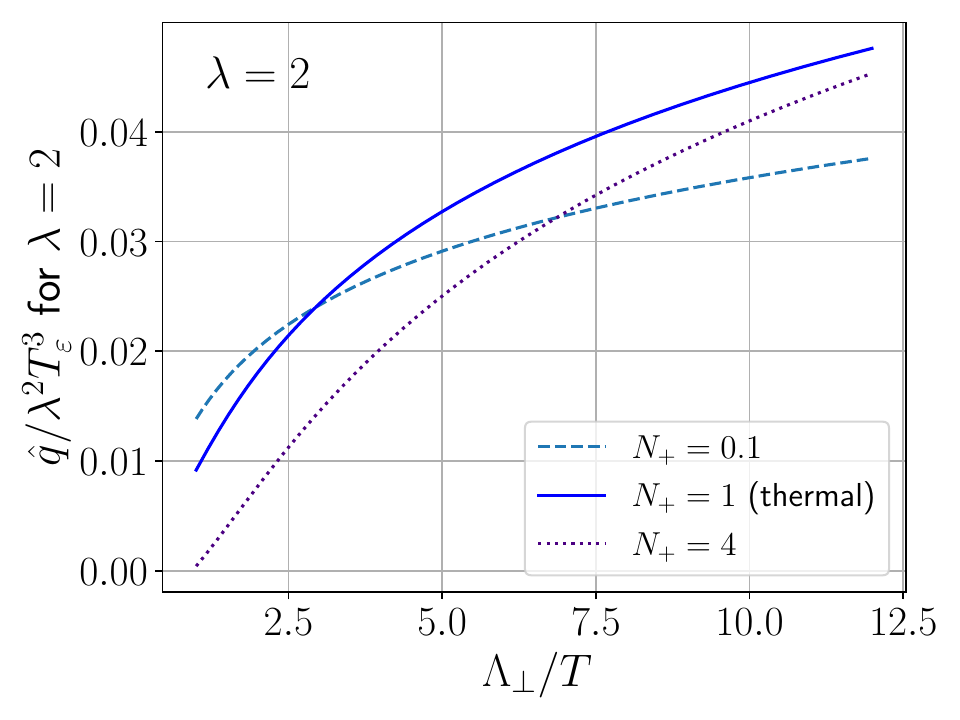}
    \caption{\label{fig:scaled_thermal_comparison}Jet momentum broadening parameter $\qhat$ for a scaled thermal distribution with various scaling coefficients $\NOg$ and coupling $\lambda=2$ as a function of the transverse momentum cutoff $\Lambda_\perp$.}
\end{figure}

Let us finally apply our analytical results and conclusions of this section to the initial stages in heavy-ion collisions, and in particular to the bottom-up thermalization scenario, whose over- and under-occupied stages we wish to qualitatively understand using the scaled thermal distribution as a toy model. 
Although we have reported recently in \res \cite{Boguslavski:2023alu, Boguslavski:2023fdm} of numerical kinetic theory simulations during bottom-up thermalization where we have studied transport coefficients including the jet quenching parameter $\qhat$,
in the present work, we are able to provide more insight into its pre-equilibrium features by using our analytical expressions.

In \fig\ref{fig:scaled_thermal_comparison} we compare $\qhat/(\lambda^2\Teps^3)$ for scaled thermal distributions 
representing an under-occupied system ($\NOg=0.1$), thermal equilibrium ($\NOg=1$) and over-occupancy ($\NOg=4$), and plot its value as a function of the momentum cutoff $\Lambda_\perp$ in the large $\Lambda_\perp$ region (Eq.~\eqref{eq:qhat_scaled_individual_components_combined}). 
We find that for under-occupied systems and small cutoff, $\qhat$ is larger than its thermal value, whereas for large cutoffs this is reversed. This confirms our 
numerical simulation results of bottom-up thermalization in \re \cite{Boguslavski:2023alu}.
One implication is that for relatively small cutoffs $\Lambda_\perp \gtrsim T$, which are comparable to the momentum carried by gluons in the plasma, collisional momentum broadening is more efficient in under-occupied plasmas than in thermal equilibrium at the same energy density. While our analytical result for large cutoffs but small occupancies agrees with the numerical evaluation of $\qhat$ in \eqref{eq:qhat_formula}, this does not imply that jets experience less broadening than in thermal equilibrium. In fact, as explained in \re \cite{Boguslavski:2023alu} and discussed in \se \ref{sec:momentum_cutoffs}, the momentum cutoff should be rather taken as a function of the jet energy and plasma temperature. It turns out that for realistic models of the momentum cutoff like \eqref{eq:kinematic_cutoff} and \eqref{eq:LPM_cutoff}, $\qhat$ exceeds the thermal value for the same energy density even in the under-occupied phases of bottom-up.

{The main difference between the scaled thermal distributions and Ref.~\cite{Boguslavski:2023alu} is that in the latter the system is characterized by a large momentum anisotropy, while our scaled thermal distribution is isotropic.} 
For over-occupied systems similar to those encountered initially in the bottom-up scenario,
our analytical study { using the scaled thermal distribution} suggests 
that $\qhat$ is always smaller than its thermal value. However, in \cite{Boguslavski:2023alu} we do not find this specific ordering during the early over-occupied stage in our numerical simulations. We take this as a numerical indication that a scaled thermal distribution does not describe this stage accurately. 

In the underoccupied phase, on the other hand, we see an enhancement of $\qhat$ for small $\lperp$ and a suppression at large $\lperp$, compared to a thermal system at the same energy density. This is consistent with our observation in Ref.~\cite{Boguslavski:2023alu} and indicates that using a scaled thermal distribution as a model of the underoccupied stage of bottom-up is a better approximation.

\section{Summary and conclusions}
\label{sec:concl}
In this paper we have generalized the calculation of the jet quenching parameter $\qhat$ to an anisotropic non-equilibrium plasma, using QCD effective kinetic theory. We describe in detail the treatment of the phase space needed for the calculation and implement the integration numerically in a kinetic theory simulation. This generalizes the usual form of $\qhat$ to a tensor that encodes momentum broadening in different directions relative to the jet, which is important for non-equilibrium systems. In Ref.~\cite{Boguslavski:2023alu} we used the expressions obtained here to study $\qhat$ during the initial stages of heavy-ion collisions.

We use an isotropic hard-thermal loop (HTL) screening prescription as well as a simple approximation thereof and provide a formula for finite jet energy and arbitrary jet angle with respect to the beam axis. Additionally, we provide expressions in the limit of infinite jet momentum, in which a transverse momentum cutoff needs to be introduced to render $\qhat$ finite. 
As a part of the derivation, we have investigated different screening prescriptions for $t$-channel gluon exchange. We give an explicit expression for the HTL form of the matrix element entering $\qhat$ and compare this full HTL matrix element to a simple screening prescription that is typically employed in EKT simulations. In particular, we find that matching the simple screening prescription to HTL in the case of transverse momentum broadening requires regulating the gluon propagator by a scale $(\xi m_D)^2$ with $\xi=e^{1/3}/2$. This value is different  than the value used in previous studies for the elastic scattering kernel. Even with this matching value of $\xi$, for small momentum cutoffs $\Lambda_\perp$ there are in some cases sizeable deviations up to 30\% in the values of $\qhat$ in thermal systems.

We also study in detail the leading logarithmic behavior of the scattering term in the Boltzmann equation in the forward scattering limit. 
 We show explicitly how a logarithmic divergence in $p$ arises from the integral of the scattering matrix element at large $p$. Due to this divergence, in  eikonal limit $p\to\infty$ a 
momentum transfer cutoff $\Lambda_\perp > \qperp$ must be introduced. We show how, conversely,  $\qhat$ at $p\to \infty$ depends logarithmically on the cutoff, $\qhat = a\ln\lperp+b$, and we find a simple expression for the coefficient $a$.

We then move to study the value of $\qhat$ in specific cases. We first recover known results in limiting values for the cutoff $\lperp$ in a thermal distribution. By evaluating $\qhat$ numerically in a thermal distribution for arbitrary values of $\lperp$, we provide an explicit interpolation formula in \eq \eqref{eq:empirical_qhat3} with fitted coefficients listed in \eqref{eq:tilde_ac_empfit} and \tab \ref{tab:fitted_empirical3}, that smoothly interpolates between analytical results at small and large momentum cutoffs. Our formula provides an accurate approximation of $\qhat$ in thermal equilibrium for all cutoffs $\Lambda_\perp$ and various couplings $0.5 \leq \lambda \leq 20$.

As a background for our study of bottom-up thermalization in Ref.~\cite{Boguslavski:2023alu}, we then study toy models for aspects of the thermalization process.
 We first confirm 
that for a maximally anisotropic plasma with no longitudinal momentum, only terms linear in the distribution function can contribute to $\qhat^{zz}$. This feature is clearly visible in the simulation of Ref.~\cite{Boguslavski:2023alu}, where $\qhat^{zz}$ becomes larger than $\qhat^{yy}$ when the system transitions from the overoccupied to the underoccupied regime. 
We then calculate $\qhat$ analytically for a scaled thermal distribution to obtain improved insight into the under- and over-occupied plasma dynamics 
during the initial stages. 
Generalizing previous results in thermal equilibrium, we derive and discuss analytic formulas of the components entering $\qhat$ as functions of the bosonic and fermionic scaling occupancies $N_\pm$. We discuss their range of validity and compare to our numerically computed $\qhat$. 
We observe that at large cutoffs, the ratio  $\qhat/(\lambda^2T_\varepsilon^2),$ with a Landau-matched temperature $T_\varepsilon$,  grows slowly with $N_+$. 
 At small cutoffs, the ratio  $\qhat/(\lambda^2T_\varepsilon^2)$   decreases rapidly with $N_+$.
This implies that for under-occupied systems $N_+ \ll 1$ the value of $\qhat$ exceeds the value of a thermal system with the same energy density at small cutoffs, and is smaller at large cutoffs. This provides further insight into the calculation during bottom-up thermalization in Ref.~\cite{Boguslavski:2023alu}. 

Our computation of $\qhat$ with full isotropic HTL self-energies goes beyond typical screening approximations with a constant mass regulator $\xi m_D$ usually employed in EKT simulations and is leading-order accurate for isotropic systems. However, we note that our screening formulation does not capture the dynamics of plasma instabilities consistently, since these have not been incorporated into the EKT implementations yet. 
There have been recent efforts to tackle the problem of anisotropic screening \cite{Hauksson:2021okc, Hauksson:2023dwh, Zhao:2023mrz}, and we leave the task of including it into $\qhat$ to future studies. 

Although this paper focuses on transverse momentum broadening, the integral expression \eqref{eq:qhat_formula} can also be used to describe longitudinal momentum broadening, $\qhat^L=\qhat^{33}$ and collisional energy loss, which we plan to study. In future work, we also want to investigate the differences between the finite $p$ expressions of $\qhat$ and the $p\to\infty$ limit with cutoff $\Lambda_\perp$. 
Our expressions have already been used in \cite{Boguslavski:2023alu} to study the initial stages in heavy-ion collisions. 
Together with the present work, this paves the way for further kinetic theory studies of jet quenching in anisotropic and pre-equilibrium systems.

\begin{acknowledgments}
The authors would like to thank X.~Du, S.~Hauksson, A.~Ipp, J.G.~Milhano, D.I.~M\"uller, A.~Sadofyev, C.~Salgado, and B.~Wu for valuable discussions. 
KB and FL are grateful to the Austrian Science Fund (FWF) for support under project DOI 10.55776/P34455, and FL is additionally supported by the Doctoral Program Particles and Interactions [DOI 10.55776/W1252].
TL and JP are supported by the Academy of Finland, the Centre of Excellence in Quark Matter (project 346324)
and project 321840  and by the European Research Council under project ERC-2018-ADG-835105 YoctoLHC.
This work was funded in part by the Knut and Alice Wallenberg foundation, contract number 2017.0036. 
This work was also supported under the European Union's Horizon 2020 research and innovation by the STRONG-2020 project (grant agreement No.~824093). The content of this article does not reflect the official opinion of the European Union and responsibility for the information and views expressed therein lies entirely with the authors.
\end{acknowledgments}

\vspace*{\baselineskip}


\appendix

\section{Boundaries of the phase-space integration \label{app:qhat_derivation_integral_boundaries}}

\begin{widetext}
Here we give a detailed derivation of the boundaries for the phase-space integrals of $\qhat$. We start with \eqref{eq:qhat_general}, which we can rewrite using 4-dimensional integrals,
\begin{align}
\hat q^{ij} &= \frac{1}{2p\nu_a}\sum_{bcd}\int\frac{\dd[4]{K}\dd[4]{P'}\dd[4]{K'}}{(2\pi)^5}\qperp^i\qperp^j\delta^{(4)}(P+K-P'-K')\left|\mathcal M_{cd}^{ab}(\vb p,\vb k;\vb p',\vb k')\right|^2 \nonumber\\
&\qquad \qquad \qquad \times\delta(K^2)\delta(P'^2)\delta(K'^2)\theta(K^0)\theta(P'^0)\theta(K'^0)f^b(\vb k)\left[1\pm f^d(\vb k')\right]\theta(p'-k'). 
\end{align}
\end{widetext}
We eliminate the $K'$ integration using the energy-momentum conserving delta function, but
will still write $K'$ or $\vb k'$ as a short notation for $P+K-P'$ or $\vb p+\vb k-\vb p'$.
We will proceed similarly as in \cite{Arnold:2003zc}. We introduce $Q^\mu=(\omega, \vb q)^\mu$ as in \eqref{eq:definition_Q} via
\begin{align}
\begin{split}
Q=P'-P\quad \Leftrightarrow\quad & \vb q = \vb p' - \vb p = \vb k - \vb k'\\
& \omega = p' - p = k - k'. 
\end{split}
\end{align}
Note that unlike the external momenta $P,P',K,K'$, the momentum transfer $Q$ is not  light-like, i.e., $Q^2=\vb q^2-\omega^2\geq 0.$ 
Then we have 
$\dd[4] K\dd[4]P'=\dd[4]K\dd[4]Q$ and thus
\begin{align}
	\hat q^{ij} &= \frac{1}{2p\nu_a}\sum_{bcd}\int\frac{\dd[4]{K}\dd[4]{Q}}{(2\pi)^5}\qperp^i\qperp^j\left|\mathcal M_{cd}^{ab}(\vb p,\vb k;\vb p',\vb k')\right|^2\nonumber\\
	&\qquad \times \theta(p'-k')\theta(K^0)\theta(P^0 + \omega)\theta(K^0 - \omega)\nonumber\\
	&\qquad\times \delta(K^2)\delta\left((P+Q)^2\right)\delta\left((K-Q)^2\right) \nonumber\\
	&\qquad\times f^b(\vb k)\left[1\pm f^d(\vb k')\right].
\end{align}
Using $P^2=K^2=0$, $P\cdot Q=-p\omega + pq\cos\theta_{pq}$ and $K\cdot Q=-k\omega + kq\cos\theta_{kq}$, we can rewrite the last two delta functions as
\begin{align}
\begin{split}
	&\delta\left((P+Q)^2\right)\delta\left((K-Q)^2\right)\\
	&=\frac{1}{4pkq^2}\delta\left(\cos\theta_{pq}-\frac{\omega}{q}-\frac{\omega^2-q^2}{2pq}\right)\\
	&\quad\qquad\times\delta\left(\cos\theta_{kq}-\frac{\omega}{q}+\frac{\omega^2-q^2}{2kq}\right).
\end{split}
\end{align}
Because of this expression, it is beneficial to perform the $\vec q$ integral in a coordinate frame in which $\theta_{pq}$ is its polar angle and the $\vec k$ integral in a frame in which $\theta_{kq}$ is its polar angle. This is one of the reasons for introducing the different coordinate systems in \se\ref{sec:coordinate_systems}.
The delta function only contributes if its argument becomes zero, which restricts the integration region to the one indicated in Eq.~\eqref{eq:phase_space_relation_omega_q_k},
\begin{align}
|\omega| < q, && p >\frac{q-\omega}{2}, && k >\frac{q+\omega}{2}. 
\end{align}
Subsequently performing the $K^0$ integral yields
\begin{align}
	\hat q^{ij} &= \frac{1}{16p^2\nu_a}\sum_{bcd}\int\frac{\dd[3]{\vb k}\dd[3]{\vb q}\dd{\omega}}{(2\pi)^5 q^2 k^2 }\qperp^i\qperp^j\left|\mathcal M_{cd}^{ab}(\vb p,\vb k;\vb p',\vb k')\right|^2 \nonumber\\
	&\times \theta(p'-k')\theta\left(p - \frac{q-\omega}{2}\right)\theta\left(k - \frac{q+\omega}{2}\right)\theta(q-|\omega|) \nonumber\\
	&\times\delta\left(\cos\theta_{pq}-\frac{\omega}{q}-\frac{\omega^2-q^2}{2pq}\right)\delta\left(\cos\theta_{kq}-\frac{\omega}{q}+\frac{\omega^2-q^2}{2kq}\right)\nonumber\\
	&\times f^b(\vb k)\left[1\pm f^d(\vb k-\vec q)\right].\label{eq:qhat_der_with_delta}
\end{align}

The $\vec k$ and $\vec q$ integrals are now performed in spherical coordinates, where the polar angle integral is performed using the delta function and the azimuthal and radial integrals remain. For the radial integrals there are different equivalent possibilities that implement the required conditions:

\begin{align}
\int_{0}^\infty\dd{q}\int_{\max\left(-q, q-2p, \frac{1}{3}(q-2p)\right)}^q\dd{\omega}\int_{\frac{q+\omega}{2}}^{p+2\omega}\dd{k}, \label{eq:boundaries_2}\\
\int_0^\infty\dd{k}\int_{-\frac{p-k}{2}}^k\dd{\omega}\int_{|\omega|}^{\min(p+p',k+k')}\dd{q}, \label{eq:boundaries_1}\\
\int_0^\infty\dd{k}\int_0^{\frac{p-k}{2}}\dd{k'}\int_{|k-k'|}^{\min(p+p',k+k')}\dd{q}. \label{eq:boundaries_1b}
\end{align}


\section{Isotropic HTL matrix element}
\label{app:full_htl_matrix_el}

Here we derive explicitly the expression for the full isotropic HTL matrix element \eqref{eq:full_htl_matrix_element}, which is needed for infrared-sensitive contributions in the matrix elements arising from soft gluon exchange. We start with Eq.~\eqref{eq:prescription_screening}, 
\begin{align}
	\Mhtl = \left|G_R(P-P')_{\mu\nu}(P+P')^\mu (K+K')^\nu\right|^2, \label{eq:prescription_screening_HTL}
\end{align}
where we insert the HTL retarded propagator in strict Coulomb gauge\footnote{By Coulomb gauge we mean using $\partial_i A^i$ as the gauge function and by strict we mean enforcing it strictly, i.e. $\partial_iA^i=0$, which amounts to setting $\xi=0$ in the Faddeev-Popov procedure \cite{Bellac:2011kqa}.} \cite{Ghiglieri:2020dpq}
\begin{align}
    \GLR(Q)&=\frac{i}{q^2+\PiLR(\omega/q)},\\
    G^{ij}_R(Q)=\left(\delta^{ij}-\frac{q^iq^j}{q^2}\right)\GTR(Q)&=\frac{-i\left(\delta^{ij}-\frac{q^iq^j}{q^2}\right)}{q^2-\omega^2+\PiTR(\omega/q)}
\end{align}
with $x=\omega/q$ and the self-energies
\begin{align}
    \RE\PiLR(x)&=m_D^2\left(1-\frac{x}{2}\ln\left|\frac{x+1}{x-1}\right|\right)\\
    \IM\PiLR(x)&=\frac{xm_D^2\pi}{2}\theta(1-|x|)\\
    \RE \PiTR(x)&=\frac{m_D^2}{2}-\frac{1}{2}(1-x^2)\RE\PiLR\\
    \IM\PiTR(x)&=-\frac{1}{2}(1-x^2)\IM\PiLR.
\end{align}
Due to $|x|=|\omega|/q < 1$ in our case (c.f., \eqref{eq:phase_space_relation_omega_q_k}), their imaginary parts are always nonzero. Note that  $G_R(-Q)$ corresponds to the advanced propagator, which has a different imaginary part in the self-energy, $\IM\Pi_R(-Q)=-\IM\Pi_R(Q)$.
Let us further abbreviate 
\begin{align}
    &\GLR(-Q)=: z_L=\frac{i}{A+Bi}, && \GTR(-Q)=:z_T=\frac{-i}{C+Di},\\
    &A = q^2+\RE\PiLR(x), && B =\IM\PiLR(-x),\\
    &C = q ^2-\omega^2+\RE\PiTR(x), && D =  \IM\PiTR(-x).
\end{align}
It will turn out that $B$ and $D$ only appear quadratically or as a product. Thus we do not need to distinguish them from $\IM\Pi_R(x)$. 
We can now split the retarded propagator in \eqref{eq:prescription_screening_HTL} into its temporal and spatial parts and use the expressions for $\vb p$, $\vb q$, and $\vb k$ in the $q$-frame, i.e., using their parametrizations \eqref{eq:qframe_p}, \eqref{eq:qframe_q} and \eqref{eq:qframe_k}, to obtain 
\begin{align}
\begin{split}
    \Mhtl&=\left|c_1z_L+c_2z_T\right|^2\\
    &=c_1^2|z_L|^2+c_2^2|z_T|^2+c_1c_2(z_L\bar z_T+\bar z_L z_T),
\end{split}
\end{align}
where $\bar z$ means taking the complex conjugate of $z$ and $c_1=(2p+\omega)(2k-\omega)$ and $c_2=4pk\sin\theta_{pq}\sin\theta_{kq}\cos\phikq$.

This leads to $|z_L|^2=|\GLR(Q)|^2=(A^2+B^2)^{-1}$, $|z_T|^2=|\GTR(Q)|^2=(C^2+D^2)^{-1}$ and
\begin{align}
	\bar z_L z_T+ z_L\bar z_T&=-2(AC+BD)|z_L|^2|z_T|^2, 
\end{align}
eventually yielding
\begin{align}
 \Mhtl=c_1^2|z_L|^2+c_2^2|z_T|^2-2c_1c_2|z_L|^2|z_T|^2(AC+BD).
\end{align}
For isotropic distributions $f(k)$ the last term is proportional to $\cos\phikq$ and may therefore be dropped.

We also need the rescaled matrix element $\tilde M = \lim_{p\to\infty}\Mhtl/p^2$ in the limit $p\to\infty$. We obtain it by scaling out $p$ 
(see \eq \eqref{eq:c_parameters_for_pinf_matrix_element_HTL})
\begin{align}
    \tilde c_1 &= \lim_{p\to\infty}c_1/p=2(2k-\omega),\tag{\ref{eq:ctilde1}}\\
    \tilde c_2 &= \lim_{p\to\infty}c_2/p=4k\sin\theta_{pq}\sin\theta_{kq}\cos\phikq, \tag{\ref{eq:ctilde2}}
\end{align}
which yields
\begin{align}
    \label{eq:full_htl_matrix_element_APP}
    \tildeMhtl =\tilde c_1^2|z_L|^2+\tilde c_2^2|z_T|^2-2\tilde c_1\tilde c_2|z_L|^2|z_T|^2(AC+BD).
\end{align}
Similarly as before, for isotropic distributions $f(k)$ the last term does not contribute.

\subsection{Sum rule\label{app:sum-rule}}

In this appendix, we show that one can analytically perform the $\omega$ integral over the HTL matrix element \eqref{eq:full_htl_matrix_element_APP},
\begin{align}
    \int_{-\infty}^\infty\frac{\dd{\omega}}{q}\tildeMhtl,
\end{align}
using the sum rule from \re \cite{Aurenche:2002pd}. The latter allows us to evaluate the integral over a spectral function
\begin{align}
\begin{split}
    &\int_0^1\frac{\dd{x}}{x}\frac{2\IM\Pi(x)}{(z+\RE\Pi(x))^2+(\IM \Pi(x))^2}\\
    &\qquad=\pi\left[\frac{1}{z+\RE\Pi(\infty)}-\frac{1}{z+\RE\Pi(0)}\right],
    \end{split}\label{eq:AGZ-sumrule}
\end{align}
provided that the function $\Pi(x)$ fulfills the conditions $\IM\Pi(0)=0$, $\IM\Pi(x)=0$ for $x\geq 1$ and $\RE\Pi(x)\geq 0 $ for $x\geq 1$. 

Our analysis applies to the small $\Lambda_\perp$ limit, i.e., we assume only soft momentum transfer $\qperp \ll k$, with $k\sim T$. Additionally, we take $p\to\infty$, which means $q^2=\qperp^2+\omega^2$, and consider isotropic distributions, for which we may neglect the term in the matrix element $\sim\cos\phikq$. The matrix element then reads
\begin{align}
    \tildeMhtl &= 16k^2\left(\left|\GLR\right|^2+\left(1-\frac{\omega^2}{q^2}\right)^2\cos^2\phikq\left|\GTR\right|^2\right),
\end{align}
where, as compared to \eqref{eq:full_htl_matrix_element} we have neglected the term linear in $\cos\phikq$ and odd powers of $\omega$, and also used $\omega^2/k^2\ll 1$ and $\qperp^2/(2kq)\ll \omega/q$. Note that only the condition $|\omega|\ll k$ allows us to perform the $\omega$ integral without taking the distribution functions into account since otherwise $\omega$ also appears in the Bose-enhancement factor, even in the isotropic case where $k'=k-\omega$ (see Eq.~\eqref{eq:qhat_formula}). We have checked numerically that including $\omega^2/k^2$ leads to only little changes. This is because although $|\omega|$ can become arbitrarily large in the integral, these regions are suppressed by the large denominators appearing in the propagators.

To use the sum rule \eqref{eq:AGZ-sumrule}, we write $|G_R|^2$ in terms of the self-energy $\Pi_R$, and expand the fraction with the imaginary part of the self-energy,
\begin{subequations}
\begin{align}
    \left|\GLR\right|^2&=\frac{2q}{\omega m_D^2\pi}\frac{\IM\PiLR(x)}{\left(q^2+\RE\PiLR\right)^2+\left(\IM\PiLR\right)^2}\\
    \left|\GTR\right|^2&=\frac{4q}{\omega m_D^2\pi(1-x^2)}\frac{\IM\PiTR(x)}{\left(\qperp^2+\RE\PiTR\right)^2+\left(\IM\PiTR\right)^2}
\end{align}
\end{subequations}
A similar trick is used in \cite{Ghiglieri:2015ala} to rewrite $|G_R|^2$ in terms of the spectral function $\rho$.
Together with 
the substitution $\frac{\dd{\omega}}{\omega}=\frac{\dd{x}}{x(1-x^2)}$, this results in
\begin{align}
    &\int_{-\infty}^\infty\frac{\dd{\omega}}{q}\tildeMhtl \\
    &\quad = \frac{32 k^2}{m_D^2}\left[\frac{1-2\cos^2\phikq}{\qperp^2+\frac{m_D^2}{3}}-\frac{1}{\qperp^2+m_D^2}+\frac{2\cos^2\phikq}{\qperp^2}\right]\nonumber.
\end{align}
For the longitudinal propagator $|\GLR|^2$ the factor $(1-x^2)$ from the coordinate transformation needs to be absorbed into the self-energy $\tildePiLR(x)=(1-x^2)\PiLR(x)$.
The relevant limits read
\begin{subequations}
\begin{align}
    \RE\PiTR(0)&=0,&\RE\tildePiLR(0)&=m_D^2,\\
    \RE\PiTR(\infty)&=\frac{m_D^2}{3},&\RE\tildePiLR(\infty)&=\frac{m_D^2}{3}.
\end{align}
\end{subequations}
Integrating over $\phikq$ and $\qperp$ leads to the cancellation of the terms involving the plasmon mass and the familiar result
\begin{align}
    &\int_0^{\Lambda_\perp}\dd{\qperp}\qperp^3\int_0^{2\pi}\dd{\phikq}\int_{-\infty}^\infty\frac{\dd{\omega}}{q}\Mhtl\nonumber\\
    &\quad = \frac{32k^2}{m_D^2}\int_0^{\Lambda_\perp}\dd{\qperp}\qperp^3 \frac{m_D^2}{\qperp^2(\qperp^2+m_D^2)}\\
    &\quad = 16 k^2 \ln\left(1+\frac{\Lambda_\perp^2}{m_D^2}\right).\label{eq:sumrule_result_final}
\end{align}
A similar result is obtained in \cite{Aurenche:2002pd, Caron-Huot:2008zna} in thermal equilibrium, where also the integral over the distribution functions $f(k)(1+f(k))$ is automatically included. Here, we have shown explicitly that the matrix element itself gives rise to the form \eqref{eq:sumrule_result_final}. In the soft limit, this enables us to perform the integral over the distribution function separately, allowing a straightforward generalization to non-equilibrium systems.
Finally, we note that we use this sum rule in \se \ref{sec:screening} to set the $\xi$ parameter in the approximate $\xi$-screening prescription.


\section{Behavior of $\hat q$ for large jet momenta} \label{app:behav_large_p}

In this appendix, we study the behavior of the jet quenching parameter $\qhat$ for large jet momentum $p$ and in particular, how to correctly perform the limit $p\to\infty$. We will show that taking the term at a na\"ive leading order in $1/p$ in the integrand of $\qhat^{ij}$ is indeed sufficient to obtain the correct leading-order contribution. 
This is not trivial since $p$ also appears in the integration boundaries and this analysis requires caution.
Thus, there are two possible sources for large-$p$ contributions to $\qhat$: the integrand and the integral boundaries. Our strategy here will be to expand the integrand in orders of $1/p$ and then perform the integrals.

We will illustrate the large $p$ behavior of $\qhat$ using the gluonic matrix element, 
\begin{align}
    \hat q &\sim \int_0^\infty\dd{k}\!\!\int_{-\frac{p-k}{2}}^k\!\!\!\dd{\omega}\!\!\int_{|\omega|}^{\text{min}(p+p',k+k')}\!\!\!\!\!\!\!\dd{q} \label{eq:qhat_largep_integrand}\\
	&\qquad  \times 
    q^2(1-v_{pq}^2)\frac{\left|\mathcal M^{gg}_{gg}\right|^2}{p^2}f_b(\vb k)\left(1\pm f_d(\vb k')\right)\nonumber,
\end{align}
The distribution function $f(\vb k)$ provides an upper limit for the momentum of the plasma constituent $k$, which we assume to be much smaller than the jet momentum $p$, thus $k \ll p$. Thus the minimum of $(p+p',k+k') = (2p+\omega, 2k-\omega)$ is always $2k-\omega$, because 
$2k-\omega < 2p+\omega$ for $\omega > k-p$, which
is always fulfilled due to the lower boundary of the $\omega$-integral, $\omega > \frac{k-p}{2}$. Then the only $p$-dependence in the integration boundaries of Eq.~\eqref{eq:qhat_largep_integrand} comes from the lower limit of the $\omega$ integral.

Therefore, we will be interested in the region $\omega < 0$, where $|\omega|$ is very large. In particular, we will assume $|\omega|>\lxi\gg k$ with a new scale $\lxi$, which will lead to simplifications in the matrix element. Additionally, the \emph{Bose-enhanced} term $\qhatff$ that includes $f(\vb k) f(\vb k')$ does not contribute in this limit, since $f(\vb k')\approx 0$. Focusing only on relevant terms, i.e., disregarding the $k$-integral since it cannot contribute to any large $p$ behavior, we analyze
\begin{align}
    \hat q &\sim \int_{-\frac{p-k}{2}}^{-\lxi}\!\!\!\dd{\omega}\!\!\int_{|\omega|}^{2k-\omega}\!\!\!\!\!\!\!\dd{q} \label{eq:qhat_largep_integrand2}
    q^2(1-v_{pq}^2)\frac{\left|\mathcal M^{gg}_{gg}\right|^2}{p^2}.
\end{align}

\subsection{The integrand of $\qhat$ for large $p$}

Now let us expand the integrand in \eqref{eq:qhat_largep_integrand2} for large $p$ explicitly.

\subsubsection{The limit of $v_{pq}$}

First, we consider $v_{pq}=\cos\theta_{pq}$ in the limit $p\to\infty$. Our starting point is Eq.~\eqref{eq:vpq},
\begin{align}
v_{pq}=\frac{1}{q}\left(\omega+\frac{\omega^2-q^2}{2p}\right).
\label{eq:vpq_v2}
\end{align}
Using \eq \eqref{eq:phase_space_relation_omega_q_k}, i.e., $q<2k-\omega$, one has
\begin{align}
    \frac{q^2-\omega^2}{2p}< \frac{4k(k-\omega)}{2p}\ll |\omega|,
\end{align}
which leads
for $p\to\infty$ to,
\begin{align}
v_{pq}=\frac{\omega}{q}. \label{eq:vpq_ptoinf}
\end{align}
However, considering the term $1-v_{pq}^2$ that appears in \eqref{eq:qhat_largep_integrand2} is more subtle, because the seemingly leading term in a $1/p$ expansion, $\omega/q$, is close to unity. This leads to the corrections
\begin{align}
    1-v_{pq}^2&=1-\frac{\omega^2}{q^2}+\frac{\omega(q^2-\omega^2)}{pq^2}+\dots \nonumber \\
    &=\left(1-\frac{\omega^2}{q^2}\right)\left(1+\frac{\omega}{p}+\dots\right),\label{eq:one_minus_vpq_limit}
\end{align}
where the correction term $\omega/p$ can in principle become large at the lower boundary of the $\omega$ integral, $\omega > -(p-k)/2$.

\subsubsection{Matrix element for large $|\omega|$}

For large $|\omega|$, and therefore also large $q$, we do not need to take into account screening effects $\mathcal{O}(m_D)$ in the matrix element, such that \eqref{eq:full_HTL_finite_p_matrix_element_replacement} reduces to $M_{\mathrm{screen}} \approx M_0=(s-u)^2/t^2$. The contribution from the transverse propagator in the sum in \eqref{eq:full_HTL_finite_p_matrix_element_replacement} is negligible for large $|\omega|$, and we are left with
\begin{align}
    \frac{\left|\mathcal M^{gg}_{gg}\right|^2}{g^4p^2}=16 \dA\CA^2 \frac{\omega^2}{q^4}\left(1+\frac{\omega}{p}+\dots\right).
    \label{eq:limit_gluonic_matrix_element}
\end{align}
Collecting the pieces, we can rewrite the relevant integrand in \eqref{eq:qhat_largep_integrand2} as
\begin{align}
    &q^2(1-v_{pq}^2)\frac{\left|\mathcal M^{gg}_{gg}\right|^2}{g^4p^2} \nonumber \\
    & \qquad= 16\dA\CA^2 \frac{(q^2-\omega^2)\omega^2}{q^4}\left(1 + \frac{2\omega}{p}+\dots\right).\label{eq:qhat_largep_integrand_explicit}
\end{align}

\subsection{Integral over a more generic integrand}
The integrand relevant for determining the large $p$ dominant behavior for $\qhat$ is a sum of terms $q^n\omega^m$, as can be seen from \eqref{eq:qhat_largep_integrand_explicit}.
Let us therefore analyse a general integrand of this form and define the integral $I_{nm}$,
\begin{align}
I_{nm}(p)=\int_{-\frac{p-k}{2}}^{-\lxi}\dd{\omega}\int_{-\omega}^{2k-\omega}\dd{q}\,q^n\omega^m. \label{eq:analysis_integrand_1}
\end{align}
Although $|\omega|\gg 2k$, we cannot neglect $2k$ in the upper integration boundary of the $q$-integral, which additionally complicates our analysis.
With $\otild=-\omega>0$ we get rid of additional minus signs and obtain for $n\neq -1$
\begin{align}
I_{nm}(p)&=(-1)^m\int_{\lxi}^{\frac{p-k}{2}}\dd{\otild}\,\int_{\otild}^{2k+\otild}\dd{q}\, q^nx^m,\\
&=\frac{(-1)^m}{n+1}\int_{\lxi}^{\frac{p-k}{2}}\dd{\otild}\,x^m\left[(2k+\otild)^{n+1}-\otild^{n+1}\right].
\end{align}
We expand the first term in a power series using the Binomial series 
\begin{align}
(x+y)^r=\sum_{k=0}^\infty\begin{pmatrix}
r \\ k
\end{pmatrix} x^{r-k}y^k,\quad x,y\in\mathbb R, \label{eq:binomial_series}
\end{align}
with $ |x| > |y|$ and $  r\in\mathbb C$.
We thus obtain
\begin{align}
I_{nm}&=\frac{(-1)^m}{n+1}\int_{\lxi}^{\frac{p-k}{2}}\dd{\otild}\\
&\qquad\times \otild^m\left[\sum_{j=0}^\infty \begin{pmatrix}n+1 \\ j\end{pmatrix}\otild^{n+1-j}(2k)^j-\otild^{n+1}\right] \nonumber \\
&=\frac{(-1)^m}{n+1}\!\!\int_{\lxi}^{\frac{p-k}{2}}\!\!\!\!\!\!\!\dd{\otild}\!\sum_{j=1}^\infty \!\begin{pmatrix}n+1 \\ j\end{pmatrix}\! \otild^{n+m+1-j}(2k)^j \\
&=\!\frac{(-1)^m}{n+1}\!\Bigg(\!\!\!\sum_{\substack{j=1\\j\neq n+m+2}}^\infty \!\!\!\!\!\!\!\begin{pmatrix}n+1 \\ j\end{pmatrix}\!\left.\frac{\otild^{n+m+2-j}}{n+m+2-j}\right|_{\otild=\lxi}^{\frac{p-k}{2}}(2k)^j\nonumber\\
&\quad+\begin{pmatrix}n+1 \\ n+m+2\end{pmatrix}\ln\left(\frac{p-k}{2\lxi}\right)(2k)^{n+m+2}\Bigg).
\end{align}
Since we are interested in the behavior at large $p$, we drop the lower boundary $\otild=\lxi$ and take only the leading-order (LO) terms with the largest powers of $p$ into account. 
Those are obtained for $j=1$, for which the (generalized) binomial coefficient yields $n+1$. It will be useful to also consider the next-to-leading order (NLO) terms in $p$. We obtain, up to an additive constant,
\begin{subequations} 
\begin{align}
\ILO_{nm}&\simeq\begin{cases}\frac{(-1)^m (2k)}{n+m+1}\left(\frac{p-k}{2}\right)^{n+m+1}, & n+m+1 \neq 0\\
(-1)^m(2k)\ln(p), & n+m+1 = 0\end{cases} \label{eq:ILO}\\
\INLO_{nm}&\simeq\begin{cases}\frac{(-1)^m (2k)^2}{(n+1)(n+m)}\begin{pmatrix}n+1 \\ 2\end{pmatrix}\left(\frac{p-k}{2}\right)^{n+m}, & n+m \neq 0\\
\frac{(-1)^m}{n+1}\begin{pmatrix}n+1 \\ 2\end{pmatrix}(2k)^2\ln(p), & n+m = 0.\end{cases}\label{eq:INLO}
\end{align}
\end{subequations}
Note that the inclusion of $k$ in $(p-k)^{n+m+1}$ in the LO term is because it will contribute at NLO.

\subsection{Behavior of $\qhat$ for large $p$\label{app:qhat_large_p}}

Gathering the results of the previous sections and applying them to the integral of $\qhat$ in \eqref{eq:qhat_largep_integrand2} with the integrand \eqref{eq:qhat_largep_integrand_explicit}, we obtain
\begin{subequations}
\begin{align}
    \qhatLO &\sim I_{-2,2} - I_{-4,4} = (2k)^2\ln p + \text{ const}, \label{eq:qhat_LO_in_p}\\
    \qhatNLO &\sim \frac{1}{p}\left(I_{-2,3}-I_{-4,5}\right) = \text{const}+\mathcal O\left(\frac{1}{p}\right).
\end{align}
\end{subequations}

Note that here with $\mathrm{NLO}$ we denote the terms proportional to $1/p$ in the integrand. For both cases, $n+m$ is constant, thus the leading terms \eqref{eq:ILO} cancel and we need the next-to-leading terms \eqref{eq:INLO}. Indeed we observe that, due to the logarithmic enhancement of the leading-order contributions $\qhatLO$, the next-to-leading order contributions $\qhatNLO$ become negligible
\footnote{This is not a trivial statement: for an integrand $q^n\omega^m(1+\omega/p)$ with $n+m>0$, we would obtain $\frac{\qhatNLO}{\qhatLO}=\frac{\frac{1}{p}\left(a p^{n+m+2} + b\right)}{cp^{n+m+1}+e} \sim \frac{a}{c} + \mathcal O\left(\frac{1}{p}\right)$, thus the ratio $\text{NLO}/\text{LO}$ does not necessarily become $0$ for $p\to\infty$. This implies that multiplying the LO term with $\omega/p$ and integrating over it yields a term of the same order as the LO term.} 
for sufficiently large $p$, and $\qhat$ can be written in the form of Eq.~\eqref{eq:qhat_largep_behavior},
\begin{align}
\qhat(p\gg Q) \simeq a_p\ln p + b_p. 
\end{align}
Therefore, for sufficiently large jet momenta $p$, it is in principle enough to expand the matrix element and the integrand for large $p$ and take only the leading-order contribution in $p$. Note, however, that to obtain the constant term it is not enough to use the leading large $p$ behavior, but one must use the full matrix element.

Let us now see how we can obtain the coefficient of the logarithm, $a_p$.
Until now, we have not considered the exact form of the distribution function $f(\vb k)$ and merely used that it provides an upper cutoff for the $k$ integral. The numerical value of the coefficient $a_p$ will depend on the exact form of $f(\vb k)$.
 
Let us consider a gluon jet scattering off a gluon in the plasma and start with Eq.~\eqref{eq:qhat_formula},
\begin{align}
\begin{split}
    \qhat&\simeq\frac{16g^4\CA^2}{2^{10}\pi^5}\int_0^{2\pi}\dd\phipq\int_0^{2\pi}\dd\phikq\int_0^\infty\dd{k}\\
    &\qquad\times\int_{-\frac{p-k}{2}}^k\dd{\omega}\int_{|\omega|}^{2k-\omega}\dd{q}
     f(\vb k)\frac{\omega^2(q^2-\omega^2)}{q^4},
\end{split} \label{eq:qhat_calculate_ap1}
\end{align}
where we have taken the leading term in the large $p$ integrand \eqref{eq:qhat_largep_integrand_explicit} that leads to the logarithmic behavior. Additionally, as explained below Eq.~\eqref{eq:qhat_largep_integrand}, it is sufficient to use $2k-\omega$ as the upper boundary of the $q$-integral.

In comparison to the general integrand we analysed in Eq.~\eqref{eq:analysis_integrand_1}, there appears also the distribution function $f(\vb k)=f(k,v_k)$, and the angle $v_k=\cos\theta_k$ depends on $v_{kq}=\cos\theta_{kq}$ and $v_{pq}=\cos\theta_{pq}$ as well, which are functions of $\omega$ and $q$, as in Eq.~\eqref{eq:vk_explicit}. For the large $p$ behavior, we are interested in the region $|\omega|\gg k$ and $q\sim |\omega|$, which renders $v_{pq}\to -1$. For $v_{kq}$, however, we cannot make a definite statement, since it changes from $-1$ to $1$ when $q$ varies between its integration boundaries $|\omega|<q<2k-\omega$.
Therefore, for anisotropic systems, without knowing the explicit form of $f(\vb k)$, we cannot calculate the coefficient $a_p$. We will thus restrict ourselves to isotropic distributions $f(k)$ here that only depend on the magnitude of $\vb k$. Then, the $\omega$ and $q$ integrations in Eq.~\eqref{eq:qhat_calculate_ap1} are given by Eq.~\eqref{eq:qhat_LO_in_p}.

For an isotropic plasma consisting of quarks and gluons with distributions $f_q$ and $f_g$, the coefficient $a_p$ is then given by Eq.~\eqref{eq:qhat_ap_coefficient},
\begin{align}
\begin{split}
    a_p/\CR&=\frac{\CA g^4}{4\pi^3}\int_0^\infty \dd{k}k^2f_g(k)\\
    &\quad+\sum_{f}\frac{\dF\CF g^4}{4\pi^3\dA}\int_0^\infty \dd{k} k^2 f_q(k).
    \end{split}
\end{align}
In thermal equilibrium, this reduces to
\begin{align}
    \frac{a_p^{\mrm{eq}}}{\CR}=\frac{g^4\zeta(3)T^3}{2\pi^3}\left(\NC + \frac{3}{4}\nf\right).\label{eq:qhat_ap_coefficient_equilibrium}
\end{align}

This logarithmic behavior of $\qhatLO$ implies that the limit $p\to\infty$ requires a UV cutoff to render $\qhat$ finite. This is typically done by restricting the transverse momentum transfer, $\qperp < \Lambda_\perp$. We will show in the next section that in this limit $\qhat$ is finite and we only need to consider the leading-order term in $p$ in the integrand.

\subsection{Large $p$ behavior combined with a momentum cutoff $\qperp< \Lambda_\perp$}

Using a transverse momentum cutoff, $\qperp < \Lambda_\perp \ll p$, we can retrace the steps in the previous sections. For large $p$, apart from factors $\mathcal O(1/p)$, this amounts to
$\qperp^2 = q^2-\omega^2<\Lambda_\perp^2$ (c.f., \eqref{eq:qperp_cutoff}), and thus we modify $I_{nm}$ to
\begin{align}
I_{nm}(p)=\int_{-\frac{p-k}{2}}^{-\lxi}\!\!\dd{\omega}\int_{-\omega}^{\sqrt{\omega^2+\Lambda_\perp^2}}\!\!\!\dd{q}\,q^n\omega^m.\label{eq:analysis_integrand_2v2}
\end{align}
The upper limit in the $q$ integral replaces $2k-\omega$ in Eq.~\eqref{eq:analysis_integrand_1} for sufficiently large $\lxi$ because $\sqrt{\omega^2+\Lambda_\perp^2}<2k-\omega$ for $-\omega > \frac{\Lambda_\perp^2}{4k} - k$,
which, since $-\omega > \lxi$, can always be fulfilled by choosing 
\begin{align}
\lxi >\frac{\Lambda_\perp^2}{4k}-k. \label{eq:qperp_limit_xi_condition}
\end{align}
Similarly as before, with $\otild=-\omega$ and for $n\neq -1$, we obtain
\begin{align}
I_{nm}(p)&=\frac{(-1)^m}{n+1}\!\int_{\lxi}^{\frac{p-k}{2}}\!\!\!\!\dd{\otild} \otild^m\left[\left(\otild^2+\Lambda_\perp^2\right)^{\frac{n+1}{2}}-\otild^{n+1}\right].
\end{align}
For the convergence of the Binomial series \eqref{eq:binomial_series}, we need to check that $\Lambda_\perp^2 < \otild^2$, which follows from \eqref{eq:qperp_limit_xi_condition},
\footnote{We know that $\Lambda_\perp^2 < 4k\lxi\left(1+\frac{k}{\lxi}\right)\approx 4k\lxi$ and thus $\Lambda_\perp^2/\lxi^2 < \frac{4k}{\lxi}\ll 1$, which makes $\Lambda_\perp < \lxi$ and thus $\Lambda_\perp < \otild$.}
and thus,
\begin{align}
&I_{nm}\!=\!\frac{(-1)^m}{n+1}\!\int_{\lxi}^{\frac{p-k}{2}}\!\!\!\!\!\dd{\otild}\!\sum_{j=1}^\infty \!\begin{pmatrix}(n+1)/2 \\ j\end{pmatrix}\!\otild^{n+m+1-2j}(\Lambda_\perp)^{2j}\\
&~=\frac{(-1)^m}{n+1}\Bigg[\!\!\!\!\!\!\!\!\!\sum_{\substack{j=1\\j\neq (n+m+2)/2}}^\infty \!\!\!\!\!\!\!\!\!\!\!\begin{pmatrix}(n+1)/2 \\ j\end{pmatrix}\!\!\left.\frac{\otild^{n+m+2-2j}}{n+m+2-2j}\right|_{\otild=\lxi}^{\frac{p-k}{2}}\!\!\Lambda_\perp^{2j}\nonumber\\
&\qquad+\begin{pmatrix}(n+1)/2 \\ (n+m+2)/2\end{pmatrix}\ln\left(\frac{p-k}{2\lxi}\right)\Lambda_\perp^{n+m+2}\Bigg].
\end{align}
We now obtain, up to an additive constant (in $p$), 
\begin{align}
I^{\mathrm{LO}}_{nm}&=\begin{cases}\frac{(-1)^m \Lambda_\perp^2}{2(n+m)}\left(\frac{p-k}{2}\right)^{n+m}, & n+m \neq 0\\
(-1)^m\Lambda_\perp^2\ln(p) , & n+m = 0\end{cases} \\
I^{\mathrm{NLO}}_{nm}&=\begin{cases}\frac{(-1)^m \Lambda_\perp^4B}{(n+1)(n+m-2)} \left(\frac{p-k}{2}\right)^{n+m-2}, & n+m-2 \neq 0\\
\frac{(-1)^m}{n+1}B\Lambda_\perp^4\ln(p) , & n+m -2 = 0,\end{cases} 
\end{align}
where $B=\begin{pmatrix}(n+1)/2 \\ 2\end{pmatrix}$.
In this case, denoting again with $\mathrm{NLO}$ the terms proportional to $1/p$ in the integrand, we obtain
\begin{subequations}
    \begin{align}
        \qhatLO &\sim I_{-2,2} - I_{-4,4} \sim\,\text{const}\,+ \mathcal O\left(\frac{1}{p^{2}}\right) , \\
        \qhatNLO &\sim \frac{1}{p}\left(I_{-2,3}-I_{-4,5}\right) \sim \frac{\text{const}}{p}+\mathcal O\left(\frac{1}{p^2}\right).
    \end{align}
\end{subequations}
Thus, with a transverse momentum cutoff $\Lambda_\perp$ in place, we can explicitly take the limit of infinite jet momentum, $p\to\infty$, and obtain a finite jet quenching parameter $\qhat$. Moreover, it is then sufficient to take the leading-order terms in the integrand (in particular the matrix elements) in an expansion in $1/p$.

\subsection{Large momentum cutoff $\Lambda_\perp$ behavior}
\label{sec:qhat_behavior_large_cutoff}

Removing the momentum cutoff, i.e. taking $\Lambda_\perp\to\infty$, leads to a divergent jet quenching parameter $\qhat$. It is reasonable to assume that this divergence will be logarithmic, as it is for finite but large jet momentum. We will now show this explicitly and calculate the coefficient of the logarithm.

We continue working in the limit $p\to\infty$, and thus take only the leading-order terms in $p$ in the integrand into account. For that we start with Eq.~\eqref{eq:qhat_der_with_delta} with explicit step-functions and perform the coordinate transformation in Eq.~\eqref{eq:coordinate_trafo_differentials_to_qperp} to integrate over $\vb \qperp$,
\begin{align}
\begin{split}
    \qhat &\sim \int_0^\infty\dd{k}\int_{q_\perp < \Lambda_\perp}\dd[2]{\vb{\qperp}}\int_{-\infty}^{k-\frac{\qperp^2}{4k}}\dd{\omega}\\
    &\qquad\qquad\times\qperp^2 f(\vb k)\left(1\pm f(\vb k')\right)\frac{\left|\mathcal M^{gg}_{gg}\right|^2}{p^2 \sqrt{\qperp^2+\omega^2}},
\end{split}
\end{align}
where we have not included the angular integrals.

The behavior for large $\Lambda_\perp$ is of course dominated by the upper integration boundary of the $\qperp$ integral. For large $\Lambda_\perp$, we now split the integral into a part $0 < q_\perp < \lxi$ and $\lxi < \qperp < \Lambda_\perp$. We can choose the scale $\lxi \gg k$, such that $\qperp\gg k$, since the momentum $k$ has a natural upper cutoff coming from the distribution function $f(\vb k)$. Then, also $|\omega|>\frac{\qperp^2}{4k}-k$ must be very large and thus $f(\vb k')\approx 0$. Hence we arrive at the scale separation
\begin{align}
    |\omega|\gg \qperp\gg k. \label{eq:large_lperp_scale_separation}
\end{align}
Using this for the integrand \eqref{eq:qhat_largep_integrand_explicit}, we obtain
\begin{align}
\begin{split}
    \qhat &\sim \int_\lxi^{\Lambda_\perp}\dd{\qperp}q_\perp\int_{-\infty}^{k-\frac{\qperp^2}{4k}}\dd{\omega}\qperp^2 
   \frac{1}{|\omega|^3}\\
   &\approx 8k^2\int_\lxi^{\Lambda_\perp}\dd{\qperp}\frac{q_\perp^3}{q_\perp^4}= 8k^2\ln\Lambda_\perp+\text{const}\label{eq:qhat_integral_to_make_log_lambda}
\end{split}
\end{align}
We have thus shown that we can write $\qhat$ in that limit as in Eq.~\eqref{eq:qhat_behavior_large_cutoff},
\begin{align}
    \qhat(\Lambda_\perp \gg Q) \simeq a_{\lperp}\ln\Lambda_\perp +b_{\lperp}.
\end{align}

It is also possible to determine the coefficient of $\ln\Lambda_\perp$, similarly as in \app\ref{app:qhat_large_p}.
For anisotropic systems, the distribution function depends implicitly on $\omega$ and $\qperp$ via the angle $\cos\theta_k=v_k$, see Eq.~\eqref{eq:vk_explicit}.
With the scale separation \eqref{eq:large_lperp_scale_separation}, we find $v_{pq}\to -1$, similarly as in \app\ref{app:qhat_large_p}. Additionally,
$v_{kq}$ changes from $v_{kq}\to -1$ for $|\omega|\to\infty$ to $v_{kq}\to 1$ for $|\omega|\to \qperp^2/4k -k$, and similarly for $v_{k'q}\to -1$.
Note that $v_{kq}$ decreases rather quickly for increasing $|\omega|$,
thus we cannot make predictions for arbitrary anisotropic systems, and assume an isotropic distribution for calculating $a_{\lperp}$.
With this approximation, we can now perform the integral in Eq.~\eqref{eq:qhat_integral_to_make_log_lambda}.

For a plasma consisting of quarks and gluons with distributions $f_q$ and $f_g$, we then obtain the coefficient
\begin{align}
\begin{split}
    \frac{a_{\lperp}}{\CR}&=\frac{\CA g^4}{2\pi^3}\int_0^\infty \dd{k}k^2f_g(k)\\
    &\quad+\sum_{f}\frac{\dF\CF g^4}{2\pi^3\dA}\int_0^\infty \dd{k} k^2 f_f(k),
    \end{split}
\end{align}
which reduces for thermal distributions to
\begin{align}
    \frac{a_{\lperp}^{\mrm{eq}}}{\CR}=\frac{g^4\zeta(3)T^3}{\pi^3}\left(\NC + \frac{3}{4}\nf\right).\label{eq:qhat_a_coefficient_equilibrium}
\end{align}
This nicely agrees with Eq.~\eqref{eq:qhat_hard_arnold} that stems from \cite{Arnold:2008vd}.

\section{Monte-Carlo sampling}\label{app:monte_carlo}

In this appendix, we describe the implementation of the Monte-Carlo integration. Our implementation can be found in \cite{kurkela_2023_10409474}.

We evaluate the five-dimensional integral in \eqref{eq:qhat_formula} with the measures \eqref{eq:qhat_pinf_measure_klast}, \eqref{eq:qhat_pinf_measure_klast2} and \eqref{eq:qhat_pinf_measure_qlast} using Monte-Carlo integration with importance sampling. For the angular parts, we sample uniformly from $0$ to $2\pi$.

The other three integrals can be summarized into three categories. The first one corresponds to \eqref{eq:qhat_pinf_measure_klast2} and is given by
\begin{align}
	I_1=\int_{\kmin}^{\kmax}\!\dd{k}\!\!\int_{\kmin}^\infty\!\dd{k'}\!\!\int_{|k-k'|}^{\min\left(k+k',\sqrt{(k-k')^2+\Lambda_\perp^2}\right)}\!\!\!\dd{q}, 
\end{align}
where we have implemented the boundaries $\kmin < k< \kmax$ coming for the finite discretization of the distribution function $f(\vb k)$, and similarly $\kmin < k'$, and also used the $\qperp$ cutoff, $q^2-\omega^2<\Lambda_\perp^2$. We can rewrite this equivalently as 
\begin{align}
	I_1=\int_{\kmin}^{\kmax}\!\dd{k}\!\!\int_{-\infty}^{k-\kmin}\!\dd{\omega}\!\!\int_{|\omega|}^{\min\left(2k-\omega,\sqrt{\omega^2+\Lambda_\perp^2}\right)}\!\!\!\dd{q},
\end{align}
which corresponds to \eqref{eq:qhat_pinf_measure_klast}.

For the different order of integration as in \eqref{eq:qhat_pinf_measure_qlast} the conditions $\kmin < k$ and $\kmin < k'$ amount to 
\begin{align}
	\begin{split}
	I_2=\int_0^\infty\!\!\!\!\dd{q}\!\!\int_{-q}^{\min(q,2\kmax-q,\kmax-\kmin)}\!\!\!\!\!\!\!\!\dd{\omega}\!\!\int_{\max\left(\frac{q+\omega}{2},\kmin,\kmin+\omega\right)}^{\kmax}\!\!\!\!\!\!\dd{k},
	\end{split}
\end{align}
where due to the $\qperp$ cutoff, for $q>\Lambda_\perp$ we cut the following region $(-\sqrt{q^2-\Lambda_\perp^2},\sqrt{q^2-\Lambda_\perp^2})$ out of the integration region of the $\omega$ integral.

For the $q$ integral we sample $q$ from a probability distribution $\sim (q+m)^{-2}$, where $m$ is the gluon asymptotic mass $m=m_D/\sqrt{2}$. Thus we perform the integral according to 
\begin{align}
\begin{split}
	\int_{\qmin}^{\qmax}\dd{q}f(q)&\approx\frac{\qmax-\qmin}{N(\qmax+m)(\qmin+m)}\\
    &\qquad \times\sum_{i=1}^N(q_i+m)^2f(q_i),
\end{split}
\end{align}
where $q_i=(1/(\qmin+m)-y_i)^{-1}-m$ and $y_i$ is sampled uniformly in $(0,(\qmax-\qmin)(\qmax+m)^{-1}(\qmin+m)^{-1})$.

We sample $\omega$ uniformly and $k$ from a $1/k$ distribution,
\begin{align}
	\int_{\kmin}^{\kmax} \dd{k}f(k)\approx \ln\left(\frac{\kmax}{\kmin}\right)\frac{1}{N}\sum_{i=1}^N k_i f(k_i),
\end{align}
where $k_i=\kmin e^{r_i}$
and $r_i$ is sampled uniformly in $(0,\ln\kmax/\kmin)$. 


\section{Details on $\qhat$ calculations in toy models}
\label{app:qhat_models}

\subsection{Extremely anisotropic distribution}
\label{app:qhatff_special_dist}

We now calculate $\qhatff^{ij}$ for a jet going in the $x$ direction and the distribution \eqref{eq:special_distribution}, which we rewrite as
\begin{align}
	f(\vec k)&=AQ^3\delta(k_z)\delta(k_x^2+k_y^2-\tilde k^2) 
    \label{eq:app_fk_ktilde}
\end{align}
We do not need to enforce a momentum cutoff $\qperp<\lperp$, since $\qhatff$ is finite even for $p\to\infty$. Hence, we can assume that the transverse momentum cutoff $\lperp$ is sufficiently large, at least $\lperp>2\tilde k$, as we will see later.

We first note that the expression for $\qhatff^{zz}$ is proportional to $\int (q^z)^2\delta(k_z)\delta(k_z')$. The delta functions enforce $k_z=k_z'=0$ and thus also $q_z=k_z-k_z'=0$ and $\qhatff^{zz}=0$. Note that we have obtained this result without using a specific form of the matrix element.

For $\qhatff^{yy}$ we use \eqref{eq:qhat_der_with_delta} and insert $f(\vec k)$ in \eqref{eq:app_fk_ktilde}, obtaining
\begin{align}
	&\qhatff^{yy} = \frac{A^2Q^6}{16p^2\nu}\int\frac{\dd[3]{\vb k}\dd[3]{\vb q}\dd{\omega}}{(2\pi)^5 q^2 k^2 }q^yq^y\left|\mathcal M(\vb p,\vb k;\vb p',\vb k')\right|^2 \nonumber\\
	& \times \delta\left(\cos\theta_{pq}-\frac{\omega}{q}-\frac{\omega^2-q^2}{2pq}\right)\delta\left(\cos\theta_{kq}-\frac{\omega}{q}+\frac{\omega^2-q^2}{2kq}\right)\nonumber\\
	&\times\theta(p'-k')\theta\left(p - \frac{q-\omega}{2}\right)\theta\left(k - \frac{q+\omega}{2}\right)\theta(q-|\omega|)\nonumber\\
	&\times\delta(k_z)\delta(q_z)\,{\delta(k^2-\tilde k^2)}
	\,{\delta\left((\vb k-\vb q)^2-\tilde k^2\right)}
	.
\end{align}
The delta functions can be rewritten as $\delta\left((\vb k-\vb q)^2-\tilde k^2\right)=\delta(\omega^2-2\omega\tilde k)=\frac{1}{2\tilde k}\left(\delta(\omega)+\delta(\omega-2\tilde k)\right)$ and $\delta(k^2-\tilde k^2)= \frac{1}{2\tilde k}\delta(k-\tilde k)$.
We can thus integrate out $k_z$ and $q_z$ and rewrite the remaining integrations as a two-dimensional integral. For $p\to\infty$ the theta functions containing $p$ and $p'$ do not need to be written explicitly. The $\delta(\omega-2k)$ term vanishes because then the third Heaviside function becomes $\theta(-q/2)=0$. Integrating over $\omega$ enforces $\omega=0$. Due to the considered large-$p$ limit, the first delta function becomes $\delta(\cos\theta_{pq})=q\delta(q_x)$, 
and integrating out $q_x$ as well, we arrive at
\begin{align}
	\qhatff^{yy} &= \frac{A^2Q^6}{16p^2\nu}\int\frac{\dd{k_x}\dd{k_y}\dd{q_y}}{(2\pi)^5 q_y k^2 }\left(q_y\right)^2\left|\mathcal M(\vb p,\vb k;\vb p',\vb k')\right|^2 \\
	&\qquad \times\delta\left(\cos\theta_{kq}-\frac{|q_y|}{2 \tilde k}\right)\theta\left(\tilde k - \frac{|q_y|}{2}\right)\frac{1}{4\tilde k^2}\delta(k-\tilde k).\nonumber
\end{align}
Effectively, $\vb q$ is parallel to the $y-$axis and $\vb k$ lies in the $x-y$ plane with length $\tilde k$ and $k_y=|q_y|/2$. For the matrix element we need $q=|q_y|$, $k=\tilde k$ and $\phikq$, which is the polar angle of $\vb k$ in a frame, in which $\vb q$ points in the $z$ direction and $\vb p$ lies in the $x-z$ plane, see \se\ref{sec:coordinate_systems}. In our case, $\vb q$ is orthogonal to $\vb p$, thus we perform the $\vb k$ integration in a frame, in which $\vb q = q \vb e_{z^3}$, $\vb p=-p\vb e_{x^3}$. Since $\vb k$ must lie in the $\vb p - \vb q$ plane, we obtain
\begin{align}
	\phikq\in\left\{0,\pi\right\}.\label{eq:cond_phikq}
\end{align}
We get a factor $2$ from the symmetry $q_y\leftrightarrow -q_y$ and insert the gluonic matrix element from \tab \ref{tab:p-inf_matrix_el} with the approximation \eqref{eq:approximated_matrix_element}, and sum over the possible values of $\cos\phikq$, 1 and -1. Thus we obtain, for a momentum cutoff $\lperp>2\tilde k$,
\begin{align}
	\qhatff^{yy}&=\frac{\dA \CA  ^2 A^2g^4Q^6}{2^6\pi^5 \nu \tilde k^3}\int_{0}^{2\tilde k}\dd{q}q\\
	&\qquad\times
	\frac{\left(2\tilde k - \sqrt{4\tilde k ^2-q^2}\right)^2+\left(2\tilde k + \sqrt{4\tilde k ^2-q^2}\right)^2}{(q^2+\xi^2m_D^2)^2}\nonumber
     \\
	&=\frac{\dA \CA  ^2 A^2g^4Q^6}{(2\pi)^5 \nu \tilde k^3}\int_{0}^{2\tilde k}\dd{q}q
	\frac{8\tilde k^2-q^2}{(q^2+\xi^2m_D^2)^2}
	\label{eq:special_case_to_integrate}
\end{align}
The integral over $q$ can be performed analytically, which yields Eq.~\eqref{eq:qhatffyy_special}.


\subsection{Evaluating $\qhat$ for the scaled thermal distribution}
\label{app:scaled_thermal_technical_details}

	Here we describe briefly a few technical details used in the evaluation of \eqref{eq:qhat_formula} with \eqref{eq:qhat_pinf_measure_qlast} for the scaled thermal distribution \eqref{eq:scaled_bose_fermi_combined} in \se\ref{sec:scaled-thermal-distribution}.
	As independent parameters we have
	\begin{itemize}
		\item the coupling $\lambda = g^2 \NC  $,
		\item the temperature $T$,
		\item the occupancy $\NOg$ (and $\NOq$, but here we consider a purely gluonic plasma).
	\end{itemize}
 All dimensionful quantities are given in terms of the temperature $T$.
 Then we have two independent parameters left, on which $\qhat$ will depend.
	Every matrix element comes with a factor $g^4\sim\lambda^2$, which we can factor out. Since the Debye mass scales with $\NOg\lambda$, the scaled matrix element then depends only on this combination,
	\begin{align}
		\left| \mathcal M^{ab}_{cd}(\NOg ,\lambda)\right|^2 = \lambda^2\left|\tilde M^{ab}_{cd}(\NOg \lambda)\right|^2.
	\end{align} 
	Then \eq \eqref{eq:qhat_formula} becomes
	\begin{align}
		\begin{split}
			\qhat = \sum_{bcd}&\int\dd{\tilde\Gamma} q^2\sin^2\theta_{pq}\frac{|\tilde M^{ab}_{cd}(\NOg \lambda)|^2}{p^2}\\
			&\times\lambda^2 f_b(k)\left(1\pm f_d(k-\omega)\right),
		\end{split}
	\end{align}
	where $\dd{\tilde\Gamma}$ denotes the integration measure and constant factors present in \eqref{eq:qhat_formula}.
	
 In a purely gluonic plasma, we can now consider 
 $\qhatf/\lambda$ and $\qhatff$ separately as functions of $\NOg\lambda$ as in Eq.~\eqref{eq:scaled_thermal_splitting},
	\begin{align}
		\qhat = \lambda\left(\frac{\qhatf}{\lambda}\right)(\NOg \lambda)+ \qhatff (\NOg \lambda), \tag{\ref{eq:scaled_thermal_splitting}}
	\end{align}
	with
	\begin{align}
		&\lambda \left(\frac{\qhatf}{\lambda}\right) = \lambda\int\dd{\Gamma}q^2\sin^2\theta_{pq}\frac{|\tilde M^{gg}_{gg}(\NOg \lambda)|^2}{p^2}\lambda f_g(k),\\ 
		&~~ \qhatff = \int\dd{\Gamma}q^2\sin^2\theta_{pq}\frac{|\tilde M^{gg}_{gg}(\NOg \lambda)|^2}{p^2}\lambda^2 f_g(k) f_g(k-\omega).
	\end{align}
	Note that $f$ also contains a factor $\NOg $, which, together with $\lambda$ combines to the $\NOg \lambda$ behavior in \eqref{eq:scaled_thermal_splitting}.
	Thus we can obtain $\qhatf$ and $\qhatff$ numerically with one independent parameter, $\NOg \lambda$, and then add the second independent parameter later.

\bibliography{qhatlong}

\begin{thebibliography}{93}%
\makeatletter
\providecommand \@ifxundefined [1]{%
 \@ifx{#1\undefined}
}%
\providecommand \@ifnum [1]{%
 \ifnum #1\expandafter \@firstoftwo
 \else \expandafter \@secondoftwo
 \fi
}%
\providecommand \@ifx [1]{%
 \ifx #1\expandafter \@firstoftwo
 \else \expandafter \@secondoftwo
 \fi
}%
\providecommand \natexlab [1]{#1}%
\providecommand \enquote  [1]{``#1''}%
\providecommand \bibnamefont  [1]{#1}%
\providecommand \bibfnamefont [1]{#1}%
\providecommand \citenamefont [1]{#1}%
\providecommand \href@noop [0]{\@secondoftwo}%
\providecommand \href [0]{\begingroup \@sanitize@url \@href}%
\providecommand \@href[1]{\@@startlink{#1}\@@href}%
\providecommand \@@href[1]{\endgroup#1\@@endlink}%
\providecommand \@sanitize@url [0]{\catcode `\\12\catcode `\$12\catcode
  `\&12\catcode `\#12\catcode `\^12\catcode `\_12\catcode `\%12\relax}%
\providecommand \@@startlink[1]{}%
\providecommand \@@endlink[0]{}%
\providecommand \url  [0]{\begingroup\@sanitize@url \@url }%
\providecommand \@url [1]{\endgroup\@href {#1}{\urlprefix }}%
\providecommand \urlprefix  [0]{URL }%
\providecommand \Eprint [0]{\href }%
\providecommand \doibase [0]{http://dx.doi.org/}%
\providecommand \selectlanguage [0]{\@gobble}%
\providecommand \bibinfo  [0]{\@secondoftwo}%
\providecommand \bibfield  [0]{\@secondoftwo}%
\providecommand \translation [1]{[#1]}%
\providecommand \BibitemOpen [0]{}%
\providecommand \bibitemStop [0]{}%
\providecommand \bibitemNoStop [0]{.\EOS\space}%
\providecommand \EOS [0]{\spacefactor3000\relax}%
\providecommand \BibitemShut  [1]{\csname bibitem#1\endcsname}%
\let\auto@bib@innerbib\@empty
\bibitem [{\citenamefont {Baier}\ \emph {et~al.}(2001)\citenamefont {Baier},
  \citenamefont {Mueller}, \citenamefont {Schiff},\ and\ \citenamefont
  {Son}}]{Baier:2000sb}%
  \BibitemOpen
  \bibfield  {author} {\bibinfo {author} {\bibfnamefont {R.}~\bibnamefont
  {Baier}}, \bibinfo {author} {\bibfnamefont {Alfred~H.}\ \bibnamefont
  {Mueller}}, \bibinfo {author} {\bibfnamefont {D.}~\bibnamefont {Schiff}}, \
  and\ \bibinfo {author} {\bibfnamefont {D.~T.}\ \bibnamefont {Son}},\
  }\bibfield  {title} {\enquote {\bibinfo {title} {{'Bottom up' thermalization
  in heavy ion collisions}},}\ }\href {\doibase 10.1016/S0370-2693(01)00191-5}
  {\bibfield  {journal} {\bibinfo  {journal} {Phys. Lett. B}\ }\textbf
  {\bibinfo {volume} {502}},\ \bibinfo {pages} {51--58} (\bibinfo {year}
  {2001})},\ \Eprint {http://arxiv.org/abs/hep-ph/0009237}
  {arXiv:hep-ph/0009237} \BibitemShut {NoStop}%
\bibitem [{\citenamefont {Gelis}\ \emph {et~al.}(2010)\citenamefont {Gelis},
  \citenamefont {Iancu}, \citenamefont {Jalilian-Marian},\ and\ \citenamefont
  {Venugopalan}}]{Gelis:2010nm}%
  \BibitemOpen
  \bibfield  {author} {\bibinfo {author} {\bibfnamefont {Francois}\
  \bibnamefont {Gelis}}, \bibinfo {author} {\bibfnamefont {Edmond}\
  \bibnamefont {Iancu}}, \bibinfo {author} {\bibfnamefont {Jamal}\ \bibnamefont
  {Jalilian-Marian}}, \ and\ \bibinfo {author} {\bibfnamefont {Raju}\
  \bibnamefont {Venugopalan}},\ }\bibfield  {title} {\enquote {\bibinfo {title}
  {{The Color Glass Condensate}},}\ }\href {\doibase
  10.1146/annurev.nucl.010909.083629} {\bibfield  {journal} {\bibinfo
  {journal} {Ann. Rev. Nucl. Part. Sci.}\ }\textbf {\bibinfo {volume} {60}},\
  \bibinfo {pages} {463--489} (\bibinfo {year} {2010})},\ \Eprint
  {http://arxiv.org/abs/1002.0333} {arXiv:1002.0333 [hep-ph]} \BibitemShut
  {NoStop}%
\bibitem [{\citenamefont {Kurkela}\ \emph {et~al.}(2016)\citenamefont
  {Kurkela}, \citenamefont {Lappi},\ and\ \citenamefont
  {Peuron}}]{Kurkela:2016mhu}%
  \BibitemOpen
  \bibfield  {author} {\bibinfo {author} {\bibfnamefont {Aleksi}\ \bibnamefont
  {Kurkela}}, \bibinfo {author} {\bibfnamefont {Tuomas}\ \bibnamefont {Lappi}},
  \ and\ \bibinfo {author} {\bibfnamefont {Jarkko}\ \bibnamefont {Peuron}},\
  }\bibfield  {title} {\enquote {\bibinfo {title} {{Time evolution of
  linearized gauge field fluctuations on a real-time lattice}},}\ }\href
  {\doibase 10.1140/epjc/s10052-016-4523-9} {\bibfield  {journal} {\bibinfo
  {journal} {Eur. Phys. J. C}\ }\textbf {\bibinfo {volume} {76}},\ \bibinfo
  {pages} {688} (\bibinfo {year} {2016})},\ \Eprint
  {http://arxiv.org/abs/1610.01355} {arXiv:1610.01355 [hep-lat]} \BibitemShut
  {NoStop}%
\bibitem [{\citenamefont {Boguslavski}\ \emph {et~al.}(2018)\citenamefont
  {Boguslavski}, \citenamefont {Kurkela}, \citenamefont {Lappi},\ and\
  \citenamefont {Peuron}}]{Boguslavski:2018beu}%
  \BibitemOpen
  \bibfield  {author} {\bibinfo {author} {\bibfnamefont {K.}~\bibnamefont
  {Boguslavski}}, \bibinfo {author} {\bibfnamefont {A.}~\bibnamefont
  {Kurkela}}, \bibinfo {author} {\bibfnamefont {T.}~\bibnamefont {Lappi}}, \
  and\ \bibinfo {author} {\bibfnamefont {J.}~\bibnamefont {Peuron}},\
  }\bibfield  {title} {\enquote {\bibinfo {title} {{Spectral function for
  overoccupied gluodynamics from real-time lattice simulations}},}\ }\href
  {\doibase 10.1103/PhysRevD.98.014006} {\bibfield  {journal} {\bibinfo
  {journal} {Phys. Rev. D}\ }\textbf {\bibinfo {volume} {98}},\ \bibinfo
  {pages} {014006} (\bibinfo {year} {2018})},\ \Eprint
  {http://arxiv.org/abs/1804.01966} {arXiv:1804.01966 [hep-ph]} \BibitemShut
  {NoStop}%
\bibitem [{\citenamefont {Boguslavski}\ \emph {et~al.}(2021)\citenamefont
  {Boguslavski}, \citenamefont {Kurkela}, \citenamefont {Lappi},\ and\
  \citenamefont {Peuron}}]{Boguslavski:2021buh}%
  \BibitemOpen
  \bibfield  {author} {\bibinfo {author} {\bibfnamefont {K.}~\bibnamefont
  {Boguslavski}}, \bibinfo {author} {\bibfnamefont {A.}~\bibnamefont
  {Kurkela}}, \bibinfo {author} {\bibfnamefont {T.}~\bibnamefont {Lappi}}, \
  and\ \bibinfo {author} {\bibfnamefont {J.}~\bibnamefont {Peuron}},\
  }\bibfield  {title} {\enquote {\bibinfo {title} {{Broad excitations in a 2+1D
  overoccupied gluon plasma}},}\ }\href {\doibase 10.1007/JHEP05(2021)225}
  {\bibfield  {journal} {\bibinfo  {journal} {JHEP}\ }\textbf {\bibinfo
  {volume} {05}},\ \bibinfo {pages} {225} (\bibinfo {year} {2021})},\ \Eprint
  {http://arxiv.org/abs/2101.02715} {arXiv:2101.02715 [hep-ph]} \BibitemShut
  {NoStop}%
\bibitem [{\citenamefont {Boguslavski}\ \emph {et~al.}(2022)\citenamefont
  {Boguslavski}, \citenamefont {Lappi}, \citenamefont {Mace},\ and\
  \citenamefont {Schlichting}}]{Boguslavski:2021kdd}%
  \BibitemOpen
  \bibfield  {author} {\bibinfo {author} {\bibfnamefont {Kirill}\ \bibnamefont
  {Boguslavski}}, \bibinfo {author} {\bibfnamefont {Tuomas}\ \bibnamefont
  {Lappi}}, \bibinfo {author} {\bibfnamefont {Mark}\ \bibnamefont {Mace}}, \
  and\ \bibinfo {author} {\bibfnamefont {S\"oren}\ \bibnamefont
  {Schlichting}},\ }\bibfield  {title} {\enquote {\bibinfo {title} {{Spectral
  function of fermions in a highly occupied non-Abelian plasma}},}\ }\href
  {\doibase 10.1016/j.physletb.2022.136963} {\bibfield  {journal} {\bibinfo
  {journal} {Phys. Lett. B}\ }\textbf {\bibinfo {volume} {827}},\ \bibinfo
  {pages} {136963} (\bibinfo {year} {2022})},\ \Eprint
  {http://arxiv.org/abs/2106.11319} {arXiv:2106.11319 [hep-ph]} \BibitemShut
  {NoStop}%
\bibitem [{\citenamefont {Arnold}\ \emph
  {et~al.}(2003{\natexlab{a}})\citenamefont {Arnold}, \citenamefont {Moore},\
  and\ \citenamefont {Yaffe}}]{Arnold:2002zm}%
  \BibitemOpen
  \bibfield  {author} {\bibinfo {author} {\bibfnamefont {Peter~Brockway}\
  \bibnamefont {Arnold}}, \bibinfo {author} {\bibfnamefont {Guy~D.}\
  \bibnamefont {Moore}}, \ and\ \bibinfo {author} {\bibfnamefont {Laurence~G.}\
  \bibnamefont {Yaffe}},\ }\bibfield  {title} {\enquote {\bibinfo {title}
  {{Effective kinetic theory for high temperature gauge theories}},}\ }\href
  {\doibase 10.1088/1126-6708/2003/01/030} {\bibfield  {journal} {\bibinfo
  {journal} {JHEP}\ }\textbf {\bibinfo {volume} {01}},\ \bibinfo {pages} {030}
  (\bibinfo {year} {2003}{\natexlab{a}})},\ \Eprint
  {http://arxiv.org/abs/hep-ph/0209353} {arXiv:hep-ph/0209353} \BibitemShut
  {NoStop}%
\bibitem [{\citenamefont {Kurkela}\ and\ \citenamefont
  {Zhu}(2015)}]{Kurkela:2015qoa}%
  \BibitemOpen
  \bibfield  {author} {\bibinfo {author} {\bibfnamefont {Aleksi}\ \bibnamefont
  {Kurkela}}\ and\ \bibinfo {author} {\bibfnamefont {Yan}\ \bibnamefont
  {Zhu}},\ }\bibfield  {title} {\enquote {\bibinfo {title} {{Isotropization and
  hydrodynamization in weakly coupled heavy-ion collisions}},}\ }\href
  {\doibase 10.1103/PhysRevLett.115.182301} {\bibfield  {journal} {\bibinfo
  {journal} {Phys. Rev. Lett.}\ }\textbf {\bibinfo {volume} {115}},\ \bibinfo
  {pages} {182301} (\bibinfo {year} {2015})},\ \Eprint
  {http://arxiv.org/abs/1506.06647} {arXiv:1506.06647 [hep-ph]} \BibitemShut
  {NoStop}%
\bibitem [{\citenamefont {Romatschke}\ and\ \citenamefont
  {Romatschke}(2019)}]{Romatschke:2017ejr}%
  \BibitemOpen
  \bibfield  {author} {\bibinfo {author} {\bibfnamefont {Paul}\ \bibnamefont
  {Romatschke}}\ and\ \bibinfo {author} {\bibfnamefont {Ulrike}\ \bibnamefont
  {Romatschke}},\ }\href {\doibase 10.1017/9781108651998} {\emph {\bibinfo
  {title} {{Relativistic Fluid Dynamics In and Out of Equilibrium}}}},\
  Cambridge Monographs on Mathematical Physics\ (\bibinfo  {publisher}
  {Cambridge University Press},\ \bibinfo {year} {2019})\ \Eprint
  {http://arxiv.org/abs/1712.05815} {arXiv:1712.05815 [nucl-th]} \BibitemShut
  {NoStop}%
\bibitem [{\citenamefont {Denicol}\ \emph {et~al.}(2012)\citenamefont
  {Denicol}, \citenamefont {Niemi}, \citenamefont {Molnar},\ and\ \citenamefont
  {Rischke}}]{Denicol:2012cn}%
  \BibitemOpen
  \bibfield  {author} {\bibinfo {author} {\bibfnamefont {G.~S.}\ \bibnamefont
  {Denicol}}, \bibinfo {author} {\bibfnamefont {H.}~\bibnamefont {Niemi}},
  \bibinfo {author} {\bibfnamefont {E.}~\bibnamefont {Molnar}}, \ and\ \bibinfo
  {author} {\bibfnamefont {D.~H.}\ \bibnamefont {Rischke}},\ }\bibfield
  {title} {\enquote {\bibinfo {title} {{Derivation of transient relativistic
  fluid dynamics from the Boltzmann equation}},}\ }\href {\doibase
  10.1103/PhysRevD.85.114047} {\bibfield  {journal} {\bibinfo  {journal} {Phys.
  Rev. D}\ }\textbf {\bibinfo {volume} {85}},\ \bibinfo {pages} {114047}
  (\bibinfo {year} {2012})},\ \bibinfo {note} {[Erratum: Phys.Rev.D 91, 039902
  (2015)]},\ \Eprint {http://arxiv.org/abs/1202.4551} {arXiv:1202.4551
  [nucl-th]} \BibitemShut {NoStop}%
\bibitem [{\citenamefont {Rezzolla}\ and\ \citenamefont
  {Zanotti}(2013)}]{Rezzolla:2013}%
  \BibitemOpen
  \bibfield  {author} {\bibinfo {author} {\bibfnamefont {Luciano}\ \bibnamefont
  {Rezzolla}}\ and\ \bibinfo {author} {\bibfnamefont {Olindo}\ \bibnamefont
  {Zanotti}},\ }\href {\doibase 10.1093/acprof:oso/9780198528906.001.0001}
  {\emph {\bibinfo {title} {{Relativistic Hydrodynamics}}}}\ (\bibinfo
  {publisher} {Oxford University Press},\ \bibinfo {year} {2013})\BibitemShut
  {NoStop}%
\bibitem [{\citenamefont {Ambrus}\ \emph {et~al.}(2023)\citenamefont {Ambrus},
  \citenamefont {Schlichting},\ and\ \citenamefont
  {Werthmann}}]{Ambrus:2022qya}%
  \BibitemOpen
  \bibfield  {author} {\bibinfo {author} {\bibfnamefont {Victor~E.}\
  \bibnamefont {Ambrus}}, \bibinfo {author} {\bibfnamefont {S.}~\bibnamefont
  {Schlichting}}, \ and\ \bibinfo {author} {\bibfnamefont {C.}~\bibnamefont
  {Werthmann}},\ }\bibfield  {title} {\enquote {\bibinfo {title} {{Establishing
  the Range of Applicability of Hydrodynamics in High-Energy Collisions}},}\
  }\href {\doibase 10.1103/PhysRevLett.130.152301} {\bibfield  {journal}
  {\bibinfo  {journal} {Phys. Rev. Lett.}\ }\textbf {\bibinfo {volume} {130}},\
  \bibinfo {pages} {152301} (\bibinfo {year} {2023})},\ \Eprint
  {http://arxiv.org/abs/2211.14356} {arXiv:2211.14356 [hep-ph]} \BibitemShut
  {NoStop}%
\bibitem [{\citenamefont {Citron}\ \emph {et~al.}(2019)\citenamefont {Citron}
  \emph {et~al.}}]{Citron:2018lsq}%
  \BibitemOpen
  \bibfield  {author} {\bibinfo {author} {\bibfnamefont {Z.}~\bibnamefont
  {Citron}} \emph {et~al.},\ }\bibfield  {title} {\enquote {\bibinfo {title}
  {{Report from Working Group 5}: {Future physics opportunities for
  high-density QCD at the LHC with heavy-ion and proton beams}},}\ }\href
  {\doibase 10.23731/CYRM-2019-007.1159} {\bibfield  {journal} {\bibinfo
  {journal} {CERN Yellow Rep. Monogr.}\ }\textbf {\bibinfo {volume} {7}},\
  \bibinfo {pages} {1159--1410} (\bibinfo {year} {2019})},\ \Eprint
  {http://arxiv.org/abs/1812.06772} {arXiv:1812.06772 [hep-ph]} \BibitemShut
  {NoStop}%
\bibitem [{\citenamefont {Ipp}\ \emph {et~al.}(2020{\natexlab{a}})\citenamefont
  {Ipp}, \citenamefont {M\"uller},\ and\ \citenamefont {Schuh}}]{Ipp:2020mjc}%
  \BibitemOpen
  \bibfield  {author} {\bibinfo {author} {\bibfnamefont {Andreas}\ \bibnamefont
  {Ipp}}, \bibinfo {author} {\bibfnamefont {David~I.}\ \bibnamefont
  {M\"uller}}, \ and\ \bibinfo {author} {\bibfnamefont {Daniel}\ \bibnamefont
  {Schuh}},\ }\bibfield  {title} {\enquote {\bibinfo {title} {{Anisotropic
  momentum broadening in the 2+1D Glasma: analytic weak field approximation and
  lattice simulations}},}\ }\href {\doibase 10.1103/PhysRevD.102.074001}
  {\bibfield  {journal} {\bibinfo  {journal} {Phys. Rev. D}\ }\textbf {\bibinfo
  {volume} {102}},\ \bibinfo {pages} {074001} (\bibinfo {year}
  {2020}{\natexlab{a}})},\ \Eprint {http://arxiv.org/abs/2001.10001}
  {arXiv:2001.10001 [hep-ph]} \BibitemShut {NoStop}%
\bibitem [{\citenamefont {Ipp}\ \emph {et~al.}(2020{\natexlab{b}})\citenamefont
  {Ipp}, \citenamefont {M\"uller},\ and\ \citenamefont {Schuh}}]{Ipp:2020nfu}%
  \BibitemOpen
  \bibfield  {author} {\bibinfo {author} {\bibfnamefont {Andreas}\ \bibnamefont
  {Ipp}}, \bibinfo {author} {\bibfnamefont {David~I.}\ \bibnamefont
  {M\"uller}}, \ and\ \bibinfo {author} {\bibfnamefont {Daniel}\ \bibnamefont
  {Schuh}},\ }\bibfield  {title} {\enquote {\bibinfo {title} {{Jet momentum
  broadening in the pre-equilibrium Glasma}},}\ }\href {\doibase
  10.1016/j.physletb.2020.135810} {\bibfield  {journal} {\bibinfo  {journal}
  {Phys. Lett. B}\ }\textbf {\bibinfo {volume} {810}},\ \bibinfo {pages}
  {135810} (\bibinfo {year} {2020}{\natexlab{b}})},\ \Eprint
  {http://arxiv.org/abs/2009.14206} {arXiv:2009.14206 [hep-ph]} \BibitemShut
  {NoStop}%
\bibitem [{\citenamefont {Carrington}\ \emph
  {et~al.}(2022{\natexlab{a}})\citenamefont {Carrington}, \citenamefont
  {Czajka},\ and\ \citenamefont {Mrowczynski}}]{Carrington:2021dvw}%
  \BibitemOpen
  \bibfield  {author} {\bibinfo {author} {\bibfnamefont {Margaret~E.}\
  \bibnamefont {Carrington}}, \bibinfo {author} {\bibfnamefont {Alina}\
  \bibnamefont {Czajka}}, \ and\ \bibinfo {author} {\bibfnamefont {Stanislaw}\
  \bibnamefont {Mrowczynski}},\ }\bibfield  {title} {\enquote {\bibinfo {title}
  {{Jet quenching in glasma}},}\ }\href {\doibase
  10.1016/j.physletb.2022.137464} {\bibfield  {journal} {\bibinfo  {journal}
  {Phys. Lett. B}\ }\textbf {\bibinfo {volume} {834}},\ \bibinfo {pages}
  {137464} (\bibinfo {year} {2022}{\natexlab{a}})},\ \Eprint
  {http://arxiv.org/abs/2112.06812} {arXiv:2112.06812 [hep-ph]} \BibitemShut
  {NoStop}%
\bibitem [{\citenamefont {Carrington}\ \emph
  {et~al.}(2022{\natexlab{b}})\citenamefont {Carrington}, \citenamefont
  {Czajka},\ and\ \citenamefont {Mrowczynski}}]{Carrington:2022bnv}%
  \BibitemOpen
  \bibfield  {author} {\bibinfo {author} {\bibfnamefont {Margaret~E.}\
  \bibnamefont {Carrington}}, \bibinfo {author} {\bibfnamefont {Alina}\
  \bibnamefont {Czajka}}, \ and\ \bibinfo {author} {\bibfnamefont {Stanislaw}\
  \bibnamefont {Mrowczynski}},\ }\bibfield  {title} {\enquote {\bibinfo {title}
  {{Transport of hard probes through glasma}},}\ }\href {\doibase
  10.1103/PhysRevC.105.064910} {\bibfield  {journal} {\bibinfo  {journal}
  {Phys. Rev. C}\ }\textbf {\bibinfo {volume} {105}},\ \bibinfo {pages}
  {064910} (\bibinfo {year} {2022}{\natexlab{b}})},\ \Eprint
  {http://arxiv.org/abs/2202.00357} {arXiv:2202.00357 [nucl-th]} \BibitemShut
  {NoStop}%
\bibitem [{\citenamefont {Avramescu}\ \emph {et~al.}(2023)\citenamefont
  {Avramescu}, \citenamefont {B\u{a}ran}, \citenamefont {Greco}, \citenamefont
  {Ipp}, \citenamefont {M\"uller},\ and\ \citenamefont
  {Ruggieri}}]{Avramescu:2023qvv}%
  \BibitemOpen
  \bibfield  {author} {\bibinfo {author} {\bibfnamefont {Dana}\ \bibnamefont
  {Avramescu}}, \bibinfo {author} {\bibfnamefont {Virgil}\ \bibnamefont
  {B\u{a}ran}}, \bibinfo {author} {\bibfnamefont {Vincenzo}\ \bibnamefont
  {Greco}}, \bibinfo {author} {\bibfnamefont {Andreas}\ \bibnamefont {Ipp}},
  \bibinfo {author} {\bibfnamefont {David.~I.}\ \bibnamefont {M\"uller}}, \
  and\ \bibinfo {author} {\bibfnamefont {Marco}\ \bibnamefont {Ruggieri}},\
  }\bibfield  {title} {\enquote {\bibinfo {title} {{Simulating jets and heavy
  quarks in the glasma using the colored particle-in-cell method}},}\ }\href
  {\doibase 10.1103/PhysRevD.107.114021} {\bibfield  {journal} {\bibinfo
  {journal} {Phys. Rev. D}\ }\textbf {\bibinfo {volume} {107}},\ \bibinfo
  {pages} {114021} (\bibinfo {year} {2023})},\ \Eprint
  {http://arxiv.org/abs/2303.05599} {arXiv:2303.05599 [hep-ph]} \BibitemShut
  {NoStop}%
\bibitem [{\citenamefont {Andres}\ \emph {et~al.}(2023)\citenamefont {Andres},
  \citenamefont {Apolin\'ario}, \citenamefont {Dominguez}, \citenamefont
  {Martinez},\ and\ \citenamefont {Salgado}}]{Andres:2022bql}%
  \BibitemOpen
  \bibfield  {author} {\bibinfo {author} {\bibfnamefont {Carlota}\ \bibnamefont
  {Andres}}, \bibinfo {author} {\bibfnamefont {Liliana}\ \bibnamefont
  {Apolin\'ario}}, \bibinfo {author} {\bibfnamefont {Fabio}\ \bibnamefont
  {Dominguez}}, \bibinfo {author} {\bibfnamefont {Marcos~Gonzalez}\
  \bibnamefont {Martinez}}, \ and\ \bibinfo {author} {\bibfnamefont
  {Carlos~A.}\ \bibnamefont {Salgado}},\ }\bibfield  {title} {\enquote
  {\bibinfo {title} {{Medium-induced radiation with vacuum propagation in the
  pre-hydrodynamics phase}},}\ }\href {\doibase 10.1007/JHEP03(2023)189}
  {\bibfield  {journal} {\bibinfo  {journal} {JHEP}\ }\textbf {\bibinfo
  {volume} {03}},\ \bibinfo {pages} {189} (\bibinfo {year} {2023})},\ \Eprint
  {http://arxiv.org/abs/2211.10161} {arXiv:2211.10161 [hep-ph]} \BibitemShut
  {NoStop}%
\bibitem [{\citenamefont {Boguslavski}\ \emph
  {et~al.}(2024{\natexlab{a}})\citenamefont {Boguslavski}, \citenamefont
  {Kurkela}, \citenamefont {Lappi}, \citenamefont {Lindenbauer},\ and\
  \citenamefont {Peuron}}]{Boguslavski:2023alu}%
  \BibitemOpen
  \bibfield  {author} {\bibinfo {author} {\bibfnamefont {Kirill}\ \bibnamefont
  {Boguslavski}}, \bibinfo {author} {\bibfnamefont {Aleksi}\ \bibnamefont
  {Kurkela}}, \bibinfo {author} {\bibfnamefont {Tuomas}\ \bibnamefont {Lappi}},
  \bibinfo {author} {\bibfnamefont {Florian}\ \bibnamefont {Lindenbauer}}, \
  and\ \bibinfo {author} {\bibfnamefont {Jarkko}\ \bibnamefont {Peuron}},\
  }\bibfield  {title} {\enquote {\bibinfo {title} {{Jet momentum broadening
  during initial stages in heavy-ion collisions}},}\ }\href {\doibase
  10.1016/j.physletb.2024.138525} {\bibfield  {journal} {\bibinfo  {journal}
  {Phys. Lett. B}\ }\textbf {\bibinfo {volume} {850}},\ \bibinfo {pages}
  {138525} (\bibinfo {year} {2024}{\natexlab{a}})},\ \Eprint
  {http://arxiv.org/abs/2303.12595} {arXiv:2303.12595 [hep-ph]} \BibitemShut
  {NoStop}%
\bibitem [{\citenamefont {Hauksson}\ and\ \citenamefont
  {Iancu}(2023)}]{Hauksson:2023tze}%
  \BibitemOpen
  \bibfield  {author} {\bibinfo {author} {\bibfnamefont {Sigtryggur}\
  \bibnamefont {Hauksson}}\ and\ \bibinfo {author} {\bibfnamefont {Edmond}\
  \bibnamefont {Iancu}},\ }\bibfield  {title} {\enquote {\bibinfo {title} {{Jet
  polarisation in an anisotropic medium}},}\ }\href {\doibase
  10.1007/JHEP08(2023)027} {\bibfield  {journal} {\bibinfo  {journal} {JHEP}\
  }\textbf {\bibinfo {volume} {08}},\ \bibinfo {pages} {027} (\bibinfo {year}
  {2023})},\ \Eprint {http://arxiv.org/abs/2303.03914} {arXiv:2303.03914
  [hep-ph]} \BibitemShut {NoStop}%
\bibitem [{\citenamefont {Baier}\ \emph
  {et~al.}(1998{\natexlab{a}})\citenamefont {Baier}, \citenamefont
  {Dokshitzer}, \citenamefont {Mueller},\ and\ \citenamefont
  {Schiff}}]{Baier:1998kq}%
  \BibitemOpen
  \bibfield  {author} {\bibinfo {author} {\bibfnamefont {R.}~\bibnamefont
  {Baier}}, \bibinfo {author} {\bibfnamefont {Yuri~L.}\ \bibnamefont
  {Dokshitzer}}, \bibinfo {author} {\bibfnamefont {Alfred~H.}\ \bibnamefont
  {Mueller}}, \ and\ \bibinfo {author} {\bibfnamefont {D.}~\bibnamefont
  {Schiff}},\ }\bibfield  {title} {\enquote {\bibinfo {title} {{Medium induced
  radiative energy loss: Equivalence between the BDMPS and Zakharov
  formalisms}},}\ }\href {\doibase 10.1016/S0550-3213(98)00546-X} {\bibfield
  {journal} {\bibinfo  {journal} {Nucl. Phys. B}\ }\textbf {\bibinfo {volume}
  {531}},\ \bibinfo {pages} {403--425} (\bibinfo {year}
  {1998}{\natexlab{a}})},\ \Eprint {http://arxiv.org/abs/hep-ph/9804212}
  {arXiv:hep-ph/9804212} \BibitemShut {NoStop}%
\bibitem [{\citenamefont {Baier}\ \emph
  {et~al.}(1998{\natexlab{b}})\citenamefont {Baier}, \citenamefont
  {Dokshitzer}, \citenamefont {Mueller},\ and\ \citenamefont
  {Schiff}}]{Baier:1998yf}%
  \BibitemOpen
  \bibfield  {author} {\bibinfo {author} {\bibfnamefont {R.}~\bibnamefont
  {Baier}}, \bibinfo {author} {\bibfnamefont {Yuri~L.}\ \bibnamefont
  {Dokshitzer}}, \bibinfo {author} {\bibfnamefont {Alfred~H.}\ \bibnamefont
  {Mueller}}, \ and\ \bibinfo {author} {\bibfnamefont {D.}~\bibnamefont
  {Schiff}},\ }\bibfield  {title} {\enquote {\bibinfo {title} {{Radiative
  energy loss of high-energy partons traversing an expanding QCD plasma}},}\
  }\href {\doibase 10.1103/PhysRevC.58.1706} {\bibfield  {journal} {\bibinfo
  {journal} {Phys. Rev. C}\ }\textbf {\bibinfo {volume} {58}},\ \bibinfo
  {pages} {1706--1713} (\bibinfo {year} {1998}{\natexlab{b}})},\ \Eprint
  {http://arxiv.org/abs/hep-ph/9803473} {arXiv:hep-ph/9803473} \BibitemShut
  {NoStop}%
\bibitem [{\citenamefont {Zakharov}(2001)}]{Zakharov:2000iz}%
  \BibitemOpen
  \bibfield  {author} {\bibinfo {author} {\bibfnamefont {B.~G.}\ \bibnamefont
  {Zakharov}},\ }\bibfield  {title} {\enquote {\bibinfo {title} {{On the energy
  loss of high-energy quarks in a finite size quark - gluon plasma}},}\ }\href
  {\doibase 10.1134/1.1358417} {\bibfield  {journal} {\bibinfo  {journal} {JETP
  Lett.}\ }\textbf {\bibinfo {volume} {73}},\ \bibinfo {pages} {49--52}
  (\bibinfo {year} {2001})},\ \Eprint {http://arxiv.org/abs/hep-ph/0012360}
  {arXiv:hep-ph/0012360} \BibitemShut {NoStop}%
\bibitem [{\citenamefont {Arnold}(2009)}]{Arnold:2008iy}%
  \BibitemOpen
  \bibfield  {author} {\bibinfo {author} {\bibfnamefont {Peter~Brockway}\
  \bibnamefont {Arnold}},\ }\bibfield  {title} {\enquote {\bibinfo {title}
  {{Simple Formula for High-Energy Gluon Bremsstrahlung in a Finite, Expanding
  Medium}},}\ }\href {\doibase 10.1103/PhysRevD.79.065025} {\bibfield
  {journal} {\bibinfo  {journal} {Phys. Rev. D}\ }\textbf {\bibinfo {volume}
  {79}},\ \bibinfo {pages} {065025} (\bibinfo {year} {2009})},\ \Eprint
  {http://arxiv.org/abs/0808.2767} {arXiv:0808.2767 [hep-ph]} \BibitemShut
  {NoStop}%
\bibitem [{\citenamefont {Andres}\ \emph {et~al.}(2020)\citenamefont {Andres},
  \citenamefont {Armesto}, \citenamefont {Niemi}, \citenamefont {Paatelainen},\
  and\ \citenamefont {Salgado}}]{Andres:2019eus}%
  \BibitemOpen
  \bibfield  {author} {\bibinfo {author} {\bibfnamefont {Carlota}\ \bibnamefont
  {Andres}}, \bibinfo {author} {\bibfnamefont {N\'estor}\ \bibnamefont
  {Armesto}}, \bibinfo {author} {\bibfnamefont {Harri}\ \bibnamefont {Niemi}},
  \bibinfo {author} {\bibfnamefont {Risto}\ \bibnamefont {Paatelainen}}, \ and\
  \bibinfo {author} {\bibfnamefont {Carlos~A.}\ \bibnamefont {Salgado}},\
  }\bibfield  {title} {\enquote {\bibinfo {title} {{Jet quenching as a probe of
  the initial stages in heavy-ion collisions}},}\ }\href {\doibase
  10.1016/j.physletb.2020.135318} {\bibfield  {journal} {\bibinfo  {journal}
  {Phys. Lett. B}\ }\textbf {\bibinfo {volume} {803}},\ \bibinfo {pages}
  {135318} (\bibinfo {year} {2020})},\ \Eprint
  {http://arxiv.org/abs/1902.03231} {arXiv:1902.03231 [hep-ph]} \BibitemShut
  {NoStop}%
\bibitem [{\citenamefont {Adhya}\ \emph {et~al.}(2020)\citenamefont {Adhya},
  \citenamefont {Salgado}, \citenamefont {Spousta},\ and\ \citenamefont
  {Tywoniuk}}]{Adhya:2019qse}%
  \BibitemOpen
  \bibfield  {author} {\bibinfo {author} {\bibfnamefont {Souvik~Priyam}\
  \bibnamefont {Adhya}}, \bibinfo {author} {\bibfnamefont {Carlos~A.}\
  \bibnamefont {Salgado}}, \bibinfo {author} {\bibfnamefont {Martin}\
  \bibnamefont {Spousta}}, \ and\ \bibinfo {author} {\bibfnamefont {Konrad}\
  \bibnamefont {Tywoniuk}},\ }\bibfield  {title} {\enquote {\bibinfo {title}
  {{Medium-induced cascade in expanding media}},}\ }\href {\doibase
  10.1007/JHEP07(2020)150} {\bibfield  {journal} {\bibinfo  {journal} {JHEP}\
  }\textbf {\bibinfo {volume} {07}},\ \bibinfo {pages} {150} (\bibinfo {year}
  {2020})},\ \Eprint {http://arxiv.org/abs/1911.12193} {arXiv:1911.12193
  [hep-ph]} \BibitemShut {NoStop}%
\bibitem [{\citenamefont {Huss}\ \emph
  {et~al.}(2021{\natexlab{a}})\citenamefont {Huss}, \citenamefont {Kurkela},
  \citenamefont {Mazeliauskas}, \citenamefont {Paatelainen}, \citenamefont
  {van~der Schee},\ and\ \citenamefont {Wiedemann}}]{Huss:2020dwe}%
  \BibitemOpen
  \bibfield  {author} {\bibinfo {author} {\bibfnamefont {Alexander}\
  \bibnamefont {Huss}}, \bibinfo {author} {\bibfnamefont {Aleksi}\ \bibnamefont
  {Kurkela}}, \bibinfo {author} {\bibfnamefont {Aleksas}\ \bibnamefont
  {Mazeliauskas}}, \bibinfo {author} {\bibfnamefont {Risto}\ \bibnamefont
  {Paatelainen}}, \bibinfo {author} {\bibfnamefont {Wilke}\ \bibnamefont
  {van~der Schee}}, \ and\ \bibinfo {author} {\bibfnamefont {Urs~Achim}\
  \bibnamefont {Wiedemann}},\ }\bibfield  {title} {\enquote {\bibinfo {title}
  {{Discovering Partonic Rescattering in Light Nucleus Collisions}},}\ }\href
  {\doibase 10.1103/PhysRevLett.126.192301} {\bibfield  {journal} {\bibinfo
  {journal} {Phys. Rev. Lett.}\ }\textbf {\bibinfo {volume} {126}},\ \bibinfo
  {pages} {192301} (\bibinfo {year} {2021}{\natexlab{a}})},\ \Eprint
  {http://arxiv.org/abs/2007.13754} {arXiv:2007.13754 [hep-ph]} \BibitemShut
  {NoStop}%
\bibitem [{\citenamefont {Huss}\ \emph
  {et~al.}(2021{\natexlab{b}})\citenamefont {Huss}, \citenamefont {Kurkela},
  \citenamefont {Mazeliauskas}, \citenamefont {Paatelainen}, \citenamefont
  {van~der Schee},\ and\ \citenamefont {Wiedemann}}]{Huss:2020whe}%
  \BibitemOpen
  \bibfield  {author} {\bibinfo {author} {\bibfnamefont {Alexander}\
  \bibnamefont {Huss}}, \bibinfo {author} {\bibfnamefont {Aleksi}\ \bibnamefont
  {Kurkela}}, \bibinfo {author} {\bibfnamefont {Aleksas}\ \bibnamefont
  {Mazeliauskas}}, \bibinfo {author} {\bibfnamefont {Risto}\ \bibnamefont
  {Paatelainen}}, \bibinfo {author} {\bibfnamefont {Wilke}\ \bibnamefont
  {van~der Schee}}, \ and\ \bibinfo {author} {\bibfnamefont {Urs~Achim}\
  \bibnamefont {Wiedemann}},\ }\bibfield  {title} {\enquote {\bibinfo {title}
  {{Predicting parton energy loss in small collision systems}},}\ }\href
  {\doibase 10.1103/PhysRevC.103.054903} {\bibfield  {journal} {\bibinfo
  {journal} {Phys. Rev. C}\ }\textbf {\bibinfo {volume} {103}},\ \bibinfo
  {pages} {054903} (\bibinfo {year} {2021}{\natexlab{b}})},\ \Eprint
  {http://arxiv.org/abs/2007.13758} {arXiv:2007.13758 [hep-ph]} \BibitemShut
  {NoStop}%
\bibitem [{\citenamefont {Arnold}\ and\ \citenamefont
  {Xiao}(2008)}]{Arnold:2008vd}%
  \BibitemOpen
  \bibfield  {author} {\bibinfo {author} {\bibfnamefont {Peter~Brockway}\
  \bibnamefont {Arnold}}\ and\ \bibinfo {author} {\bibfnamefont {Wei}\
  \bibnamefont {Xiao}},\ }\bibfield  {title} {\enquote {\bibinfo {title}
  {{High-energy jet quenching in weakly-coupled quark-gluon plasmas}},}\ }\href
  {\doibase 10.1103/PhysRevD.78.125008} {\bibfield  {journal} {\bibinfo
  {journal} {Phys. Rev. D}\ }\textbf {\bibinfo {volume} {78}},\ \bibinfo
  {pages} {125008} (\bibinfo {year} {2008})},\ \Eprint
  {http://arxiv.org/abs/0810.1026} {arXiv:0810.1026 [hep-ph]} \BibitemShut
  {NoStop}%
\bibitem [{\citenamefont {Caron-Huot}(2009)}]{Caron-Huot:2008zna}%
  \BibitemOpen
  \bibfield  {author} {\bibinfo {author} {\bibfnamefont {Simon}\ \bibnamefont
  {Caron-Huot}},\ }\bibfield  {title} {\enquote {\bibinfo {title} {{O(g) plasma
  effects in jet quenching}},}\ }\href {\doibase 10.1103/PhysRevD.79.065039}
  {\bibfield  {journal} {\bibinfo  {journal} {Phys. Rev. D}\ }\textbf {\bibinfo
  {volume} {79}},\ \bibinfo {pages} {065039} (\bibinfo {year} {2009})},\
  \Eprint {http://arxiv.org/abs/0811.1603} {arXiv:0811.1603 [hep-ph]}
  \BibitemShut {NoStop}%
\bibitem [{\citenamefont {Aurenche}\ \emph {et~al.}(2002)\citenamefont
  {Aurenche}, \citenamefont {Gelis},\ and\ \citenamefont
  {Zaraket}}]{Aurenche:2002pd}%
  \BibitemOpen
  \bibfield  {author} {\bibinfo {author} {\bibfnamefont {P.}~\bibnamefont
  {Aurenche}}, \bibinfo {author} {\bibfnamefont {F.}~\bibnamefont {Gelis}}, \
  and\ \bibinfo {author} {\bibfnamefont {H.}~\bibnamefont {Zaraket}},\
  }\bibfield  {title} {\enquote {\bibinfo {title} {{A Simple sum rule for the
  thermal gluon spectral function and applications}},}\ }\href {\doibase
  10.1088/1126-6708/2002/05/043} {\bibfield  {journal} {\bibinfo  {journal}
  {JHEP}\ }\textbf {\bibinfo {volume} {05}},\ \bibinfo {pages} {043} (\bibinfo
  {year} {2002})},\ \Eprint {http://arxiv.org/abs/hep-ph/0204146}
  {arXiv:hep-ph/0204146} \BibitemShut {NoStop}%
\bibitem [{\citenamefont {Maldacena}(1998)}]{Maldacena:1997re}%
  \BibitemOpen
  \bibfield  {author} {\bibinfo {author} {\bibfnamefont {Juan~Martin}\
  \bibnamefont {Maldacena}},\ }\bibfield  {title} {\enquote {\bibinfo {title}
  {{The Large N limit of superconformal field theories and supergravity}},}\
  }\href {\doibase 10.1023/A:1026654312961} {\bibfield  {journal} {\bibinfo
  {journal} {Adv. Theor. Math. Phys.}\ }\textbf {\bibinfo {volume} {2}},\
  \bibinfo {pages} {231--252} (\bibinfo {year} {1998})},\ \Eprint
  {http://arxiv.org/abs/hep-th/9711200} {arXiv:hep-th/9711200} \BibitemShut
  {NoStop}%
\bibitem [{\citenamefont {Witten}(1998)}]{Witten:1998qj}%
  \BibitemOpen
  \bibfield  {author} {\bibinfo {author} {\bibfnamefont {Edward}\ \bibnamefont
  {Witten}},\ }\bibfield  {title} {\enquote {\bibinfo {title} {{Anti-de Sitter
  space and holography}},}\ }\href {\doibase 10.4310/ATMP.1998.v2.n2.a2}
  {\bibfield  {journal} {\bibinfo  {journal} {Adv. Theor. Math. Phys.}\
  }\textbf {\bibinfo {volume} {2}},\ \bibinfo {pages} {253--291} (\bibinfo
  {year} {1998})},\ \Eprint {http://arxiv.org/abs/hep-th/9802150}
  {arXiv:hep-th/9802150} \BibitemShut {NoStop}%
\bibitem [{\citenamefont {Liu}\ \emph {et~al.}(2006)\citenamefont {Liu},
  \citenamefont {Rajagopal},\ and\ \citenamefont {Wiedemann}}]{Liu:2006ug}%
  \BibitemOpen
  \bibfield  {author} {\bibinfo {author} {\bibfnamefont {Hong}\ \bibnamefont
  {Liu}}, \bibinfo {author} {\bibfnamefont {Krishna}\ \bibnamefont
  {Rajagopal}}, \ and\ \bibinfo {author} {\bibfnamefont {Urs~Achim}\
  \bibnamefont {Wiedemann}},\ }\bibfield  {title} {\enquote {\bibinfo {title}
  {{Calculating the jet quenching parameter from AdS/CFT}},}\ }\href {\doibase
  10.1103/PhysRevLett.97.182301} {\bibfield  {journal} {\bibinfo  {journal}
  {Phys. Rev. Lett.}\ }\textbf {\bibinfo {volume} {97}},\ \bibinfo {pages}
  {182301} (\bibinfo {year} {2006})},\ \Eprint
  {http://arxiv.org/abs/hep-ph/0605178} {arXiv:hep-ph/0605178} \BibitemShut
  {NoStop}%
\bibitem [{\citenamefont {Zhang}\ \emph {et~al.}(2013)\citenamefont {Zhang},
  \citenamefont {Hou},\ and\ \citenamefont {Ren}}]{Zhang:2012jd}%
  \BibitemOpen
  \bibfield  {author} {\bibinfo {author} {\bibfnamefont {Zi-qiang}\
  \bibnamefont {Zhang}}, \bibinfo {author} {\bibfnamefont {De-fu}\ \bibnamefont
  {Hou}}, \ and\ \bibinfo {author} {\bibfnamefont {Hai-cang}\ \bibnamefont
  {Ren}},\ }\bibfield  {title} {\enquote {\bibinfo {title} {{The finite 't
  Hooft coupling correction on jet quenching parameter in a $\mathcal N=4$
  Super Yang-Mills Plasma}},}\ }\href {\doibase 10.1007/JHEP01(2013)032}
  {\bibfield  {journal} {\bibinfo  {journal} {JHEP}\ }\textbf {\bibinfo
  {volume} {01}},\ \bibinfo {pages} {032} (\bibinfo {year} {2013})},\ \Eprint
  {http://arxiv.org/abs/1210.5187} {arXiv:1210.5187 [hep-th]} \BibitemShut
  {NoStop}%
\bibitem [{\citenamefont {Kumar}\ \emph {et~al.}(2022)\citenamefont {Kumar},
  \citenamefont {Majumder},\ and\ \citenamefont {Weber}}]{Kumar:2020wvb}%
  \BibitemOpen
  \bibfield  {author} {\bibinfo {author} {\bibfnamefont {Amit}\ \bibnamefont
  {Kumar}}, \bibinfo {author} {\bibfnamefont {Abhijit}\ \bibnamefont
  {Majumder}}, \ and\ \bibinfo {author} {\bibfnamefont {Johannes~Heinrich}\
  \bibnamefont {Weber}},\ }\bibfield  {title} {\enquote {\bibinfo {title} {{Jet
  transport coefficient q\textasciicircum{} in lattice QCD}},}\ }\href
  {\doibase 10.1103/PhysRevD.106.034505} {\bibfield  {journal} {\bibinfo
  {journal} {Phys. Rev. D}\ }\textbf {\bibinfo {volume} {106}},\ \bibinfo
  {pages} {034505} (\bibinfo {year} {2022})},\ \Eprint
  {http://arxiv.org/abs/2010.14463} {arXiv:2010.14463 [hep-lat]} \BibitemShut
  {NoStop}%
\bibitem [{\citenamefont {Moore}\ \emph {et~al.}(2021)\citenamefont {Moore},
  \citenamefont {Schlichting}, \citenamefont {Schlusser},\ and\ \citenamefont
  {Soudi}}]{Moore:2021jwe}%
  \BibitemOpen
  \bibfield  {author} {\bibinfo {author} {\bibfnamefont {Guy~D.}\ \bibnamefont
  {Moore}}, \bibinfo {author} {\bibfnamefont {Soeren}\ \bibnamefont
  {Schlichting}}, \bibinfo {author} {\bibfnamefont {Niels}\ \bibnamefont
  {Schlusser}}, \ and\ \bibinfo {author} {\bibfnamefont {Ismail}\ \bibnamefont
  {Soudi}},\ }\bibfield  {title} {\enquote {\bibinfo {title} {{Non-perturbative
  determination of collisional broadening and medium induced radiation in QCD
  plasmas}},}\ }\href {\doibase 10.1007/JHEP10(2021)059} {\bibfield  {journal}
  {\bibinfo  {journal} {JHEP}\ }\textbf {\bibinfo {volume} {10}},\ \bibinfo
  {pages} {059} (\bibinfo {year} {2021})},\ \Eprint
  {http://arxiv.org/abs/2105.01679} {arXiv:2105.01679 [hep-ph]} \BibitemShut
  {NoStop}%
\bibitem [{\citenamefont {Song}\ \emph {et~al.}(2023)\citenamefont {Song},
  \citenamefont {Grishmanovskii},\ and\ \citenamefont
  {Soloveva}}]{Song:2022wil}%
  \BibitemOpen
  \bibfield  {author} {\bibinfo {author} {\bibfnamefont {Taesoo}\ \bibnamefont
  {Song}}, \bibinfo {author} {\bibfnamefont {Ilia}\ \bibnamefont
  {Grishmanovskii}}, \ and\ \bibinfo {author} {\bibfnamefont {Olga}\
  \bibnamefont {Soloveva}},\ }\bibfield  {title} {\enquote {\bibinfo {title}
  {{Soft gluon emission from heavy quark scattering in strongly interacting
  quark-gluon plasma}},}\ }\href {\doibase 10.1103/PhysRevD.107.036009}
  {\bibfield  {journal} {\bibinfo  {journal} {Phys. Rev. D}\ }\textbf {\bibinfo
  {volume} {107}},\ \bibinfo {pages} {036009} (\bibinfo {year} {2023})},\
  \Eprint {http://arxiv.org/abs/2210.04010} {arXiv:2210.04010 [nucl-th]}
  \BibitemShut {NoStop}%
\bibitem [{\citenamefont {Grishmanovskii}\ \emph {et~al.}(2022)\citenamefont
  {Grishmanovskii}, \citenamefont {Song}, \citenamefont {Soloveva},
  \citenamefont {Greiner},\ and\ \citenamefont
  {Bratkovskaya}}]{Grishmanovskii:2022tpb}%
  \BibitemOpen
  \bibfield  {author} {\bibinfo {author} {\bibfnamefont {Ilia}\ \bibnamefont
  {Grishmanovskii}}, \bibinfo {author} {\bibfnamefont {Taesoo}\ \bibnamefont
  {Song}}, \bibinfo {author} {\bibfnamefont {Olga}\ \bibnamefont {Soloveva}},
  \bibinfo {author} {\bibfnamefont {Carsten}\ \bibnamefont {Greiner}}, \ and\
  \bibinfo {author} {\bibfnamefont {Elena}\ \bibnamefont {Bratkovskaya}},\
  }\bibfield  {title} {\enquote {\bibinfo {title} {{Exploring jet transport
  coefficients by elastic scattering in the strongly interacting quark-gluon
  plasma}},}\ }\href {\doibase 10.1103/PhysRevC.106.014903} {\bibfield
  {journal} {\bibinfo  {journal} {Phys. Rev. C}\ }\textbf {\bibinfo {volume}
  {106}},\ \bibinfo {pages} {014903} (\bibinfo {year} {2022})},\ \Eprint
  {http://arxiv.org/abs/2204.01561} {arXiv:2204.01561 [nucl-th]} \BibitemShut
  {NoStop}%
\bibitem [{\citenamefont {Ghiglieri}\ \emph {et~al.}(2016)\citenamefont
  {Ghiglieri}, \citenamefont {Moore},\ and\ \citenamefont
  {Teaney}}]{Ghiglieri:2015ala}%
  \BibitemOpen
  \bibfield  {author} {\bibinfo {author} {\bibfnamefont {Jacopo}\ \bibnamefont
  {Ghiglieri}}, \bibinfo {author} {\bibfnamefont {Guy~D.}\ \bibnamefont
  {Moore}}, \ and\ \bibinfo {author} {\bibfnamefont {Derek}\ \bibnamefont
  {Teaney}},\ }\bibfield  {title} {\enquote {\bibinfo {title} {{Jet-Medium
  Interactions at NLO in a Weakly-Coupled Quark-Gluon Plasma}},}\ }\href
  {\doibase 10.1007/JHEP03(2016)095} {\bibfield  {journal} {\bibinfo  {journal}
  {JHEP}\ }\textbf {\bibinfo {volume} {03}},\ \bibinfo {pages} {095} (\bibinfo
  {year} {2016})},\ \Eprint {http://arxiv.org/abs/1509.07773} {arXiv:1509.07773
  [hep-ph]} \BibitemShut {NoStop}%
\bibitem [{\citenamefont {Schlichting}\ and\ \citenamefont
  {Soudi}(2021)}]{Schlichting:2020lef}%
  \BibitemOpen
  \bibfield  {author} {\bibinfo {author} {\bibfnamefont {Soeren}\ \bibnamefont
  {Schlichting}}\ and\ \bibinfo {author} {\bibfnamefont {Ismail}\ \bibnamefont
  {Soudi}},\ }\bibfield  {title} {\enquote {\bibinfo {title} {{Medium-induced
  fragmentation and equilibration of highly energetic partons}},}\ }\href
  {\doibase 10.1007/JHEP07(2021)077} {\bibfield  {journal} {\bibinfo  {journal}
  {JHEP}\ }\textbf {\bibinfo {volume} {07}},\ \bibinfo {pages} {077} (\bibinfo
  {year} {2021})},\ \Eprint {http://arxiv.org/abs/2008.04928} {arXiv:2008.04928
  [hep-ph]} \BibitemShut {NoStop}%
\bibitem [{\citenamefont {Mehtar-Tani}\ \emph {et~al.}(2023)\citenamefont
  {Mehtar-Tani}, \citenamefont {Schlichting},\ and\ \citenamefont
  {Soudi}}]{Mehtar-Tani:2022zwf}%
  \BibitemOpen
  \bibfield  {author} {\bibinfo {author} {\bibfnamefont {Y.}~\bibnamefont
  {Mehtar-Tani}}, \bibinfo {author} {\bibfnamefont {S.}~\bibnamefont
  {Schlichting}}, \ and\ \bibinfo {author} {\bibfnamefont {I.}~\bibnamefont
  {Soudi}},\ }\bibfield  {title} {\enquote {\bibinfo {title} {{Jet
  thermalization in QCD kinetic theory}},}\ }\href {\doibase
  10.1007/JHEP05(2023)091} {\bibfield  {journal} {\bibinfo  {journal} {JHEP}\
  }\textbf {\bibinfo {volume} {05}},\ \bibinfo {pages} {091} (\bibinfo {year}
  {2023})},\ \Eprint {http://arxiv.org/abs/2209.10569} {arXiv:2209.10569
  [hep-ph]} \BibitemShut {NoStop}%
\bibitem [{\citenamefont {Burke}\ \emph {et~al.}(2014)\citenamefont {Burke}
  \emph {et~al.}}]{JET:2013cls}%
  \BibitemOpen
  \bibfield  {author} {\bibinfo {author} {\bibfnamefont {Karen~M.}\
  \bibnamefont {Burke}} \emph {et~al.} (\bibinfo {collaboration} {JET}),\
  }\bibfield  {title} {\enquote {\bibinfo {title} {{Extracting the jet
  transport coefficient from jet quenching in high-energy heavy-ion
  collisions}},}\ }\href {\doibase 10.1103/PhysRevC.90.014909} {\bibfield
  {journal} {\bibinfo  {journal} {Phys. Rev. C}\ }\textbf {\bibinfo {volume}
  {90}},\ \bibinfo {pages} {014909} (\bibinfo {year} {2014})},\ \Eprint
  {http://arxiv.org/abs/1312.5003} {arXiv:1312.5003 [nucl-th]} \BibitemShut
  {NoStop}%
\bibitem [{\citenamefont {Cao}\ \emph {et~al.}(2021{\natexlab{a}})\citenamefont
  {Cao} \emph {et~al.}}]{JETSCAPE:2021ehl}%
  \BibitemOpen
  \bibfield  {author} {\bibinfo {author} {\bibfnamefont {S.}~\bibnamefont
  {Cao}} \emph {et~al.} (\bibinfo {collaboration} {JETSCAPE}),\ }\bibfield
  {title} {\enquote {\bibinfo {title} {{Determining the jet transport
  coefficient $\hat q$ from inclusive hadron suppression measurements using
  Bayesian parameter estimation}},}\ }\href {\doibase
  10.1103/PhysRevC.104.024905} {\bibfield  {journal} {\bibinfo  {journal}
  {Phys. Rev. C}\ }\textbf {\bibinfo {volume} {104}},\ \bibinfo {pages}
  {024905} (\bibinfo {year} {2021}{\natexlab{a}})},\ \Eprint
  {http://arxiv.org/abs/2102.11337} {arXiv:2102.11337 [nucl-th]} \BibitemShut
  {NoStop}%
\bibitem [{\citenamefont {Romatschke}\ and\ \citenamefont
  {Strickland}(2005)}]{Romatschke:2004au}%
  \BibitemOpen
  \bibfield  {author} {\bibinfo {author} {\bibfnamefont {Paul}\ \bibnamefont
  {Romatschke}}\ and\ \bibinfo {author} {\bibfnamefont {Michael}\ \bibnamefont
  {Strickland}},\ }\bibfield  {title} {\enquote {\bibinfo {title} {{Collisional
  energy loss of a heavy quark in an anisotropic quark-gluon plasma}},}\ }\href
  {\doibase 10.1103/PhysRevD.71.125008} {\bibfield  {journal} {\bibinfo
  {journal} {Phys. Rev. D}\ }\textbf {\bibinfo {volume} {71}},\ \bibinfo
  {pages} {125008} (\bibinfo {year} {2005})},\ \Eprint
  {http://arxiv.org/abs/hep-ph/0408275} {arXiv:hep-ph/0408275} \BibitemShut
  {NoStop}%
\bibitem [{\citenamefont {Romatschke}(2007)}]{Romatschke:2006bb}%
  \BibitemOpen
  \bibfield  {author} {\bibinfo {author} {\bibfnamefont {Paul}\ \bibnamefont
  {Romatschke}},\ }\bibfield  {title} {\enquote {\bibinfo {title} {{Momentum
  broadening in an anisotropic plasma}},}\ }\href {\doibase
  10.1103/PhysRevC.75.014901} {\bibfield  {journal} {\bibinfo  {journal} {Phys.
  Rev. C}\ }\textbf {\bibinfo {volume} {75}},\ \bibinfo {pages} {014901}
  (\bibinfo {year} {2007})},\ \Eprint {http://arxiv.org/abs/hep-ph/0607327}
  {arXiv:hep-ph/0607327} \BibitemShut {NoStop}%
\bibitem [{\citenamefont {Dumitru}\ \emph {et~al.}(2008)\citenamefont
  {Dumitru}, \citenamefont {Nara}, \citenamefont {Schenke},\ and\ \citenamefont
  {Strickland}}]{Dumitru:2007rp}%
  \BibitemOpen
  \bibfield  {author} {\bibinfo {author} {\bibfnamefont {Adrian}\ \bibnamefont
  {Dumitru}}, \bibinfo {author} {\bibfnamefont {Yasushi}\ \bibnamefont {Nara}},
  \bibinfo {author} {\bibfnamefont {Bjoern}\ \bibnamefont {Schenke}}, \ and\
  \bibinfo {author} {\bibfnamefont {Michael}\ \bibnamefont {Strickland}},\
  }\bibfield  {title} {\enquote {\bibinfo {title} {{Jet broadening in unstable
  non-Abelian plasmas}},}\ }\href {\doibase 10.1103/PhysRevC.78.024909}
  {\bibfield  {journal} {\bibinfo  {journal} {Phys. Rev. C}\ }\textbf {\bibinfo
  {volume} {78}},\ \bibinfo {pages} {024909} (\bibinfo {year} {2008})},\
  \Eprint {http://arxiv.org/abs/0710.1223} {arXiv:0710.1223 [hep-ph]}
  \BibitemShut {NoStop}%
\bibitem [{\citenamefont {Hauksson}\ \emph {et~al.}(2022)\citenamefont
  {Hauksson}, \citenamefont {Jeon},\ and\ \citenamefont
  {Gale}}]{Hauksson:2021okc}%
  \BibitemOpen
  \bibfield  {author} {\bibinfo {author} {\bibfnamefont {Sigtryggur}\
  \bibnamefont {Hauksson}}, \bibinfo {author} {\bibfnamefont {Sangyong}\
  \bibnamefont {Jeon}}, \ and\ \bibinfo {author} {\bibfnamefont {Charles}\
  \bibnamefont {Gale}},\ }\bibfield  {title} {\enquote {\bibinfo {title}
  {{Momentum broadening of energetic partons in an anisotropic plasma}},}\
  }\href {\doibase 10.1103/PhysRevC.105.014914} {\bibfield  {journal} {\bibinfo
   {journal} {Phys. Rev. C}\ }\textbf {\bibinfo {volume} {105}},\ \bibinfo
  {pages} {014914} (\bibinfo {year} {2022})},\ \Eprint
  {http://arxiv.org/abs/2109.04575} {arXiv:2109.04575 [hep-ph]} \BibitemShut
  {NoStop}%
\bibitem [{\citenamefont {Andres}\ \emph {et~al.}(2022)\citenamefont {Andres},
  \citenamefont {Dominguez}, \citenamefont {Sadofyev},\ and\ \citenamefont
  {Salgado}}]{Andres:2022ndd}%
  \BibitemOpen
  \bibfield  {author} {\bibinfo {author} {\bibfnamefont {Carlota}\ \bibnamefont
  {Andres}}, \bibinfo {author} {\bibfnamefont {Fabio}\ \bibnamefont
  {Dominguez}}, \bibinfo {author} {\bibfnamefont {Andrey~V.}\ \bibnamefont
  {Sadofyev}}, \ and\ \bibinfo {author} {\bibfnamefont {Carlos~A.}\
  \bibnamefont {Salgado}},\ }\bibfield  {title} {\enquote {\bibinfo {title}
  {{Jet broadening in flowing matter: Resummation}},}\ }\href {\doibase
  10.1103/PhysRevD.106.074023} {\bibfield  {journal} {\bibinfo  {journal}
  {Phys. Rev. D}\ }\textbf {\bibinfo {volume} {106}},\ \bibinfo {pages}
  {074023} (\bibinfo {year} {2022})},\ \Eprint
  {http://arxiv.org/abs/2207.07141} {arXiv:2207.07141 [hep-ph]} \BibitemShut
  {NoStop}%
\bibitem [{\citenamefont {Barata}\ \emph {et~al.}(2022)\citenamefont {Barata},
  \citenamefont {Sadofyev},\ and\ \citenamefont {Salgado}}]{Barata:2022krd}%
  \BibitemOpen
  \bibfield  {author} {\bibinfo {author} {\bibfnamefont {Jo\~ao}\ \bibnamefont
  {Barata}}, \bibinfo {author} {\bibfnamefont {Andrey~V.}\ \bibnamefont
  {Sadofyev}}, \ and\ \bibinfo {author} {\bibfnamefont {Carlos~A.}\
  \bibnamefont {Salgado}},\ }\bibfield  {title} {\enquote {\bibinfo {title}
  {{Jet broadening in dense inhomogeneous matter}},}\ }\href {\doibase
  10.1103/PhysRevD.105.114010} {\bibfield  {journal} {\bibinfo  {journal}
  {Phys. Rev. D}\ }\textbf {\bibinfo {volume} {105}},\ \bibinfo {pages}
  {114010} (\bibinfo {year} {2022})},\ \Eprint
  {http://arxiv.org/abs/2202.08847} {arXiv:2202.08847 [hep-ph]} \BibitemShut
  {NoStop}%
\bibitem [{\citenamefont {Barata}\ \emph
  {et~al.}(2023{\natexlab{a}})\citenamefont {Barata}, \citenamefont
  {Sadofyev},\ and\ \citenamefont {Wang}}]{Barata:2022utc}%
  \BibitemOpen
  \bibfield  {author} {\bibinfo {author} {\bibfnamefont {Jo\~ao}\ \bibnamefont
  {Barata}}, \bibinfo {author} {\bibfnamefont {Andrey~V.}\ \bibnamefont
  {Sadofyev}}, \ and\ \bibinfo {author} {\bibfnamefont {Xin-Nian}\ \bibnamefont
  {Wang}},\ }\bibfield  {title} {\enquote {\bibinfo {title} {{Quantum partonic
  transport in QCD matter}},}\ }\href {\doibase 10.1103/PhysRevD.107.L051503}
  {\bibfield  {journal} {\bibinfo  {journal} {Phys. Rev. D}\ }\textbf {\bibinfo
  {volume} {107}},\ \bibinfo {pages} {L051503} (\bibinfo {year}
  {2023}{\natexlab{a}})},\ \Eprint {http://arxiv.org/abs/2210.06519}
  {arXiv:2210.06519 [hep-ph]} \BibitemShut {NoStop}%
\bibitem [{\citenamefont {Barata}\ \emph
  {et~al.}(2023{\natexlab{b}})\citenamefont {Barata}, \citenamefont
  {Mayo~L\'opez}, \citenamefont {Sadofyev},\ and\ \citenamefont
  {Salgado}}]{Barata:2023qds}%
  \BibitemOpen
  \bibfield  {author} {\bibinfo {author} {\bibfnamefont {Jo\~ao}\ \bibnamefont
  {Barata}}, \bibinfo {author} {\bibfnamefont {Xo\'an}\ \bibnamefont
  {Mayo~L\'opez}}, \bibinfo {author} {\bibfnamefont {Andrey~V.}\ \bibnamefont
  {Sadofyev}}, \ and\ \bibinfo {author} {\bibfnamefont {Carlos~A.}\
  \bibnamefont {Salgado}},\ }\bibfield  {title} {\enquote {\bibinfo {title}
  {{Medium induced gluon spectrum in dense inhomogeneous matter}},}\ }\href
  {\doibase 10.1103/PhysRevD.108.034018} {\bibfield  {journal} {\bibinfo
  {journal} {Phys. Rev. D}\ }\textbf {\bibinfo {volume} {108}},\ \bibinfo
  {pages} {034018} (\bibinfo {year} {2023}{\natexlab{b}})},\ \Eprint
  {http://arxiv.org/abs/2304.03712} {arXiv:2304.03712 [hep-ph]} \BibitemShut
  {NoStop}%
\bibitem [{\citenamefont {Kuzmin}\ \emph {et~al.}(2024)\citenamefont {Kuzmin},
  \citenamefont {Mayo~L\'opez}, \citenamefont {Reiten},\ and\ \citenamefont
  {Sadofyev}}]{Kuzmin:2023hko}%
  \BibitemOpen
  \bibfield  {author} {\bibinfo {author} {\bibfnamefont {Matvey~V.}\
  \bibnamefont {Kuzmin}}, \bibinfo {author} {\bibfnamefont {Xo\'an}\
  \bibnamefont {Mayo~L\'opez}}, \bibinfo {author} {\bibfnamefont {Jared}\
  \bibnamefont {Reiten}}, \ and\ \bibinfo {author} {\bibfnamefont {Andrey~V.}\
  \bibnamefont {Sadofyev}},\ }\bibfield  {title} {\enquote {\bibinfo {title}
  {{Jet quenching in anisotropic flowing matter}},}\ }\href {\doibase
  10.1103/PhysRevD.109.014036} {\bibfield  {journal} {\bibinfo  {journal}
  {Phys. Rev. D}\ }\textbf {\bibinfo {volume} {109}},\ \bibinfo {pages}
  {014036} (\bibinfo {year} {2024})},\ \Eprint
  {http://arxiv.org/abs/2309.00683} {arXiv:2309.00683 [hep-ph]} \BibitemShut
  {NoStop}%
\bibitem [{\citenamefont {Das}\ \emph {et~al.}(2015)\citenamefont {Das},
  \citenamefont {Ruggieri}, \citenamefont {Mazumder}, \citenamefont {Greco},\
  and\ \citenamefont {Alam}}]{Das:2015aga}%
  \BibitemOpen
  \bibfield  {author} {\bibinfo {author} {\bibfnamefont {Santosh~K.}\
  \bibnamefont {Das}}, \bibinfo {author} {\bibfnamefont {Marco}\ \bibnamefont
  {Ruggieri}}, \bibinfo {author} {\bibfnamefont {Surasree}\ \bibnamefont
  {Mazumder}}, \bibinfo {author} {\bibfnamefont {Vincenzo}\ \bibnamefont
  {Greco}}, \ and\ \bibinfo {author} {\bibfnamefont {Jan-e}\ \bibnamefont
  {Alam}},\ }\bibfield  {title} {\enquote {\bibinfo {title} {{Heavy quark
  diffusion in the pre-equilibrium stage of heavy ion collisions}},}\ }\href
  {\doibase 10.1088/0954-3899/42/9/095108} {\bibfield  {journal} {\bibinfo
  {journal} {J. Phys. G}\ }\textbf {\bibinfo {volume} {42}},\ \bibinfo {pages}
  {095108} (\bibinfo {year} {2015})},\ \Eprint
  {http://arxiv.org/abs/1501.07521} {arXiv:1501.07521 [nucl-th]} \BibitemShut
  {NoStop}%
\bibitem [{\citenamefont {Mrowczynski}(2018)}]{Mrowczynski:2017kso}%
  \BibitemOpen
  \bibfield  {author} {\bibinfo {author} {\bibfnamefont {Stanislaw}\
  \bibnamefont {Mrowczynski}},\ }\bibfield  {title} {\enquote {\bibinfo {title}
  {{Heavy Quarks in Turbulent QCD Plasmas}},}\ }\href {\doibase
  10.1140/epja/i2018-12478-5} {\bibfield  {journal} {\bibinfo  {journal} {Eur.
  Phys. J. A}\ }\textbf {\bibinfo {volume} {54}},\ \bibinfo {pages} {43}
  (\bibinfo {year} {2018})},\ \Eprint {http://arxiv.org/abs/1706.03127}
  {arXiv:1706.03127 [nucl-th]} \BibitemShut {NoStop}%
\bibitem [{\citenamefont {Sun}\ \emph {et~al.}(2019)\citenamefont {Sun},
  \citenamefont {Coci}, \citenamefont {Das}, \citenamefont {Plumari},
  \citenamefont {Ruggieri},\ and\ \citenamefont {Greco}}]{Sun:2019fud}%
  \BibitemOpen
  \bibfield  {author} {\bibinfo {author} {\bibfnamefont {Yifeng}\ \bibnamefont
  {Sun}}, \bibinfo {author} {\bibfnamefont {Gabriele}\ \bibnamefont {Coci}},
  \bibinfo {author} {\bibfnamefont {Santosh~Kumar}\ \bibnamefont {Das}},
  \bibinfo {author} {\bibfnamefont {Salvatore}\ \bibnamefont {Plumari}},
  \bibinfo {author} {\bibfnamefont {Marco}\ \bibnamefont {Ruggieri}}, \ and\
  \bibinfo {author} {\bibfnamefont {Vincenzo}\ \bibnamefont {Greco}},\
  }\bibfield  {title} {\enquote {\bibinfo {title} {{Impact of Glasma on heavy
  quark observables in nucleus-nucleus collisions at LHC}},}\ }\href {\doibase
  10.1016/j.physletb.2019.134933} {\bibfield  {journal} {\bibinfo  {journal}
  {Phys. Lett. B}\ }\textbf {\bibinfo {volume} {798}},\ \bibinfo {pages}
  {134933} (\bibinfo {year} {2019})},\ \Eprint
  {http://arxiv.org/abs/1902.06254} {arXiv:1902.06254 [nucl-th]} \BibitemShut
  {NoStop}%
\bibitem [{\citenamefont {Boguslavski}\ \emph {et~al.}(2020)\citenamefont
  {Boguslavski}, \citenamefont {Kurkela}, \citenamefont {Lappi},\ and\
  \citenamefont {Peuron}}]{Boguslavski:2020tqz}%
  \BibitemOpen
  \bibfield  {author} {\bibinfo {author} {\bibfnamefont {K.}~\bibnamefont
  {Boguslavski}}, \bibinfo {author} {\bibfnamefont {A.}~\bibnamefont
  {Kurkela}}, \bibinfo {author} {\bibfnamefont {T.}~\bibnamefont {Lappi}}, \
  and\ \bibinfo {author} {\bibfnamefont {J.}~\bibnamefont {Peuron}},\
  }\bibfield  {title} {\enquote {\bibinfo {title} {{Heavy quark diffusion in an
  overoccupied gluon plasma}},}\ }\href {\doibase 10.1007/JHEP09(2020)077}
  {\bibfield  {journal} {\bibinfo  {journal} {JHEP}\ }\textbf {\bibinfo
  {volume} {09}},\ \bibinfo {pages} {077} (\bibinfo {year} {2020})},\ \Eprint
  {http://arxiv.org/abs/2005.02418} {arXiv:2005.02418 [hep-ph]} \BibitemShut
  {NoStop}%
\bibitem [{\citenamefont {Ruggieri}\ \emph {et~al.}(2022)\citenamefont
  {Ruggieri}, \citenamefont {Pooja}, \citenamefont {Prakash},\ and\
  \citenamefont {Das}}]{Ruggieri:2022kxv}%
  \BibitemOpen
  \bibfield  {author} {\bibinfo {author} {\bibfnamefont {Marco}\ \bibnamefont
  {Ruggieri}}, \bibinfo {author} {\bibnamefont {Pooja}}, \bibinfo {author}
  {\bibfnamefont {Jai}\ \bibnamefont {Prakash}}, \ and\ \bibinfo {author}
  {\bibfnamefont {Santosh~K.}\ \bibnamefont {Das}},\ }\bibfield  {title}
  {\enquote {\bibinfo {title} {{Memory effects on energy loss and diffusion of
  heavy quarks in the quark-gluon plasma}},}\ }\href {\doibase
  10.1103/PhysRevD.106.034032} {\bibfield  {journal} {\bibinfo  {journal}
  {Phys. Rev. D}\ }\textbf {\bibinfo {volume} {106}},\ \bibinfo {pages}
  {034032} (\bibinfo {year} {2022})},\ \Eprint
  {http://arxiv.org/abs/2203.06712} {arXiv:2203.06712 [hep-ph]} \BibitemShut
  {NoStop}%
\bibitem [{\citenamefont {Boguslavski}\ \emph
  {et~al.}(2024{\natexlab{b}})\citenamefont {Boguslavski}, \citenamefont
  {Kurkela}, \citenamefont {Lappi}, \citenamefont {Lindenbauer},\ and\
  \citenamefont {Peuron}}]{Boguslavski:2023fdm}%
  \BibitemOpen
  \bibfield  {author} {\bibinfo {author} {\bibfnamefont {K.}~\bibnamefont
  {Boguslavski}}, \bibinfo {author} {\bibfnamefont {A.}~\bibnamefont
  {Kurkela}}, \bibinfo {author} {\bibfnamefont {T.}~\bibnamefont {Lappi}},
  \bibinfo {author} {\bibfnamefont {F.}~\bibnamefont {Lindenbauer}}, \ and\
  \bibinfo {author} {\bibfnamefont {J.}~\bibnamefont {Peuron}},\ }\bibfield
  {title} {\enquote {\bibinfo {title} {{Heavy quark diffusion coefficient in
  heavy-ion collisions via kinetic theory}},}\ }\href {\doibase
  10.1103/PhysRevD.109.014025} {\bibfield  {journal} {\bibinfo  {journal}
  {Phys. Rev. D}\ }\textbf {\bibinfo {volume} {109}},\ \bibinfo {pages}
  {014025} (\bibinfo {year} {2024}{\natexlab{b}})},\ \Eprint
  {http://arxiv.org/abs/2303.12520} {arXiv:2303.12520 [hep-ph]} \BibitemShut
  {NoStop}%
\bibitem [{\citenamefont {Kurkela}\ and\ \citenamefont
  {Moore}(2011{\natexlab{a}})}]{Kurkela:2011ti}%
  \BibitemOpen
  \bibfield  {author} {\bibinfo {author} {\bibfnamefont {Aleksi}\ \bibnamefont
  {Kurkela}}\ and\ \bibinfo {author} {\bibfnamefont {Guy~D.}\ \bibnamefont
  {Moore}},\ }\bibfield  {title} {\enquote {\bibinfo {title} {{Thermalization
  in Weakly Coupled Nonabelian Plasmas}},}\ }\href {\doibase
  10.1007/JHEP12(2011)044} {\bibfield  {journal} {\bibinfo  {journal} {JHEP}\
  }\textbf {\bibinfo {volume} {12}},\ \bibinfo {pages} {044} (\bibinfo {year}
  {2011}{\natexlab{a}})},\ \Eprint {http://arxiv.org/abs/1107.5050}
  {arXiv:1107.5050 [hep-ph]} \BibitemShut {NoStop}%
\bibitem [{\citenamefont {Kurkela}\ and\ \citenamefont
  {Moore}(2011{\natexlab{b}})}]{Kurkela:2011ub}%
  \BibitemOpen
  \bibfield  {author} {\bibinfo {author} {\bibfnamefont {Aleksi}\ \bibnamefont
  {Kurkela}}\ and\ \bibinfo {author} {\bibfnamefont {Guy~D.}\ \bibnamefont
  {Moore}},\ }\bibfield  {title} {\enquote {\bibinfo {title} {{Bjorken Flow,
  Plasma Instabilities, and Thermalization}},}\ }\href {\doibase
  10.1007/JHEP11(2011)120} {\bibfield  {journal} {\bibinfo  {journal} {JHEP}\
  }\textbf {\bibinfo {volume} {11}},\ \bibinfo {pages} {120} (\bibinfo {year}
  {2011}{\natexlab{b}})},\ \Eprint {http://arxiv.org/abs/1108.4684}
  {arXiv:1108.4684 [hep-ph]} \BibitemShut {NoStop}%
\bibitem [{\citenamefont {Berges}\ \emph
  {et~al.}(2014{\natexlab{a}})\citenamefont {Berges}, \citenamefont
  {Boguslavski}, \citenamefont {Schlichting},\ and\ \citenamefont
  {Venugopalan}}]{Berges:2013eia}%
  \BibitemOpen
  \bibfield  {author} {\bibinfo {author} {\bibfnamefont {J.}~\bibnamefont
  {Berges}}, \bibinfo {author} {\bibfnamefont {K.}~\bibnamefont {Boguslavski}},
  \bibinfo {author} {\bibfnamefont {S.}~\bibnamefont {Schlichting}}, \ and\
  \bibinfo {author} {\bibfnamefont {R.}~\bibnamefont {Venugopalan}},\
  }\bibfield  {title} {\enquote {\bibinfo {title} {{Turbulent thermalization
  process in heavy-ion collisions at ultrarelativistic energies}},}\ }\href
  {\doibase 10.1103/PhysRevD.89.074011} {\bibfield  {journal} {\bibinfo
  {journal} {Phys. Rev. D}\ }\textbf {\bibinfo {volume} {89}},\ \bibinfo
  {pages} {074011} (\bibinfo {year} {2014}{\natexlab{a}})},\ \Eprint
  {http://arxiv.org/abs/1303.5650} {arXiv:1303.5650 [hep-ph]} \BibitemShut
  {NoStop}%
\bibitem [{\citenamefont {Berges}\ \emph
  {et~al.}(2014{\natexlab{b}})\citenamefont {Berges}, \citenamefont
  {Boguslavski}, \citenamefont {Schlichting},\ and\ \citenamefont
  {Venugopalan}}]{Berges:2013fga}%
  \BibitemOpen
  \bibfield  {author} {\bibinfo {author} {\bibfnamefont {Juergen}\ \bibnamefont
  {Berges}}, \bibinfo {author} {\bibfnamefont {Kirill}\ \bibnamefont
  {Boguslavski}}, \bibinfo {author} {\bibfnamefont {Soeren}\ \bibnamefont
  {Schlichting}}, \ and\ \bibinfo {author} {\bibfnamefont {Raju}\ \bibnamefont
  {Venugopalan}},\ }\bibfield  {title} {\enquote {\bibinfo {title} {{Universal
  attractor in a highly occupied non-Abelian plasma}},}\ }\href {\doibase
  10.1103/PhysRevD.89.114007} {\bibfield  {journal} {\bibinfo  {journal} {Phys.
  Rev. D}\ }\textbf {\bibinfo {volume} {89}},\ \bibinfo {pages} {114007}
  (\bibinfo {year} {2014}{\natexlab{b}})},\ \Eprint
  {http://arxiv.org/abs/1311.3005} {arXiv:1311.3005 [hep-ph]} \BibitemShut
  {NoStop}%
\bibitem [{\citenamefont {Cao}\ \emph {et~al.}(2021{\natexlab{b}})\citenamefont
  {Cao}, \citenamefont {Sirimanna},\ and\ \citenamefont
  {Majumder}}]{Cao:2021rpv}%
  \BibitemOpen
  \bibfield  {author} {\bibinfo {author} {\bibfnamefont {Shanshan}\
  \bibnamefont {Cao}}, \bibinfo {author} {\bibfnamefont {Chathuranga}\
  \bibnamefont {Sirimanna}}, \ and\ \bibinfo {author} {\bibfnamefont {Abhijit}\
  \bibnamefont {Majumder}},\ }\bibfield  {title} {\enquote {\bibinfo {title}
  {{The medium modification of high-virtuality partons}},}\ }\href@noop {} {\
  (\bibinfo {year} {2021}{\natexlab{b}})},\ \Eprint
  {http://arxiv.org/abs/2101.03681} {arXiv:2101.03681 [hep-ph]} \BibitemShut
  {NoStop}%
\bibitem [{\citenamefont {Tachibana}\ \emph {et~al.}(2023)\citenamefont
  {Tachibana} \emph {et~al.}}]{JETSCAPE:2023hqn}%
  \BibitemOpen
  \bibfield  {author} {\bibinfo {author} {\bibfnamefont {Y.}~\bibnamefont
  {Tachibana}} \emph {et~al.} (\bibinfo {collaboration} {JETSCAPE}),\
  }\bibfield  {title} {\enquote {\bibinfo {title} {{Hard Jet Substructure in a
  Multi-stage Approach}},}\ }\href@noop {} {\  (\bibinfo {year} {2023})},\
  \Eprint {http://arxiv.org/abs/2301.02485} {arXiv:2301.02485 [hep-ph]}
  \BibitemShut {NoStop}%
\bibitem [{\citenamefont {Mueller}(2000)}]{Mueller:1999pi}%
  \BibitemOpen
  \bibfield  {author} {\bibinfo {author} {\bibfnamefont {Alfred~H.}\
  \bibnamefont {Mueller}},\ }\bibfield  {title} {\enquote {\bibinfo {title}
  {{The Boltzmann equation for gluons at early times after a heavy ion
  collision}},}\ }\href {\doibase 10.1016/S0370-2693(00)00084-8} {\bibfield
  {journal} {\bibinfo  {journal} {Phys. Lett. B}\ }\textbf {\bibinfo {volume}
  {475}},\ \bibinfo {pages} {220--224} (\bibinfo {year} {2000})},\ \Eprint
  {http://arxiv.org/abs/hep-ph/9909388} {arXiv:hep-ph/9909388} \BibitemShut
  {NoStop}%
\bibitem [{\citenamefont {Keegan}\ \emph {et~al.}(2016)\citenamefont {Keegan},
  \citenamefont {Kurkela}, \citenamefont {Mazeliauskas},\ and\ \citenamefont
  {Teaney}}]{Keegan:2016cpi}%
  \BibitemOpen
  \bibfield  {author} {\bibinfo {author} {\bibfnamefont {Liam}\ \bibnamefont
  {Keegan}}, \bibinfo {author} {\bibfnamefont {Aleksi}\ \bibnamefont
  {Kurkela}}, \bibinfo {author} {\bibfnamefont {Aleksas}\ \bibnamefont
  {Mazeliauskas}}, \ and\ \bibinfo {author} {\bibfnamefont {Derek}\
  \bibnamefont {Teaney}},\ }\bibfield  {title} {\enquote {\bibinfo {title}
  {{Initial conditions for hydrodynamics from weakly coupled pre-equilibrium
  evolution}},}\ }\href {\doibase 10.1007/JHEP08(2016)171} {\bibfield
  {journal} {\bibinfo  {journal} {JHEP}\ }\textbf {\bibinfo {volume} {08}},\
  \bibinfo {pages} {171} (\bibinfo {year} {2016})},\ \Eprint
  {http://arxiv.org/abs/1605.04287} {arXiv:1605.04287 [hep-ph]} \BibitemShut
  {NoStop}%
\bibitem [{\citenamefont {Kurkela}\ \emph {et~al.}(2019)\citenamefont
  {Kurkela}, \citenamefont {Mazeliauskas}, \citenamefont {Paquet},
  \citenamefont {Schlichting},\ and\ \citenamefont {Teaney}}]{Kurkela:2018vqr}%
  \BibitemOpen
  \bibfield  {author} {\bibinfo {author} {\bibfnamefont {Aleksi}\ \bibnamefont
  {Kurkela}}, \bibinfo {author} {\bibfnamefont {Aleksas}\ \bibnamefont
  {Mazeliauskas}}, \bibinfo {author} {\bibfnamefont {Jean-Fran\c{c}ois}\
  \bibnamefont {Paquet}}, \bibinfo {author} {\bibfnamefont {S\"oren}\
  \bibnamefont {Schlichting}}, \ and\ \bibinfo {author} {\bibfnamefont {Derek}\
  \bibnamefont {Teaney}},\ }\bibfield  {title} {\enquote {\bibinfo {title}
  {{Effective kinetic description of event-by-event pre-equilibrium dynamics in
  high-energy heavy-ion collisions}},}\ }\href {\doibase
  10.1103/PhysRevC.99.034910} {\bibfield  {journal} {\bibinfo  {journal} {Phys.
  Rev. C}\ }\textbf {\bibinfo {volume} {99}},\ \bibinfo {pages} {034910}
  (\bibinfo {year} {2019})},\ \Eprint {http://arxiv.org/abs/1805.00961}
  {arXiv:1805.00961 [hep-ph]} \BibitemShut {NoStop}%
\bibitem [{\citenamefont {Kurkela}\ \emph {et~al.}(2021)\citenamefont
  {Kurkela}, \citenamefont {Mazeliauskas},\ and\ \citenamefont
  {T\"ornkvist}}]{Kurkela:2021ctp}%
  \BibitemOpen
  \bibfield  {author} {\bibinfo {author} {\bibfnamefont {Aleksi}\ \bibnamefont
  {Kurkela}}, \bibinfo {author} {\bibfnamefont {Aleksas}\ \bibnamefont
  {Mazeliauskas}}, \ and\ \bibinfo {author} {\bibfnamefont {Robin}\
  \bibnamefont {T\"ornkvist}},\ }\bibfield  {title} {\enquote {\bibinfo {title}
  {{Collective flow in single-hit QCD kinetic theory}},}\ }\href {\doibase
  10.1007/JHEP11(2021)216} {\bibfield  {journal} {\bibinfo  {journal} {JHEP}\
  }\textbf {\bibinfo {volume} {11}},\ \bibinfo {pages} {216} (\bibinfo {year}
  {2021})},\ \Eprint {http://arxiv.org/abs/2104.08179} {arXiv:2104.08179
  [hep-ph]} \BibitemShut {NoStop}%
\bibitem [{\citenamefont {Abraao~York}\ \emph {et~al.}(2014)\citenamefont
  {Abraao~York}, \citenamefont {Kurkela}, \citenamefont {Lu},\ and\
  \citenamefont {Moore}}]{AbraaoYork:2014hbk}%
  \BibitemOpen
  \bibfield  {author} {\bibinfo {author} {\bibfnamefont {Mark~C.}\ \bibnamefont
  {Abraao~York}}, \bibinfo {author} {\bibfnamefont {Aleksi}\ \bibnamefont
  {Kurkela}}, \bibinfo {author} {\bibfnamefont {Egang}\ \bibnamefont {Lu}}, \
  and\ \bibinfo {author} {\bibfnamefont {Guy~D.}\ \bibnamefont {Moore}},\
  }\bibfield  {title} {\enquote {\bibinfo {title} {{UV cascade in classical
  Yang-Mills theory via kinetic theory}},}\ }\href {\doibase
  10.1103/PhysRevD.89.074036} {\bibfield  {journal} {\bibinfo  {journal} {Phys.
  Rev. D}\ }\textbf {\bibinfo {volume} {89}},\ \bibinfo {pages} {074036}
  (\bibinfo {year} {2014})},\ \Eprint {http://arxiv.org/abs/1401.3751}
  {arXiv:1401.3751 [hep-ph]} \BibitemShut {NoStop}%
\bibitem [{\citenamefont {Kurkela}\ and\ \citenamefont
  {Mazeliauskas}(2019{\natexlab{a}})}]{Kurkela:2018oqw}%
  \BibitemOpen
  \bibfield  {author} {\bibinfo {author} {\bibfnamefont {Aleksi}\ \bibnamefont
  {Kurkela}}\ and\ \bibinfo {author} {\bibfnamefont {Aleksas}\ \bibnamefont
  {Mazeliauskas}},\ }\bibfield  {title} {\enquote {\bibinfo {title} {{Chemical
  equilibration in weakly coupled QCD}},}\ }\href {\doibase
  10.1103/PhysRevD.99.054018} {\bibfield  {journal} {\bibinfo  {journal} {Phys.
  Rev. D}\ }\textbf {\bibinfo {volume} {99}},\ \bibinfo {pages} {054018}
  (\bibinfo {year} {2019}{\natexlab{a}})},\ \Eprint
  {http://arxiv.org/abs/1811.03068} {arXiv:1811.03068 [hep-ph]} \BibitemShut
  {NoStop}%
\bibitem [{\citenamefont {Du}\ and\ \citenamefont
  {Schlichting}(2021{\natexlab{a}})}]{Du:2020dvp}%
  \BibitemOpen
  \bibfield  {author} {\bibinfo {author} {\bibfnamefont {Xiaojian}\
  \bibnamefont {Du}}\ and\ \bibinfo {author} {\bibfnamefont {S\"oren}\
  \bibnamefont {Schlichting}},\ }\bibfield  {title} {\enquote {\bibinfo {title}
  {{Equilibration of weakly coupled QCD plasmas}},}\ }\href {\doibase
  10.1103/PhysRevD.104.054011} {\bibfield  {journal} {\bibinfo  {journal}
  {Phys. Rev. D}\ }\textbf {\bibinfo {volume} {104}},\ \bibinfo {pages}
  {054011} (\bibinfo {year} {2021}{\natexlab{a}})},\ \Eprint
  {http://arxiv.org/abs/2012.09079} {arXiv:2012.09079 [hep-ph]} \BibitemShut
  {NoStop}%
\bibitem [{\citenamefont {Du}\ and\ \citenamefont
  {Schlichting}(2021{\natexlab{b}})}]{Du:2020zqg}%
  \BibitemOpen
  \bibfield  {author} {\bibinfo {author} {\bibfnamefont {Xiaojian}\
  \bibnamefont {Du}}\ and\ \bibinfo {author} {\bibfnamefont {S\"oren}\
  \bibnamefont {Schlichting}},\ }\bibfield  {title} {\enquote {\bibinfo {title}
  {{Equilibration of the Quark-Gluon Plasma at Finite Net-Baryon Density in QCD
  Kinetic Theory}},}\ }\href {\doibase 10.1103/PhysRevLett.127.122301}
  {\bibfield  {journal} {\bibinfo  {journal} {Phys. Rev. Lett.}\ }\textbf
  {\bibinfo {volume} {127}},\ \bibinfo {pages} {122301} (\bibinfo {year}
  {2021}{\natexlab{b}})},\ \Eprint {http://arxiv.org/abs/2012.09068}
  {arXiv:2012.09068 [hep-ph]} \BibitemShut {NoStop}%
\bibitem [{\citenamefont {Laine}\ and\ \citenamefont
  {Vuorinen}(2016)}]{Laine:2016hma}%
  \BibitemOpen
  \bibfield  {author} {\bibinfo {author} {\bibfnamefont {Mikko}\ \bibnamefont
  {Laine}}\ and\ \bibinfo {author} {\bibfnamefont {Aleksi}\ \bibnamefont
  {Vuorinen}},\ }\href {\doibase 10.1007/978-3-319-31933-9} {\emph {\bibinfo
  {title} {{Basics of Thermal Field Theory}}}},\ Vol.\ \bibinfo {volume} {925}\
  (\bibinfo  {publisher} {Springer},\ \bibinfo {year} {2016})\ \Eprint
  {http://arxiv.org/abs/1701.01554} {arXiv:1701.01554 [hep-ph]} \BibitemShut
  {NoStop}%
\bibitem [{\citenamefont {Arnold}\ and\ \citenamefont
  {Yaffe}(1998)}]{Arnold:1997gh}%
  \BibitemOpen
  \bibfield  {author} {\bibinfo {author} {\bibfnamefont {Peter~Brockway}\
  \bibnamefont {Arnold}}\ and\ \bibinfo {author} {\bibfnamefont {Laurence~G.}\
  \bibnamefont {Yaffe}},\ }\bibfield  {title} {\enquote {\bibinfo {title}
  {{Effective theories for real time correlations in hot plasmas}},}\ }\href
  {\doibase 10.1103/PhysRevD.57.1178} {\bibfield  {journal} {\bibinfo
  {journal} {Phys. Rev. D}\ }\textbf {\bibinfo {volume} {57}},\ \bibinfo
  {pages} {1178--1192} (\bibinfo {year} {1998})},\ \Eprint
  {http://arxiv.org/abs/hep-ph/9709449} {arXiv:hep-ph/9709449} \BibitemShut
  {NoStop}%
\bibitem [{\citenamefont {He}\ \emph {et~al.}(2015)\citenamefont {He},
  \citenamefont {Luo}, \citenamefont {Wang},\ and\ \citenamefont
  {Zhu}}]{He:2015pra}%
  \BibitemOpen
  \bibfield  {author} {\bibinfo {author} {\bibfnamefont {Yayun}\ \bibnamefont
  {He}}, \bibinfo {author} {\bibfnamefont {Tan}\ \bibnamefont {Luo}}, \bibinfo
  {author} {\bibfnamefont {Xin-Nian}\ \bibnamefont {Wang}}, \ and\ \bibinfo
  {author} {\bibfnamefont {Yan}\ \bibnamefont {Zhu}},\ }\bibfield  {title}
  {\enquote {\bibinfo {title} {{Linear Boltzmann Transport for Jet Propagation
  in the Quark-Gluon Plasma: Elastic Processes and Medium Recoil}},}\ }\href
  {\doibase 10.1103/PhysRevC.91.054908} {\bibfield  {journal} {\bibinfo
  {journal} {Phys. Rev. C}\ }\textbf {\bibinfo {volume} {91}},\ \bibinfo
  {pages} {054908} (\bibinfo {year} {2015})},\ \bibinfo {note} {[Erratum:
  Phys.Rev.C 97, 019902 (2018)]},\ \Eprint {http://arxiv.org/abs/1503.03313}
  {arXiv:1503.03313 [nucl-th]} \BibitemShut {NoStop}%
\bibitem [{\citenamefont {Kurkela}\ and\ \citenamefont
  {Mazeliauskas}(2019{\natexlab{b}})}]{Kurkela:2018xxd}%
  \BibitemOpen
  \bibfield  {author} {\bibinfo {author} {\bibfnamefont {Aleksi}\ \bibnamefont
  {Kurkela}}\ and\ \bibinfo {author} {\bibfnamefont {Aleksas}\ \bibnamefont
  {Mazeliauskas}},\ }\bibfield  {title} {\enquote {\bibinfo {title} {{Chemical
  Equilibration in Hadronic Collisions}},}\ }\href {\doibase
  10.1103/PhysRevLett.122.142301} {\bibfield  {journal} {\bibinfo  {journal}
  {Phys. Rev. Lett.}\ }\textbf {\bibinfo {volume} {122}},\ \bibinfo {pages}
  {142301} (\bibinfo {year} {2019}{\natexlab{b}})},\ \Eprint
  {http://arxiv.org/abs/1811.03040} {arXiv:1811.03040 [hep-ph]} \BibitemShut
  {NoStop}%
\bibitem [{\citenamefont {Arnold}\ \emph
  {et~al.}(2003{\natexlab{b}})\citenamefont {Arnold}, \citenamefont {Moore},\
  and\ \citenamefont {Yaffe}}]{Arnold:2003zc}%
  \BibitemOpen
  \bibfield  {author} {\bibinfo {author} {\bibfnamefont {Peter~Brockway}\
  \bibnamefont {Arnold}}, \bibinfo {author} {\bibfnamefont {Guy~D}\
  \bibnamefont {Moore}}, \ and\ \bibinfo {author} {\bibfnamefont {Laurence~G.}\
  \bibnamefont {Yaffe}},\ }\bibfield  {title} {\enquote {\bibinfo {title}
  {{Transport coefficients in high temperature gauge theories. 2. Beyond
  leading log}},}\ }\href {\doibase 10.1088/1126-6708/2003/05/051} {\bibfield
  {journal} {\bibinfo  {journal} {JHEP}\ }\textbf {\bibinfo {volume} {05}},\
  \bibinfo {pages} {051} (\bibinfo {year} {2003}{\natexlab{b}})},\ \Eprint
  {http://arxiv.org/abs/hep-ph/0302165} {arXiv:hep-ph/0302165} \BibitemShut
  {NoStop}%
\bibitem [{\citenamefont {Mrowczynski}(1988)}]{Mrowczynski:1988dz}%
  \BibitemOpen
  \bibfield  {author} {\bibinfo {author} {\bibfnamefont {Stanislaw}\
  \bibnamefont {Mrowczynski}},\ }\bibfield  {title} {\enquote {\bibinfo {title}
  {{Stream Instabilities of the Quark - Gluon Plasma}},}\ }\href {\doibase
  10.1016/0370-2693(88)90124-4} {\bibfield  {journal} {\bibinfo  {journal}
  {Phys. Lett. B}\ }\textbf {\bibinfo {volume} {214}},\ \bibinfo {pages} {587}
  (\bibinfo {year} {1988})},\ \bibinfo {note} {[Erratum: Phys.Lett.B 656, 273
  (2007)]}\BibitemShut {NoStop}%
\bibitem [{\citenamefont {Romatschke}\ and\ \citenamefont
  {Strickland}(2003)}]{Romatschke:2003ms}%
  \BibitemOpen
  \bibfield  {author} {\bibinfo {author} {\bibfnamefont {Paul}\ \bibnamefont
  {Romatschke}}\ and\ \bibinfo {author} {\bibfnamefont {Michael}\ \bibnamefont
  {Strickland}},\ }\bibfield  {title} {\enquote {\bibinfo {title} {{Collective
  modes of an anisotropic quark gluon plasma}},}\ }\href {\doibase
  10.1103/PhysRevD.68.036004} {\bibfield  {journal} {\bibinfo  {journal} {Phys.
  Rev. D}\ }\textbf {\bibinfo {volume} {68}},\ \bibinfo {pages} {036004}
  (\bibinfo {year} {2003})},\ \Eprint {http://arxiv.org/abs/hep-ph/0304092}
  {arXiv:hep-ph/0304092} \BibitemShut {NoStop}%
\bibitem [{\citenamefont {Iancu}\ \emph {et~al.}(2018)\citenamefont {Iancu},
  \citenamefont {Taels},\ and\ \citenamefont {Wu}}]{Iancu:2018trm}%
  \BibitemOpen
  \bibfield  {author} {\bibinfo {author} {\bibfnamefont {Edmond}\ \bibnamefont
  {Iancu}}, \bibinfo {author} {\bibfnamefont {Pieter}\ \bibnamefont {Taels}}, \
  and\ \bibinfo {author} {\bibfnamefont {Bin}\ \bibnamefont {Wu}},\ }\bibfield
  {title} {\enquote {\bibinfo {title} {{Jet quenching parameter in an expanding
  QCD plasma}},}\ }\href {\doibase 10.1016/j.physletb.2018.10.007} {\bibfield
  {journal} {\bibinfo  {journal} {Phys. Lett. B}\ }\textbf {\bibinfo {volume}
  {786}},\ \bibinfo {pages} {288--295} (\bibinfo {year} {2018})},\ \Eprint
  {http://arxiv.org/abs/1806.07177} {arXiv:1806.07177 [hep-ph]} \BibitemShut
  {NoStop}%
\bibitem [{\citenamefont {Qin}\ and\ \citenamefont
  {Majumder}(2010)}]{Qin:2009gw}%
  \BibitemOpen
  \bibfield  {author} {\bibinfo {author} {\bibfnamefont {Guang-You}\
  \bibnamefont {Qin}}\ and\ \bibinfo {author} {\bibfnamefont {Abhijit}\
  \bibnamefont {Majumder}},\ }\bibfield  {title} {\enquote {\bibinfo {title}
  {{A pQCD-based description of heavy and light flavor jet quenching}},}\
  }\href {\doibase 10.1103/PhysRevLett.105.262301} {\bibfield  {journal}
  {\bibinfo  {journal} {Phys. Rev. Lett.}\ }\textbf {\bibinfo {volume} {105}},\
  \bibinfo {pages} {262301} (\bibinfo {year} {2010})},\ \Eprint
  {http://arxiv.org/abs/0910.3016} {arXiv:0910.3016 [hep-ph]} \BibitemShut
  {NoStop}%
\bibitem [{\citenamefont {Xu}\ \emph {et~al.}(2014)\citenamefont {Xu},
  \citenamefont {Buzzatti},\ and\ \citenamefont {Gyulassy}}]{Xu:2014ica}%
  \BibitemOpen
  \bibfield  {author} {\bibinfo {author} {\bibfnamefont {Jiechen}\ \bibnamefont
  {Xu}}, \bibinfo {author} {\bibfnamefont {Alessandro}\ \bibnamefont
  {Buzzatti}}, \ and\ \bibinfo {author} {\bibfnamefont {Miklos}\ \bibnamefont
  {Gyulassy}},\ }\bibfield  {title} {\enquote {\bibinfo {title} {{Azimuthal jet
  flavor tomography with CUJET2.0 of nuclear collisions at RHIC and LHC}},}\
  }\href {\doibase 10.1007/JHEP08(2014)063} {\bibfield  {journal} {\bibinfo
  {journal} {JHEP}\ }\textbf {\bibinfo {volume} {08}},\ \bibinfo {pages} {063}
  (\bibinfo {year} {2014})},\ \Eprint {http://arxiv.org/abs/1402.2956}
  {arXiv:1402.2956 [hep-ph]} \BibitemShut {NoStop}%
\bibitem [{\citenamefont {Kumar}\ \emph {et~al.}(2023)\citenamefont {Kumar}
  \emph {et~al.}}]{JETSCAPE:2022jer}%
  \BibitemOpen
  \bibfield  {author} {\bibinfo {author} {\bibfnamefont {A.}~\bibnamefont
  {Kumar}} \emph {et~al.} (\bibinfo {collaboration} {JETSCAPE}),\ }\bibfield
  {title} {\enquote {\bibinfo {title} {{Inclusive jet and hadron suppression in
  a multistage approach}},}\ }\href {\doibase 10.1103/PhysRevC.107.034911}
  {\bibfield  {journal} {\bibinfo  {journal} {Phys. Rev. C}\ }\textbf {\bibinfo
  {volume} {107}},\ \bibinfo {pages} {034911} (\bibinfo {year} {2023})},\
  \Eprint {http://arxiv.org/abs/2204.01163} {arXiv:2204.01163 [hep-ph]}
  \BibitemShut {NoStop}%
\bibitem [{\citenamefont {Arnold}\ and\ \citenamefont
  {Dogan}(2008)}]{Arnold:2008zu}%
  \BibitemOpen
  \bibfield  {author} {\bibinfo {author} {\bibfnamefont {Peter~Brockway}\
  \bibnamefont {Arnold}}\ and\ \bibinfo {author} {\bibfnamefont {Caglar}\
  \bibnamefont {Dogan}},\ }\bibfield  {title} {\enquote {\bibinfo {title} {{QCD
  Splitting/Joining Functions at Finite Temperature in the Deep LPM Regime}},}\
  }\href {\doibase 10.1103/PhysRevD.78.065008} {\bibfield  {journal} {\bibinfo
  {journal} {Phys. Rev. D}\ }\textbf {\bibinfo {volume} {78}},\ \bibinfo
  {pages} {065008} (\bibinfo {year} {2008})},\ \Eprint
  {http://arxiv.org/abs/0804.3359} {arXiv:0804.3359 [hep-ph]} \BibitemShut
  {NoStop}%
\bibitem [{\citenamefont {Lindenbauer}(2023)}]{lindenbauer_2023_10419945}%
  \BibitemOpen
  \bibfield  {author} {\bibinfo {author} {\bibfnamefont {Florian}\ \bibnamefont
  {Lindenbauer}},\ }\href {\doibase 10.5281/zenodo.10419945} {\enquote
  {\bibinfo {title} {Data and analysis code for arxiv:2312.00447},}\ }
  (\bibinfo {year} {2023}),\ \bibinfo {note}
  {\url{https://doi.org/10.5281/zenodo.10419945}}\BibitemShut {NoStop}%
\bibitem [{\citenamefont {Virtanen}\ \emph {et~al.}(2020)\citenamefont
  {Virtanen} \emph {et~al.}}]{Virtanen:2019joe}%
  \BibitemOpen
  \bibfield  {author} {\bibinfo {author} {\bibfnamefont {Pauli}\ \bibnamefont
  {Virtanen}} \emph {et~al.},\ }\bibfield  {title} {\enquote {\bibinfo {title}
  {{SciPy 1.0--Fundamental Algorithms for Scientific Computing in Python}},}\
  }\href {\doibase 10.1038/s41592-019-0686-2} {\bibfield  {journal} {\bibinfo
  {journal} {Nature Meth.}\ }\textbf {\bibinfo {volume} {17}},\ \bibinfo
  {pages} {261} (\bibinfo {year} {2020})},\ \Eprint
  {http://arxiv.org/abs/1907.10121} {arXiv:1907.10121 [cs.MS]} \BibitemShut
  {NoStop}%
\bibitem [{\citenamefont {Hauksson}\ and\ \citenamefont
  {Gale}(2024)}]{Hauksson:2023dwh}%
  \BibitemOpen
  \bibfield  {author} {\bibinfo {author} {\bibfnamefont {Sigtryggur}\
  \bibnamefont {Hauksson}}\ and\ \bibinfo {author} {\bibfnamefont {Charles}\
  \bibnamefont {Gale}},\ }\bibfield  {title} {\enquote {\bibinfo {title}
  {{Polarized photons from the early stages of relativistic heavy-ion
  collisions}},}\ }\href {\doibase 10.1103/PhysRevC.109.034902} {\bibfield
  {journal} {\bibinfo  {journal} {Phys. Rev. C}\ }\textbf {\bibinfo {volume}
  {109}},\ \bibinfo {pages} {034902} (\bibinfo {year} {2024})},\ \Eprint
  {http://arxiv.org/abs/2306.10307} {arXiv:2306.10307 [nucl-th]} \BibitemShut
  {NoStop}%
\bibitem [{\citenamefont {Zhao}\ \emph {et~al.}(2023)\citenamefont {Zhao},
  \citenamefont {Qiu}, \citenamefont {Guo},\ and\ \citenamefont
  {Strickland}}]{Zhao:2023mrz}%
  \BibitemOpen
  \bibfield  {author} {\bibinfo {author} {\bibfnamefont {Ruizhe}\ \bibnamefont
  {Zhao}}, \bibinfo {author} {\bibfnamefont {Luhua}\ \bibnamefont {Qiu}},
  \bibinfo {author} {\bibfnamefont {Yun}\ \bibnamefont {Guo}}, \ and\ \bibinfo
  {author} {\bibfnamefont {Michael}\ \bibnamefont {Strickland}},\ }\bibfield
  {title} {\enquote {\bibinfo {title} {{Collective modes of a collisional
  anisotropic quark-gluon plasma}},}\ }\href {\doibase
  10.1103/PhysRevD.108.034023} {\bibfield  {journal} {\bibinfo  {journal}
  {Phys. Rev. D}\ }\textbf {\bibinfo {volume} {108}},\ \bibinfo {pages}
  {034023} (\bibinfo {year} {2023})},\ \Eprint
  {http://arxiv.org/abs/2306.12851} {arXiv:2306.12851 [hep-ph]} \BibitemShut
  {NoStop}%
\bibitem [{\citenamefont {Bellac}(2011)}]{Bellac:2011kqa}%
  \BibitemOpen
  \bibfield  {author} {\bibinfo {author} {\bibfnamefont {Michel~Le}\
  \bibnamefont {Bellac}},\ }\href {\doibase 10.1017/CBO9780511721700} {\emph
  {\bibinfo {title} {{Thermal Field Theory}}}},\ Cambridge Monographs on
  Mathematical Physics\ (\bibinfo  {publisher} {Cambridge University Press},\
  \bibinfo {year} {2011})\BibitemShut {NoStop}%
\bibitem [{\citenamefont {Ghiglieri}\ \emph {et~al.}(2020)\citenamefont
  {Ghiglieri}, \citenamefont {Kurkela}, \citenamefont {Strickland},\ and\
  \citenamefont {Vuorinen}}]{Ghiglieri:2020dpq}%
  \BibitemOpen
  \bibfield  {author} {\bibinfo {author} {\bibfnamefont {Jacopo}\ \bibnamefont
  {Ghiglieri}}, \bibinfo {author} {\bibfnamefont {Aleksi}\ \bibnamefont
  {Kurkela}}, \bibinfo {author} {\bibfnamefont {Michael}\ \bibnamefont
  {Strickland}}, \ and\ \bibinfo {author} {\bibfnamefont {Aleksi}\ \bibnamefont
  {Vuorinen}},\ }\bibfield  {title} {\enquote {\bibinfo {title} {{Perturbative
  Thermal QCD: Formalism and Applications}},}\ }\href {\doibase
  10.1016/j.physrep.2020.07.004} {\bibfield  {journal} {\bibinfo  {journal}
  {Phys. Rept.}\ }\textbf {\bibinfo {volume} {880}},\ \bibinfo {pages} {1--73}
  (\bibinfo {year} {2020})},\ \Eprint {http://arxiv.org/abs/2002.10188}
  {arXiv:2002.10188 [hep-ph]} \BibitemShut {NoStop}%
\bibitem [{\citenamefont {Kurkela}\ and\ \citenamefont
  {Lindenbauer}(2023)}]{kurkela_2023_10409474}%
  \BibitemOpen
  \bibfield  {author} {\bibinfo {author} {\bibfnamefont {Aleksi}\ \bibnamefont
  {Kurkela}}\ and\ \bibinfo {author} {\bibfnamefont {Florian}\ \bibnamefont
  {Lindenbauer}},\ }\href {\doibase 10.5281/zenodo.10409474} {\enquote
  {\bibinfo {title} {{EKTqhat - Effective kinetic theory solver with jet
  quenching parameter}},}\ } (\bibinfo {year} {2023}),\ \bibinfo {note}
  {\url{https://doi.org/10.5281/zenodo.10409474}}\BibitemShut {NoStop}%
\end{thebibliography}%

\end{document}